\documentclass[aps,prx,twocolumn,floatfix,longbibliography,nofootinbib,superscriptaddress]{revtex4-2}

\usepackage[utf8]{inputenc}
\usepackage{natbib}
\usepackage{graphicx}
\usepackage{tikz}
\usepackage{xcolor}
\usetikzlibrary{calc}
\usepackage[tbtags]{amsmath}

\definecolor{tiffanyblue}{rgb}{0.04, 0.73, 0.71}
\definecolor{charcoal}{rgb}{0.21, 0.27, 0.31}
\definecolor{burgundy}{rgb}{0.5, 0.0, 0.13}
\definecolor{blue(munsell)}{rgb}{0.0, 0.5, 0.69}
\definecolor{britishracinggreen}{rgb}{0.0, 0.26, 0.15}
\definecolor{dogwoodrose}{rgb}{0.84, 0.09, 0.41}
\definecolor{electricpurple}{rgb}{0.75, 0.0, 1.0}
\definecolor{electricindigo}{rgb}{0.44, 0.0, 1.0}
\definecolor{blue(ncs)}{rgb}{0.0, 0.53, 0.74}
\definecolor{brightmaroon}{rgb}{0.76, 0.13, 0.28}
\definecolor{folly}{rgb}{1.0, 0.0, 0.31}

\usepackage[colorlinks, linkcolor=tiffanyblue]{hyperref}
\hypersetup{colorlinks,allcolors=blue}
\usepackage{amssymb}
\usepackage{amsmath}
\usepackage{gensymb}
\usepackage{float}
\usepackage{amsmath}
\usepackage{tabularx,graphicx}
\usepackage{epstopdf}
\usepackage{latexsym}
\usepackage{color, colortbl}
\usepackage{psfrag}
\usepackage{bbm}
\usepackage{bm}
\usepackage{titlesec}
\usepackage{dsfont}
\usepackage{feynmp}
\usepackage{slashed}
\usepackage{multirow}
\usepackage[normalem]{ulem}
\usepackage{booktabs}

\renewcommand{\vec}[1]{\boldsymbol{#1}}
\newcommand{\bcen}{\begin{center}}
	\newcommand{\ecen}{\end{center}}
\newcommand{\btab}{\begin{tabular}}
	\newcommand{\etab}{\end{tabular}}
\newcommand{\bdes}{\begin{description}}
	\newcommand{\edes}{\end{description}}

\newcommand{\ul}{\underline}
\newcommand{\beq}{\begin{equation}}
	\newcommand{\eeq}{\end{equation}}
\newcommand{\bea}{\begin{eqnarray}}
	\newcommand{\eea}{\end{eqnarray}}

\newcommand{\half}{\frac{1}{2}}
\newcommand{\bary}{\begin{array}}
	\newcommand{\eary}{\end{array}}
\newcommand{\benum}{\begin{enumerate}}
	\newcommand{\eenum}{\end{enumerate}}
\newcommand{\bitem}{\begin{itemize}}
	\newcommand{\eitem}{\end{itemize}}

\newcommand{\bfig}{\begin{figure}}
	\newcommand{\efig}{\end{figure}}
\renewcommand{\vec}[1]{\boldsymbol{#1}}
\newcommand{\br} { \boldsymbol{r}}

\newcommand{\bzero} { {\boldsymbol{0}}}

\newcommand{\mQ} {{\bf Q}}
\newcommand{\mQp} {{\bf Q'}}

\newcommand{\bra}[1]{{\langle #1 |}}
\newcommand{\ket}[1]{| #1 \rangle}

\newcommand{\eqn}[1] {eqn.~(\ref{#1})}

\newcommand{\sect}[1] {Section~\ref{#1}}
\newcommand{\Sect}[1] {Section~\ref{#1}}
\newcommand{\fig}[1]{fig.~\ref{#1}}
\newcommand{\Fig}[1]{Fig.~\ref{#1}}

\makeatletter

\newcommand{\Rmnum}[1]{\expandafter\@slowromancap\romannumeral #1@}
\makeatother

\newcommand{\1}{1\!\!1}
\newcommand{\SP}[1]

\newcommand{\mycirc}[1]{\tikz[remember picture, baseline=(a1.base)]{\node[circle, thick, draw=electricindigo, inner sep=1.2pt] (a1) {#1};}}

\newcommand{\mycirct}[1]{\tikz[remember picture, baseline=(a3.base)]{\node[circle, thick, draw=blue(ncs), inner sep=1.2pt] (a3) {#1};}}

\newcommand{\mycircs}[1]{\tikz[remember picture, baseline=(a2.base)]{\node[circle, thick, draw=dogwoodrose, inner sep=1.2pt] (a2) {#1};}}

\begin{document}

\title{Symmetry-protected topological wire in a topological vacuum}

\author{Subrata Pachhal}
\email{pachhal@iitk.ac.in}
\affiliation{Department of Physics, Indian Institute of Technology Kanpur, Kalyanpur, UP 208016, India}

\author{Adhip Agarwala}
\email{adhip@iitk.ac.in}
\affiliation{Department of Physics, Indian Institute of Technology Kanpur, Kalyanpur, UP 208016, India}

\begin{abstract}

Symmetry-protected topological phases host gapless modes at their boundary with a featureless environment of the same dimension or a trivial vacuum. In this study, we explore their behavior in a higher-dimensional environment, which itself is non-trivial - a {\it topological vacuum}. In particular, we embed a one-dimensional topological wire within a two-dimensional Chern insulator, allowing the zero-dimensional edge modes of the wire to interplay with the surrounding chiral boundary states created by the environment. In contrast to a trivial vacuum, we show that depending on the nature of low-energy modes, the topology of the environment selectively influences the topological phase transitions of the wire. Interestingly, such selectivity leads to scenarios where the environment trivializes the wire and even induces topological character in an otherwise trivial phase - an example of `proximity-induced topology'.  Using both numerical and analytical approaches, we establish the general framework of such embedding and uncover the role of symmetries in shaping the fate of low-energy theories. Our findings will provide a deeper understanding of heterostructural topological systems, paving the way for their experimental exploration.

\end{abstract}

\maketitle

\section{INTRODUCTION}\label{sec_intro}
Study of gapless boundary modes of free-fermionic symmetry-protected topological (SPT) phases across various dimensions has emerged as one of the cornerstones in modern condensed matter research \cite{moore_tibirth_2010, hasan_topo_2010, Ryu_topocls_2010, qi_toposc_2011, shen_tibook_2012, bernevig_topobook_2013}. These unique quantum states arise within the bulk gap of topological insulators/superconductors, where they are protected and characterized by underlying discrete symmetries of the system \cite{Altland_PRB_1997, Kitaevtenfold, chiu_topoclass_2016, Ludwig_topo_2016, Agarwala_AOP_2017, Jorrit_topoclass_2017}. Their inherent robustness against local disturbances is particularly significant \cite{ halperin_roedge_1982,  xu_robust_2006}, making them promising candidates for the development of topological quantum computing \cite{kitaev_fault-tolerant_2003, nyak_topocomp_2008, pachos_introduction_2012}. The pursuit to understand these states and to pave the way for their application in quantum technology has sparked extensive theoretical \cite{wu_em_2006, Gurarie_PRB_2007, fu_mbs_2008, teo_sm_2008, Roy_PRL_2010, teo_mbs_2010, Potter_multichannel_2010, beenakker_mbs_2013, Thakurathi_MBS_2014, Thakurathi_mbs_2015, yazdani_mzmhunt_2023} and experimental \cite{misra_em_2002, hsieh_topological_2008, bernevig_nature_2017, bernevig_topmat_database, kanungo2022realizing, meier_2016, markus_hgte_2007, Alpichshev_bite_2010, roushan2009topological} investigations over the last decade. 

Although these exotic states are relatively well understood when the topological phase is in a featureless environment of the same dimension \cite{shen_tibook_2012, bernevig_topobook_2013}, recent interests have surged to realize them in heterostructure interfaces combining different dimensional subsystems \cite{teo_topodefect_2010, choy_majorana_2011, nadj-perge_proposal_2013, Soori_TISC_2013, sgiozaki_topo_2014, heimis_shiba_2014, rainis_nano_2014, jana_nano_2019, tuegel_embedded_2019, SLAGER_tbi_2019, Deb_nano_2021, velury_ts_2022, adak_tisc_2022, Saxena_nano_2022, pritam_prb_2023, nyari_topological_2023, laszloffy_topological_2023, harry_interface_2024, mondal_amzm_2024, bhowmik_fflo_2025}. A rich set of phenomena associated with various types of lower-dimensional impurities, such as vortices, defects, dislocation, disclinations, and embedded magnetic atoms on insulators/superconductors, has been predicted theoretically \cite{wnag_interface_2004, ran_em_2009, Wimmer_PRL_2010, shiozaki_gf_2012, Potter_mzmsurface_2012, Vladimir_piflux_2012, Zhang_scnanowire_2013, slager_dislocation_2014, waldimir_disc_2014, slager_gf_2015,  paulose_topo_2015, peng_mbs_2015, Sablikov_nomag_2015, neupart_shiba_2016,  Poyhonen_topo_2016, sahlberg_1dtopo_2017, Zhang_PRX_2021, kaladzhyan_tft_2018, sticlet_topological_2019, bena_imp_2019, Li_disc_2020, sedlmayr_shiba_2021, carrol_I_2021, carrol_II_2021, Miao_afm_2023, chakraborty_mbs_2024, Felix_PRB_2013, Felix_PRB_2014, Vardan_PRB_2017, braunecker_interplay_2013, vazifeh_self-organized_2013, klinovaja_topological_2013}. A significant amount of experimental efforts have been dedicated to the fabrication and study of these heterostructures, with some revealing promising indications of topological features \cite{das_zbp_2012, finck_zbp_2013, mourik_nanowire_2012, nadj-perge_observation_2014, Xu_ptsc_2014, shoman2015topological, ruby_end_2015, menard_coherent_2015, pawlak_probing_2016, feldman_high-resolution_2017, ruby_exploring_2017, wang_1dmajorana_2020, schneider_precursors_2022, liebhaber_quantum_2022,mesaros2024topologically}.

On the theoretical front, a complete classification of defects and their gapless modes in topological systems was comprehensively done in \cite{teo_topodefect_2010}. In \cite{tuegel_embedded_2019}, the idea of `embedded topological insulators' was introduced, where the role of embedding an SPT system in a trivial environment was studied. Wimmer et.~al.~\cite{Wimmer_PRL_2010} showed that even an electrostatic defect in a p-wave superconductor can host Majorana edge modes. Since then, a flurry of activity has been focused on realizing such protected qubits on defects in topological superconducting systems \cite{Sau_nanow_2013, Maska_rashba_2017, Poyhonen_topo_2016, sahlberg_1dtopo_2017, Zhang_PRX_2021, Legg_mzm_2021}. However, most of these studies deal with heterostructures where either the environment or the embedded subsystem is topologically trivial. The scenario when both of them are topological, allowing their corresponding boundary modes to interplay with each other in a composite system, has been little explored \cite{Sedlmayr_JPCM_2022}. Interestingly, the authors in \cite{Sedlmayr_JPCM_2022} had found that a topological superconducting environment can often gap out the Majorana edge modes of an embedded impurity chain.

\begin{figure}
	\centering \includegraphics[width=0.95\linewidth]{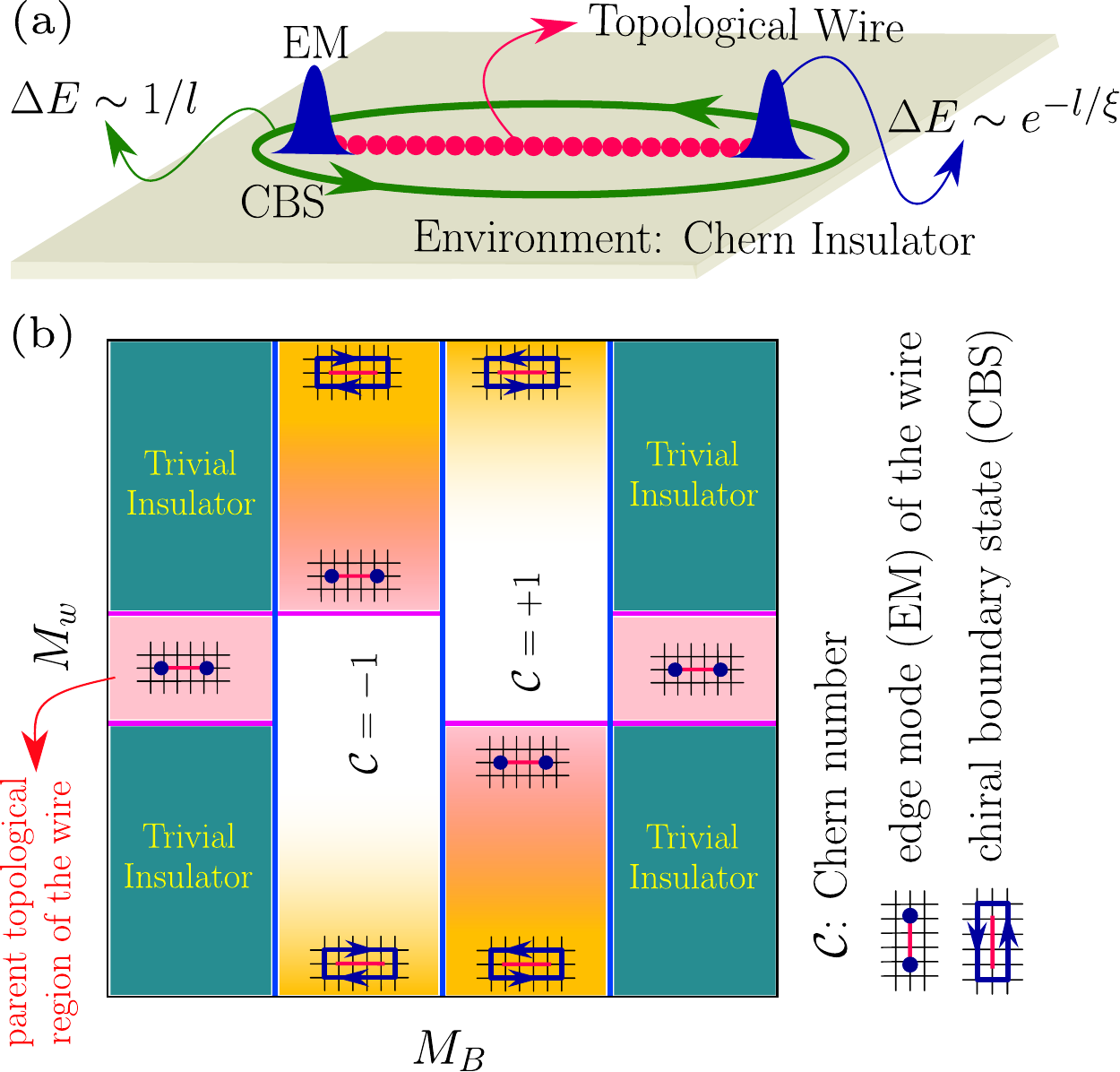}
	\caption{(a) Schematic picture of a topological wire embedded in a Chern insulating environment. The wire hosts edge modes (EM) with a gap $\Delta E$ exponentially suppressed in length ($l$) of the wire, while the Chern insulator supports chiral boundary states (CBS) around the wire with a linearly suppressed gap. (b) Schematic phase diagram with $M_w$ and $M_B$ as the microscopic parameters for the wire and the environment, respectively. {\it Vertical} phase boundaries indicate the topological criticality of the Chern insulator, while the {\it horizontal} boundaries represent the same for the wire. In a trivial environment, the wire maintains its parent topological properties. When the environment is topological ($\mathcal{C}=\pm1$), the wire exhibits features like selective disappearance of criticality, consequently losing its parent topology and environment-induced EM in an otherwise trivial regime.}
	\label{fig1}
\end{figure}

In this work, we ask, what happens when two different dimensional insulators, each exhibiting an SPT phase in its respective {\it parent} dimension, are coupled together? And what roles do the symmetries play in such systems? In particular, our study delves into the fascinating interplay between a topological wire akin to the Su-Schrieffer-Heeger (SSH) model \cite{SSH_PRB_1980} and a two-dimensional Chern insulating environment when coupled together. In general, they will act as boundaries to each other, resulting in the emergence of different dimensional gapless boundary modes in the same composite system: zero-dimensional edge mode (EM) of the wire and one-dimensional chiral boundary states (CBS) hosted by the Chern insulator (CI), schematically illustrated in \Fig{fig1}(a). These low energy modes are characterized by their gap ($\Delta E$) scaling with system size ($l$); for instance, while CBS follows $\Delta E \sim 1/l$, EM shows $\Delta E \sim \exp (-l/\xi)$ character, where $\xi$ is localization length of the system \cite{shen_tibook_2012, bernevig_topobook_2013}. Since both of them arise in the spectrum near the Fermi energy, they can often lead to non-intuitive phenomena when coupled to each other. To understand the fate of these boundary modes in such composite systems, we employ numerical exact diagonalization and find the complete phase diagram in the ($M_B$ - $M_w$) parameter space, schematically represented in \Fig{fig1}(b), where $M_B$ and $M_w$ are the respective microscopic parameter of the environment and the wire. In a trivial environment, we find that the wire retains its parent topological phase and phase transitions via bulk gap closing and reopening (topological quantum critical point - TQCP), similar to an embedded topological insulator \cite{tuegel_embedded_2019}. In contrast, when the environment has non-trivial topology characterized by Chern number $\mathcal{C}=\pm1$, some remarkable features emerge in the system, such as: (1) the topological criticalities of the wire are selectively susceptible to the topological character of the environment. For example, depending on the $\mathcal{C}$, particular TQCPs of the wire gap out. This also results in (2) a `sub-system metal' (i.e., the stable TQCP) protected by symmetries. Furthermore, (3) the topological environment washes out the parent topology of wire and induces topological character in an otherwise trivial phase - an example of `proximity-induced topology.' The parameter regime this happens is again dependent on the $\mathcal{C}$ of the environment. (4) We further show that these features are entirely symmetry-restricted and, therefore, can be generalized. We uncover the governing physics behind this process of `symmetry-protected embedding'. As an illustration of this generality, we discuss another microscopic system where the wire is now embedded in CI but on a different lattice. (5) Moreover, we show how, by selectively breaking the symmetries of the environment, we can tune the low-energy nature of these systems.  In this work, we develop a comprehensive, low-energy theory to unravel these distinctive features emerging from the interplay between multi-dimensional edge modes.

The article is organized as follows: in \sect{sec_ham}, we describe the Hamiltonian and symmetries of the composite system of study, where a topological wire is coupled to a Chern insulating environment. The phases, phase transitions, and their features observed in the numerical analysis of such systems have been discussed in \sect{sec_numerics}. To understand these numerical results, we formulate an effective low-energy theory of coupling between different dimensional subsystems in \sect{sec_edge}. In \sect{sec_spe}, we uncover the symmetry-protection of exotic behaviors of the embedded wire. The effects of symmetry-broken environments are explored in \sect{sec_symmetry}. We discuss generalizations of our results across various symmetry classes and dimensionality in \sect{sec_dis} along with a few possible scopes of experimentally realizing such composite systems. Finally, we conclude our investigation in \sect{sec_summ} by summarizing our findings and providing some interesting outlook for future studies. Computational and analytical details are pointed out in the corresponding Appendices.

\section{MODEL AND SYMMETRIES}\label{sec_ham}
Our system of interest comprises a one-dimensional topological wire coupled to a two-dimensional Chern insulating environment, as depicted in~\Fig{fig1}(a). First, we consider spinless fermions residing on a square lattice, represented by the annihilation and creation operators $c_{i\alpha}$ and $c^\dagger_{i\alpha}$ at position $\{x_i,y_i\}$, where $\alpha \equiv$ $A, B$ denotes the orbital degree of freedom. These fermions hop between nearest neighbors, governed by the Hamiltonian:
\begin{equation}
	H =  \sum_{i, \vec{\delta}}\Big( \Psi^\dagger_{i} T_{\vec{\delta}}\Psi_{i+\vec{\delta}} + \text{h.c.}  \Big)+ \sum_i \Psi^\dagger_i \Gamma_i \Psi_i,
	\label{eq_rham}
\end{equation}
where $\Psi_i = (c_{iA}~~ c_{iB} )^T$ forms the real space basis and $\vec{\delta}=(\hat{x}$, $\hat{y})$ denotes the unit vectors along $x$ and $y$ directions. The associated hopping matrices are $T_{\hat{x}}=-\half \left(\sigma_{z}+i\sigma_{x}\right)$ and $T_{\hat{y}}=-\half\left(\sigma_{z}+i\sigma_{y}\right)$, while the term
$\Gamma_i=(2-M_i)\sigma_{z}$ accounts for the orbital energy splitting at each site, with $\sigma_{x/y/z}$ representing the Pauli matrices. 

When the {\it mass} parameter of the Hamiltonian $M_i$ is uniform throughout the lattice ($M_i=M_B ~\forall i$), the system reduces to the pristine Qi-Wu-Zhang (QWZ)
model of spinless fermions described by the momentum space Hamiltonian \cite{BHZ_Sci_2006, QWZ_spinlessBHZ_2006},
\begin{align}
	\mathcal{H}_{\text{QWZ}}(\vec{k}) = &~ \sin k_x \sigma_x + \sin k_y \sigma_y \nonumber \\  & +  \left(2-M_B-\cos k_x-\cos k_y\right)\sigma_z.\label{eq_clsDHam}
\end{align}
The model given in \eqn{eq_clsDHam} lacks time-reversal symmetry (TRS) but possesses charge-conjugation symmetry (CGS) along with parity implemented by the following operations, 
\begin{align}
	\text{CGS} &: \sigma_x \mathcal{H}^{*}_{\text{QWZ}}(\vec{k}) \sigma_x = -\mathcal{H}_{\text{QWZ}}(-\vec{k}),\nonumber \\
	\text{Parity} &: \sigma_z \mathcal{H}_{\text{QWZ}}(\vec{k}) \sigma_z = \mathcal{H}_{\text{QWZ}}(-\vec{k}),\label{eq_cgsp}
\end{align}
restricting it to symmetry class D of the free-fermion tenfold classification~\cite{Altland_PRB_1997, Agarwala_AOP_2017}. The system supports two distinct topologically non-trivial phases characterized by Chern number $\mathcal{C}=-1$ for $0<M_B<2$ and $\mathcal{C}=+1$ for $2<M_B<4$, separated by Dirac-like topological quantum critical points (TQCPs) at $M_B = 0, 2, 4$ (see Appendix~\ref{apndx_bhz1} for the details about this model). In general, a non-trivial Chern number of the bulk band represents CBS on open boundaries and quantized anomalous Hall response \cite{He_QAH_2013, Liu_QAH_2016}.

\begin{figure}
	\includegraphics[width=0.8\columnwidth]{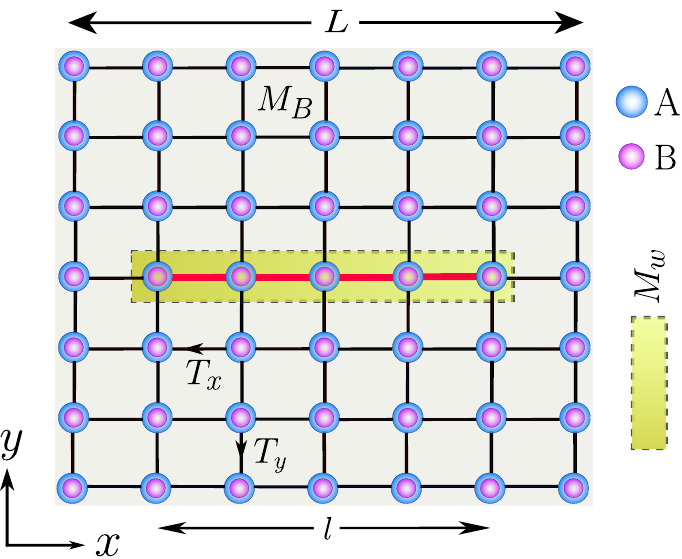}
	\caption{Schematic diagram of the hopping Hamiltonian in \eqn{eq_rham}. $T_x$ and $T_y$ are the hopping matrices of the two-orbital (A and B) QWZ model with {\it mass} parameter $M_B$. While hopping strengths are uniform, a segment is created by using a different {\it mass} $M_w$ for $l$ sites connected by {\it red} (dark gray) bonds. This forms a topological wire ({\it boxed} region) of length $l$ in a Chern insulating environment of linear size $L$.}
	\label{fig2}
\end{figure}

Now, within this Chern insulating substrate, we construct a line segment where the {\it mass} $M_i$ differs from its surrounding environment, thus creating an open wire-like system mimicking the one-dimensional counterpart of the QWZ model itself, which we refer to as the QWZ wire. In \Fig {fig2}, we schematically present such a system, which illustrates within a QWZ environment ($M_i=M_B$ on $(L^2-l)$ sites), a wire of length $l$ is created by putting a different {\it mass} $M_w$ (on $l$ sites). The Hamiltonian describing an isolated QWZ wire along $x$-direction reads,
\beq
\mathcal{H}_{\text{wire}}(k_x) = \sin k_x \sigma_x +  \left(2- M_w -\cos k_x\right)\sigma_z, \label{eq_wire}
\eeq
which exhibits a non-zero winding number in the parameter range $1<M_w<3$ with TQCPs at $M_w = 1, 3$ akin to the paradigmatic Su-Schrieffer-Heeger wire \cite{SSH_PRB_1980}. While the Chern insulating QWZ Hamiltonian is in class D, an isolated QWZ wire retains all three symmetries: TRS, CGS, and sub-lattice symmetry (SLS; also known as chiral symmetry), along with parity, as demonstrated by the following relations:
\begin{align}
	\text{TRS} &: \sigma_z \mathcal{H}^{*}_{\text{wire}}(k_x) \sigma_z = \mathcal{H}_{\text{wire}}(-k_x),\nonumber \\
	\text{CGS} &: \sigma_x \mathcal{H}^{*}_{\text{wire}}(k_x) \sigma_x = -\mathcal{H}_{\text{wire}}(-k_x),\nonumber \\
	\text{SLS} &: \sigma_y \mathcal{H}_{\text{wire}}(k_x) \sigma_y = -\mathcal{H}_{\text{wire}}(k_x),\nonumber \\
	\text{Parity} &: \sigma_z \mathcal{H}_{\text{wire}}(k_x) \sigma_z = \mathcal{H}_{\text{wire}}(-k_x).\label{eq_tcsp}
\end{align}
Thus, an isolated QWZ wire belongs to the BDI symmetry class. The phase diagram and corresponding topologically protected zero-energy edge modes of this model have been discussed in Appendix~\ref{apndx_bhz2}.

The composite system of these two Hamiltonians, as shown in \Fig{fig2} and described in \eqn{eq_rham}, manifests a topological wire in a topological vacuum. The overall system inherits the lower symmetry class, i.e., class D, which can be verified by the real space charge-conjugation operation on $H$ given in \eqn{eq_rham}. Under the operation,
\beq
\Psi_i \rightarrow \sigma_x[\Psi_i^\dagger]^T   \qquad  \Psi^\dagger_i \rightarrow [\Psi_i]^T \sigma_x \label{eq_sxpsi}
\eeq
we find $H \rightarrow H$, confirming that the composite system retains CGS and indeed belongs to class D. Thus, for any parameter value, the eigenspectrum will have $E \rightarrow -E$ symmetry. Throughout this article, we always consider the half-filling of electrons such that the Fermi energy is pinned at $E_F=0$.

\begin{figure*}
	\includegraphics[width=0.95\linewidth]{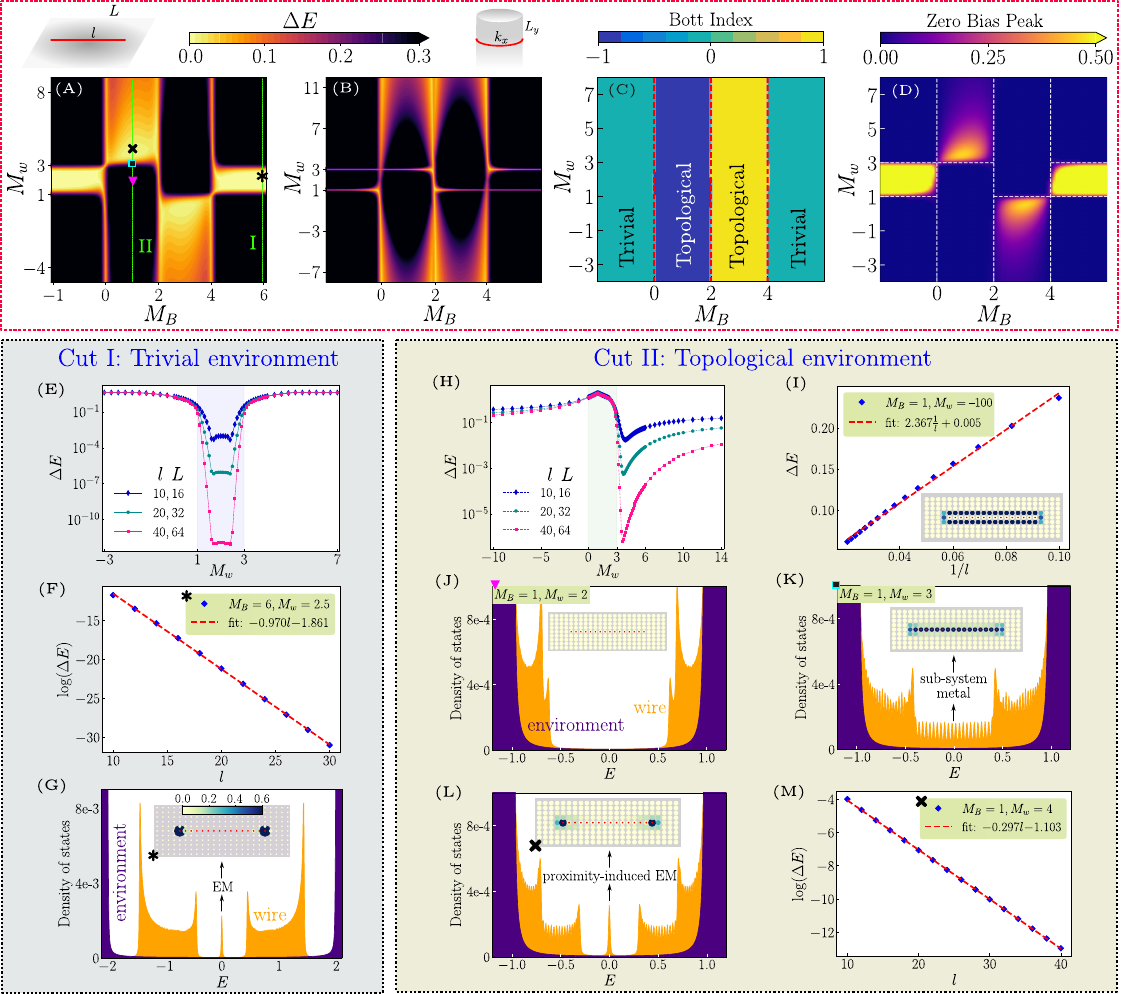}
	\caption{(A) Spectral gap ($\Delta E$) phase diagram of the composite system with $l=24$, $L=40$ (see schematic above) in the ($M_B$ - $M_w$) parameter space. The Cut I, II, and marked points are discussed below. (B) $\Delta E$ behavior in ribbon geometry where $k_x$ momenta are well defined and $y$-width is $L_y=80$ (see schematic above). (C) Bott index to characterize the two-dimensional topology of the composite system. (D) Zero-bias peak (LDOS at the wire's edges) signals EM in the wire. For both (C) and (D), $l=24$ and $L=40$. (E) Along Cut I ($M_B =6$; trivial environment), $\Delta E$ decreases as system size increases for $1<M_w<3$. (F) $\Delta E$ for $M_B=6, M_w=2.5$ ($*$ in (A)) shows exponential scaling with $l$. (G) DOS shows mid-gap states of the wire with LDOS on its two edges (see inset). (H) $\Delta E$ behavior along Cut II ($M_B =1$; topological environment with $\mathcal{C}=-1$) for different system sizes. At very large $|M_w|$ gap $\propto \frac{1}{l}$ as depicted in (I) for $M_w=100$. Inset shows LDOS of low energy states form a CBS. (J), (K) and (L) illustrate DOS for corresponding marked points in (A). In (J), the wire does not host EM, albeit being in its parent topological region. While (K) shows a one-dimensional metal within a CI, (L) depicts the proximity-induced EM in the trivial wire. In (M), we show their exponential scaling similar to parent EM as in (F).}
	\label{fig3}
\end{figure*}

\section{NUMERICAL ANALYSIS}\label{sec_numerics}
We explore the tight-binding Hamiltonian described in \eqn{eq_rham} using exact diagonalization and present our results in \Fig{fig3}. While performing numerical analysis, we always maintain periodic boundary conditions (PBC) on the CI environment (unless stated otherwise) to avoid any effects of the global outer edges of the composite system. We present the behavior of eigenspectra via spectral gap $\Delta E$, their scaling with system size, and the local density of states (LDOS; see Appendix~\ref{apndx_bhz1}), etc. 
in the ($M_B$ - $M_w$) parameter space. Note, $\Delta E$ is defined as the difference between the two closest energy eigenvalues near $E_F=0$. Further, we consider two cases (a) when $l<L$, i.e., a finite wire surrounded by the environment, and (b) ribbon geometry of width ($L_y$) where $k_x$ is a good quantum number.
In \Fig{fig3}(A) and~\ref{fig3}(B), $\Delta E$ is shown for a finite and a ribbon geometry, respectively. In \Fig{fig3}(C), we plot the Bott index (for details on Bott index, see Appendix~\ref{apndx_bott}) for the composite system, showing that the topological character of the two-dimensional environment remains the same as the pristine QWZ model. In \Fig{fig3}(D), we plot the LDOS at the two edges of the wire in the finite geometry. This is, in particular, accessible as a zero-bias peak in experiments \cite{yin_stminti_2021}. In our analysis, we consider both possible regimes: (i) trivial environment ($M_B>4$ or $M_B<0$) and (ii) topological environment ($0<M_B<4$).

\subsection{Trivial environment}\label{sec_trivial}
We first explore the behavior of $\Delta E$ at $M_B=6$ when the environment is trivial (see Cut I in \Fig{fig3}(A)). 
Between $1<M_w<3$, while $\Delta E \rightarrow 0$ in a finite geometry (\Fig{fig3}(A)), $\Delta E \neq 0$ in the ribbon geometry suggesting the existence of EM on the wire. Moreover, $\Delta E$ in this regime falls exponentially with increasing size of $l$ as seen in \Fig{fig3}(E) and~\ref{fig3}(F). The edge character of these is further borne out in the density of states (DOS) shown in \Fig{fig3}(G) and their LDOS (inset of \Fig{fig3}(G)). This is consistent also with the zero-bias peak (ZBP) as shown in \Fig{fig3}(D). Thus, when the topological wire resides in the trivial environment of the CI, it carries all the tell-tale signatures of the one-dimensional topological system as was discussed just below \eqn{eq_wire}.

\subsection{Topological environment}\label{sec_topological}
In order to study the effect of the topological environment on the wire, we now focus on $M_B=1$ (see Cut II in \Fig{fig3}(A)) where the environment has $\mathcal{C}=-1$. The behavior of $\Delta E$ with $M_w$ along Cut-II is shown in  \Fig{fig3}(H). In the extreme limit of $|M_w|$ (for e,g. $M_w=-100$), the wire essentially acts like a puncture in the system, leading to the formation of CBS on environment edges around the wire with characteristic $1/l$ gap scaling as can be seen in \Fig{fig3}(I).

As we reduce $M_w$ from its extreme value of $-100$, the first surprising observation 
lies at $M_w=1$, where the TQCP of the wire vanishes (see \Fig{fig3}(A) and~\ref{fig3}(B)). Moreover, within the range $1<M_w<3$, where the wire would have been in a topological phase if embedded in a trivial vacuum (as discussed in \sect{sec_trivial}), the presence of a non-trivial environment instead causes it to lose its inherent topological character. For a specific value $M_w=2$, \Fig{fig3}(J) shows gapped DOS along with no edge signature in LDOS. This can be seen in both the $\Delta E$ (see \Fig{fig3}(A),~\ref{fig3}(B)) and featureless ZBP (see \Fig{fig3}(D)) in this regime. Clearly, if the wire and the environment had been disjoint - each would have had its own set of boundary modes: zero-dimensional EM on the wire and one-dimensional CBS around the wire. Thus the physics of {\it hybridization between the multidimensional edge theories} gap each other out. Incidentally, at $M_w=3$, where $\Delta E \rightarrow 0$ in \Fig{fig3}(B), the DOS of the wire exhibits a one-dimensional metallic behavior (see \Fig{fig3}(K)) with LDOS showing electronic states all through the wire (inset of \Fig{fig3}(K)). This reflects that even in the presence of the topological environment, {\it at least one} of the TQCPs of the wire remains intact. The origin of this {\it sub-system metal}, its robustness to the environment, and its selectivity should again be a result of the interplay of boundary theories, as we will elaborate later.

Another surprising observation lies in the range of $3<M_w \lessapprox 5$ where the wire in a trivial vacuum would have been a trivial insulator, in the topological environment, it now hosts mid-gap states (see DOS in \Fig{fig3}(L) for $M_w=4$) which has LDOS peaked at the wire edges (inset of \Fig{fig3}(L)). These again vanish under PBC (\Fig{fig3}(B)), and have a signatures in ZBP (\Fig{fig3}(D)). Furthermore, their system size scaling with $l$ shows characteristic exponential fall as seen in \Fig{fig3}(M). The DOS, in fact, has a striking resemblance to that of a usual one-dimensional topological wire with two zero-energy modes. We term this observation as {\it proximity induced topology} where a $d$ dimensional topological environment can induce $(d-1)$ topological features in the subsystem. This is in contrast with Majorana systems where s-wave bulk superconductivity (trivial) induces p-wave superconductivity (topological) in a spin-orbit coupled nanowire under magnetic field \cite{Zhang_scnanowire_2013, das_zbp_2012, Potter_mzmsurface_2012, finck_zbp_2013, mourik_nanowire_2012, Xu_ptsc_2014, nadj-perge_observation_2014} or where a $d$ dimensional topological system can induce topological character in another $d$ dimensional system in proximity \cite{Hsieh_bpte_2016, Cheng_proximity_2019, khanna_pisc_2014, panas_proximity_2020, CIM_proximity}.

As $M_w$ is further increased $M_w \gtrapprox 5$, the proximity-induced edge modes give way to the CBS states around the whole wire since the wire again starts behaving as a boundary for the environment. Interestingly, the above phenomenology switches in the $\mathcal{C} = 1$ phase ($2<M_B<4$) where now the sub-system metal resides at $M_w=1$ and the proximity-induced EM appears in the range of $-1 \lessapprox M_w<1$ (see \Fig{fig3}(A-D)). With these observations, we will now describe the origin and understanding of these phenomena in the next few sections.

\section{UNDERSTANDING THE COMPOSITE SYSTEM}\label{sec_edge}

\begin{figure*}
	\centering	\includegraphics[width=1\linewidth]{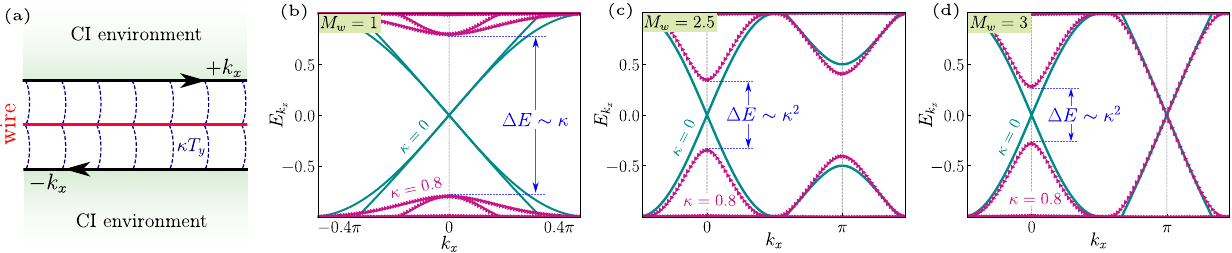}
	\caption{(a) Coupling between the wire and the environment (with $\mathcal{C}=-1$) in ribbon geometry, via $y$-directional hopping $T_y$ tuned by coupling parameter $\kappa$ (in this case $\kappa_1 = \kappa_2=\kappa$). Considering $M_B=1$, (b), (c), and (d) illustrate the low energy band structures at $M_w=1, 2.5$, and $3$, respectively, for both the isolated case ($\kappa=0$, {\it solid} line) and with finite coupling ($\kappa\neq0$, {\it dotted} line). For $M_w=1$, (b) shows that the CBS and the wire at $k_x=0$, mutually gap out with $\Delta E \sim \kappa$. (c) At $M_w=2.5$, while the wire (at $k_x=\pi$) remains gapped, the CBS (at $k_x=0$) gets gapped out as $\Delta E \sim \kappa^2$. (d) At $M_w=3$, while the CBS (at $k_x=0$) gaps out as $\Delta E \sim \kappa^2$, the wire (at $k_x =\pi$) remains gapless.}
	\label{fig_edge}
\end{figure*}

Having looked at the numerical observations in the $(M_B$ - $M_w)$ parameter space, we will now develop an understanding based on a low-energy theory of the junction between the wire and the environment. We define the junction Hamiltonian in the ribbon geometry assuming translational symmetry in $x$ direction as, 
\begin{equation}
	H_{\text{junction}} = H_{\text{wire}} + H_{\text{QWZ}} + H_{\text{coupling}}\label{eq_bhzcoupledwire}
\end{equation}
where real space form $H_{\text{wire}}$ and $H_{\text{QWZ}}$ is, 
\bea
H_{\text{wire}} &=& \sum_{i \in \text{wire}}\left(\big( \Psi^\dagger_{i} T_{x}\Psi_{i+\hat{x}} + \text{h.c.}  \big)+ \Psi^\dagger_i \left(2-M_w\right)\sigma_z \Psi_i \right)\notag \\
H_{\text{QWZ}} &=& \sum_{\substack{i\notin \text{wire} \\ (i+ \vec{\delta}) \notin \text{wire}}}\left(\big( \Psi^\dagger_{i} T_{\vec{\delta}}\Psi_{i+\vec{\delta}} + \text{h.c.}  \big) + \Psi^\dagger_i (2-M_B)\sigma_z \Psi_i\right) \notag 
\label{eq_junction}
\eea	
where $\Psi_i$ is same as defined earlier below \eqn{eq_rham}. The coupling between these two subsystems is given by,
\beq
H_{\text{coupling}} =  \sum_{\substack{i \in \text{wire} \\ (i \pm \hat{y}) \notin \text{wire}}} \tilde{T}_{y} \Big( \Psi^\dagger_{i} \Psi_{i + \hat{y}} +  \Psi^\dagger_{i - \hat{y}} \Psi_{i}\Big)  +\text{h.c.} 
\eeq
where,
\beq
\tilde{T}_y = -\frac{1}{2} \left(\kappa_1 \sigma_z + i \kappa_2 \sigma_y\right)= \half \begin{pmatrix}
	-\kappa_1 & -\kappa_2\\
	\kappa_2 & \kappa_1
\end{pmatrix}.\label{eq_kcoupmat}
\eeq
Here, we have introduced two parameters, $\kappa_1$ and $\kappa_2$, to tune the strengths of the couplings between the wire and the environment.  We define the symmetric and antisymmetric versions of them, which will be useful later. 

\beq
\kappa_s = \frac{(\kappa_1 + \kappa_2)}{2} \qquad \qquad \kappa_a = \frac{(\kappa_1 - \kappa_2)}{2} \label{eq_kska}
\eeq
In particular, $\kappa_1=\kappa_2=0$ ($\kappa_s=\kappa_a=0$) represents the complete decoupling of wire from the environment - the isolated case, while at $\kappa_1=\kappa_2=1$ ($\kappa_s=1, \kappa_a=0$) the Hamiltonian goes back to the composite system as defined in \eqn{eq_rham} with ribbon geometry. We now focus on the physics of the junction along Cut II where $M_B=1$, such the environment is topological, and the other parameters $M_w, \kappa_1, 
\kappa_2$ are tuned.

In the  $\kappa_1=\kappa_2=0$ limit, the Chern insulating environment is in a gapped state with a bulk-gap $ \Delta E \sim 2$, however given its topological character $\mathcal{C}=-1$, the wire region acts like a boundary generating two CBS at $k_x=0$, the $\Gamma$ point (see \Fig{fig_edge}(a)). Considering the wire is placed at $y=0$, while just above the wire ($y>0$), the environment has a right-moving mode, below the wire ($y<0$), it has a left-moving mode (note in $\mathcal{C}=+1$ situation flips; see Appendix~\ref{apndx_cbs}). These one-dimensional modes form a basis via, 
\begin{align}
	& c^{\dagger}_{k_x, +}|\Omega\rangle  =\ket{k_x, +}  \sim \begin{pmatrix}
		1\\
		1
	\end{pmatrix} e^{-y/\xi} e^{ik_x x} ~\text{for $y>0$},\nonumber \\
	& c^{\dagger}_{k_x, -} |\Omega \rangle = \ket{k_x, -} \sim \begin{pmatrix}
		1\\
		-1
	\end{pmatrix} e^{y/\xi} e^{ik_x x} ~\text{for $y<0$},\label{eq_cbssolutions}
\end{align}
with an effective dispersion given by,
\beq
h_{\text{CBS}} = k_x \sigma_z
\label{cbs}
\eeq
within the CBS manifold. Here $|\Omega \rangle$ is the fermionic vacuum, and $\xi$ is the CBS localization length in the bulk of the Chern insulator, which is dependent on the bulk gap. As we show below, the introduction of $\kappa_1, \kappa_2$ leads to the hybridization of these low-energy CBS states with those of the wire (tuned by $M_w$), leading to the low-energy phenomenology. For $M_w \ll 1$, the wire is trivially gapped, and even at $\kappa_1=\kappa_2=1$, the low energy states are dominated by CBS as seen in \Fig{fig3}(I).

\subsection{Vanishing of topological criticality}
\ul{\it $M_w=1$}: In the decoupled limit, at $M_w=1$ the wire itself becomes gapless at the $\Gamma$ point leading to the following low-energy Hamiltonian 
\beq
h_{\text{wire}} =  k_x \sigma_x \label{eq_w1}
\eeq
where the basis of the wire is comprised of $|k_x, A \rangle =  c^{\dagger}_{k_x, A}|\Omega \rangle$ and $|k_x, B \rangle = c^{\dagger}_{k_x, B} |\Omega \rangle$. So the complete low-energy theory can be written in terms of the basis of the CBS and wire states as follows,
\beq
H_{\text{eff}}  (\kappa_s=\kappa_a=0) = \begin{pmatrix}
	h_{\text{CBS}} & \bzero \\ 
	\bzero & h_{\text{wire}}
\end{pmatrix}
\label{decou}
\eeq
which written in a $4\times 4$ direct product basis defined using Pauli matrices $\sigma$'s and $\tau$'s, will be of the form
\bea
H_{\text{eff}} =  k_x \sigma_z \otimes \frac{(\1 + \tau_z)}{2} + k_x \sigma_x \otimes \frac{(\1 - \tau_z)}{2} 
\eea
Now projecting $H_{\text{coupling}}$ into this low energy subspace (see Appendix~\ref{apndx_soft1} for details) one finds that, 
\beq
H_{\text{eff}} = H_{\text{eff}}  (\kappa_s=\kappa_a=0) - \kappa_s \Big( \1 \otimes \tau_x + \sigma_y \otimes \tau_y \Big) \label{eq_lingap}
\eeq
Interestingly, the leading corrections of $\kappa$ perturbations anti-commute with the decoupled limit Hamiltonian in \eqn{decou}, thus mutually gapping out both the wire and the CBS sector. Consequently, $M_w=1$ criticality of the wire completely washes out due to this coupling since the electrons in CBS and wire scatter into each other as seen in numerically obtained dispersion in \Fig{fig_edge}(b) where both $\kappa=0$ and $\kappa=0.8$ are shown (for the numerical data $\kappa_s=\kappa$, $\kappa_a=0$). The effective theory further predicts that the gapping out is linear in $\kappa$ as mentioned in \Fig{fig_edge}(b) (also see  Appendix~\ref{apndx_soft1}).

\subsection{Trivialization of topological wire}
\ul{\it $M_w=2.5$}: In the regime of $1<M_w<3$ the wire at $\kappa_s=\kappa_a=0$ is expected to be in the topological regime. However, as seen in \sect{sec_topological} - this regime doesn't contain any topological edge modes in finite geometry, once coupled with the topological environment. Again referring to the decoupled effective Hamiltonian (\eqn{decou}), while the CBS states remain at the $k_x=0$, the lowest energy states for the wire is gapped with an effective dispersion of
\beq
h_{\text{wire}} = -k_x \sigma_x + \frac{1}{2}\sigma_z
\eeq
near $k_x = \pi$ (see \Fig{fig_edge}(c)). The existence of such modes, despite being gapped, enables the CBS to hybridize through virtual processes, resulting in a gap opening quadratically with $\kappa_s$ (see Appendix~\ref{apndx_soft2}). Thus the physics of hybridization renders the wire {\it smoothly} gapped all through $-\infty <M_w <3$ - leading to a trivial phase. So, in the presence of a topological environment, the parent topological phase of the wire loses its character.

\subsection{Survival of subsystem metal}
{\ul {\it $M_w=3.0$}:} We now discuss the case of $M_w=3$ where the wire in the decoupled limit is again critical. However, unlike $M_w=1.0$, the low energy modes of the wire now reside at $k_x=\pi$. When projected to the low energy subspace, the effective Hamiltonian remains similar to \eqn{decou}, where $h_{\text{CBS}}$ is given by \eqn{cbs} and 
\beq
h_{\text{wire}} = -k_x \sigma_x \label{eq_w3}
\eeq
Interestingly, the projection of $H_{\text{coupling}}$ to this subspace is identically $\it zero$. This, in fact, is a direct result of a momentum mismatch between the two low-energy theories of CBS and the wire. 

Furthermore, we find that while the leading correction to the CBS sector is,
\begin{align}
	\tilde{h}_{\text{CBS}} \propto -\kappa^2_s  \sigma_x,\label{eq_cbs2nd}
\end{align}
for the wire, the correction is,
\beq
\tilde{h}_{\text{wire}} \propto \kappa_s \kappa_a \sigma_z \label{eq_wirecorrection}
\eeq
such that the effective Hamiltonian in the low-energy manifold is
\beq
H_{\text{eff}} = H_{\text{eff}}  (\kappa_s=\kappa_a=0) +\begin{pmatrix}
	-\kappa^2_s  \sigma_x & \bzero \\ 
	\bzero &  \kappa_s \kappa_a \sigma_z
\end{pmatrix}
\eeq
The presence of $\kappa_a$ in the case of wire shows that even the second order perturbation goes to zero when $\kappa_1=\kappa_2$. This leads to the surprising observation that the wire continues to remain metallic at $M_w=3$ (see \Fig{fig_edge}(d)). Such criticality, therefore, is the robust `subsystem metal' as it exists in a topological environment. Even when $\kappa_a\neq0$, the nature of the correction term can only shift the critical value of $M_w$ where the one-dimensional metallic phase occurs. The behavior of the shift and the perturbation analysis are shown in Appendix~\ref{apndx_soft3}.

\begin{figure}
	\centering
	\includegraphics[width=1\linewidth]{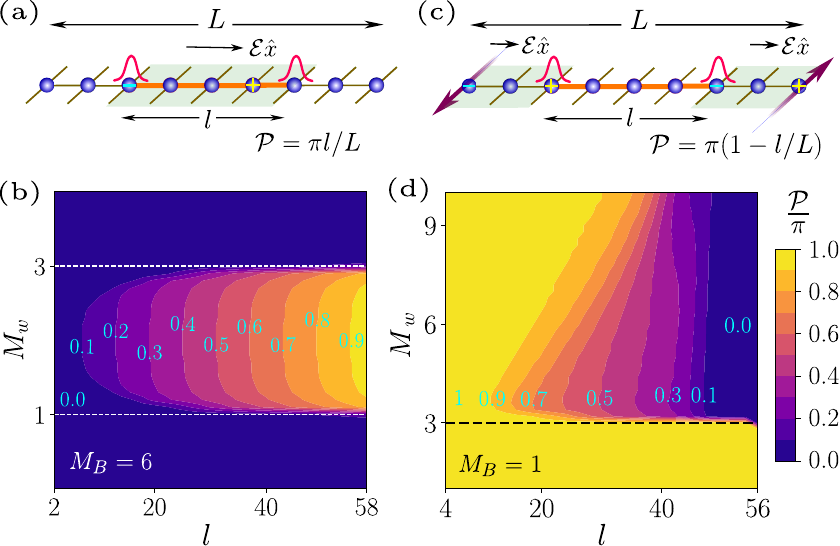}
	\caption{(a) In the trivial environment of linear size $L$, a wire of length $l$ hosts EMs which give rise to non-quantized polarization $\mathcal{P}$. A tiny electric field $\mathcal{E} \hat{x}$ is applied to calculate $\mathcal{P}$ (also see Appendix~\ref{apndx_pol}). (b) For $M_B=6$, $\mathcal{P}$ shows a distinct jump in the topological regime of the wire ($1<M_w<3$) when $l$ is sufficiently large. (c) In the topological environment, the coexistence of CBS in the outer boundary and the proximity-induced EM create signatures in $\mathcal{P}$. (d) For $M_B=1$, $\mathcal{P}$ for various $l$ show a dip immediately after $M_w=3$ revealing proximity-induced EM.}
	\label{fig_pit}
\end{figure}

\begin{figure*}
	\centering
	\includegraphics[width=0.95\linewidth]{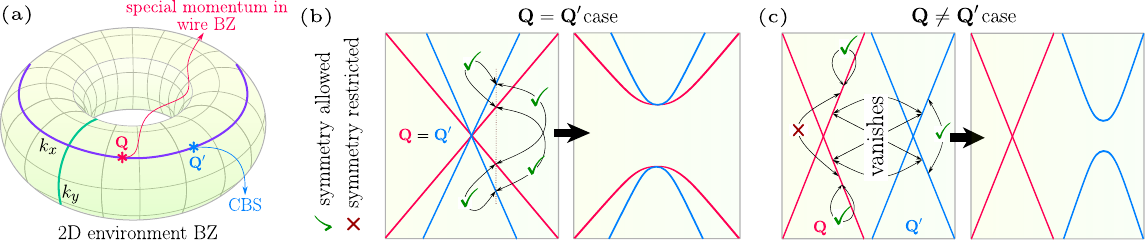}
	\caption{(a) Toroidal Brillouin zone (BZ) of the two-dimensional (2D) environment. In ribbon geometry, the embedding wire creates a one-dimensional (1D) BZ on the torus. The CBS of the environment also gets projected on the 1D BZ. $\mQ$ and $\mQp$ are the special momenta of the wire and the CBS, respectively, where their low energy theories lie. (b) In the case of $\mQ= \mQp$, symmetry allowed couplings between CBS and the wire gap each other out. (c) These couplings vanish in the case of $\mQ \neq \mQp$. Virtual processes gap out the CBS but symmetries protect the wire criticality.}
	\label{fig_idea}
\end{figure*}

\subsection{Proximity-induced topology}\label{sec_pit}

{\ul {\it $M_w \gtrapprox 3.0$}:} In the previous subsection, we had noticed that at $M_w=3$ the low-energy Hamiltonian of the wire near $k_x=\pi$ remained linear even in the presence of coupling with the CBS states. Thus, any deviations of $M_w$ around $3$ lead to an effective theory of the wire as
\beq
h_{\text{wire}} =  -k_x \sigma_x - m_w \sigma_z
\eeq
where $m_w=M_w-3$. The theory of the wire looks like a one-dimensional Dirac theory where we expect the presence of anomalous edge modes in an open boundary. Given $M_w<3$ ($m_w<0$) was already seen to be smoothly connected to a trivial limit, one can anticipate non-trivial modes for $M_w>3$ ($m_w>0$). These EMs were already seen in the ZBP (\Fig{fig3}(D)) and DOS (\Fig{fig3}(L)), which are similar to the parent EMs of the wire. We now characterize them using some of the topological markers and find characteristic differences from that of parent EMs.

The non-trivial topology of the bulk bands of the wire (see \eqn{eq_wire}), under PBC, can be characterized by unit quantized {\it winding number}. In real space, this amounts to a {\it polarization} ${\mathcal P} = \pi$. Extracting polarization in a composite system such as ours needs the application of a perturbative electric field (see \fig{fig_pit}(a)) (for details, see  Appendix~\ref{apndx_pol}). Considering a $l$ sized wire in a $L$ sized trivial environment (say $M_B \ll 0$) this would lead to a ${\mathcal P} = \frac{\pi l}{L}$ for $M_w$ between $1$ and $3$ (see \Fig{fig_pit}(a,b)). This is a direct result of the effective dipole moment that arises due to the application of the perturbative electric field on edge-localized zero-energy modes. Therefore, the natural question to pose is, whether the proximity-induced phase also has non-trivial polarization. 

At $M_B=1$, the CI environment itself has a chiral edge state (i.e., CBS) on the outer boundary, which leads to a non-trivial polarization of $\pi$. The formation of a non-trivial EM on the wire then leads to a redistribution of dipole moment into two regions of length $\frac{L-l}{2}$ (see \Fig{fig_pit}(c)). The total polarization then must behave as, 
\beq
{\mathcal P} = \pi \Big(1 - \frac{l}{L} \Big)
\eeq
Indeed, the numerical results as shown (see \Fig{fig_pit}(d)) are consistent, showing the non-trivial effect of proximity-induced topology. These proximity-induced EMs also have entanglement properties reminiscent of conventional EMs, as discussed in Appendix~\ref{apndx_mi}.

Now as $M_w$ increases further, the CBS edge modes around the wire, even while gapped, come closer to the Fermi energy since their hybridization scale is $\propto \frac{1}{M_w}$ (see Appendix~\ref{apndx_soft2}). These CBS modes eventually mix with the proximity-induced EM, rendering them featureless. The large $M_w$ region can be equated with an effective puncture having low energy CBS states. Thus the system at very large $M_w$ shows no anomalous polarization (see \Fig{fig_pit}(d)), neither a significant ZBP (see \Fig{fig3}(D)) but an LDOS around the wire similar to \Fig{fig3}(I) and gapless spectrum in \Fig{fig3}(A) signaling the formation of CBS.

\section{SYMMETRY-PROTECTED EMBEDDING}\label{sec_spe}

In this section, we show how the phenomena and nature of the low-energy theories, as discussed in the last section in the context of embedding a topological wire within a square lattice environment, are rather a direct result of the interplay of intricate symmetries of the system and are therefore general even in other environments satisfying the same symmetry structure. In particular, our analysis showed that the phenomenology of the system can effectively be captured within how the low-energy theory of the wire gets affected by the low-energy states of the environment. 

In principle, the embedding of a one-dimensional wire within a two-dimensional lattice model - effectively projects the states of a two-dimensional toroidal Brillouin zone onto a one-dimensional Brillouin zone of the wire (see \Fig{fig_idea}(a)). The wire can, in general, host a gapless theory at the momentum point $\mQ$, which is blind to the environment if the surrounding is trivially gapped. However,  when the environment is topological (in this case, a Chern insulator), it can host gapless chiral boundary modes at $\mQp$ within the same one-dimensional Brillouin zone. Thus, one can pose, given the symmetries, how these gapless theories couple with each other. 

The low energy degree of freedom for the wire consists of $c^\dagger_{k_x,A}$ and $c^\dagger_{k_x,B}$ ($k_x$ near $\mQ$) (see \eqn{eq_w1} and~(\ref{eq_w3})) while for those in the CBS $\equiv c^\dagger_{k_x,+}$ and $c^\dagger_{k_x,-}$ (see \eqn{cbs}) with ($k_x$ near $\mQp$). Relabeling the wire degree of freedom as $f^\dagger_{k,\alpha}$ ($\alpha=A, B$) and those for the CBS as $d^\dagger_{k_x, \beta}$ ($\beta=\pm 1$),  their symmetry transformation under the charge-conjugation symmetry and parity (the symmetry common to both the environment and the wire) is given by Table~\ref{tab_symm} (also see Appendix~\ref{apndx_symmetry}).

\begin{table}
	\centering
	\begin{tabular}{c c c c c} 
		\hline
		\hline 
		& ~operator~ & ~under CGS $\otimes$ parity~\\
		\midrule
		\multirow{2}{*}{wire} & ~~$f^\dagger_{k_x,A}$~ & $f_{k_x,B}$ \\
		& ~$f^\dagger_{k_x,B}$~ & $f_{k_x,A}$ \\\midrule  
		\multirow{2}{*}{~CBS~}& ~$d^\dagger_{k_x,+}$~ & $d_{k_x,+}$ \\ & ~$d^\dagger_{k_x,-}$~ & $-d_{k_x,-}$  \\[0.1cm] \hline \hline
	\end{tabular}
	\caption{Transformation of degrees of freedom of the wire and the CBS, under a combined action of charge-conjugation symmetry (CGS) and Parity.}
\label{tab_symm}
\end{table}

A generic hybridization between the two theories is therefore given by
\beq
H_{\text{hyb}} = \sum_{k_x} \sum_{\alpha \beta} \Big( V_{\alpha \beta} (k_x) f^\dagger_{k_x, \alpha} d_{k_x, \beta} + \text{h.c.} \Big) \label{eq_hyb1}
\eeq
Under the action of these symmetries, one finds that $V_{A,+} = -V^{*}_{B, +}$ and $V_{A,-} = V^{*}_{B, -}$. (see Appendix~\ref{apndx_symmetry} for proof of these conditions). This implies if $\mQ=\mQp$ such that both these states exist at any given $k$, they can generically gap out. On the other hand, if in a given momentum neighborhood of the wire, no such CBS manifold exists, the wire can only hybridize with the bulk. This hybridization can, via higher-order processes, effectively result in a coupling of the form 
\beq
H^{\text{wire}}_{\text{hyb}} = \sum_{k_x} \sum_{\alpha \alpha'} \Big( V^f_{\alpha \alpha'} (k_x) f^\dagger_{k_x, \alpha} f_{k_x, \alpha'} + \text{h.c.} \Big)\label{eq_hyb2}
\eeq
However, the symmetries (Table~\ref{tab_symm}) restricts $V^{f}_{AA} = - V^{f}_{BB}$, $V^{f}_{AB} = V^{f}_{BA}= 0$ . Similarly, a second-order effect on CBS leads to,
\beq
H^{\text{CBS}}_{\text{hyb}} = \sum_{k_x} \sum_{\beta \beta'} \Big( V^d_{\beta \beta'} (k_x) d^\dagger_{k_x, \beta} d_{k_x, \beta'} + \text{h.c.} \Big)\label{eq_hyb3}
\eeq
which allows only for $V^{d}_{+-} = (V^{d}_{+-})^*$ and $V^{d}_{++/--} = 0$. Given these couplings in general, both theories can gap out. However, due to the existence of an independent parity symmetry, any gapless theory of the wire of the CBS is guaranteed to exist only at Time-Reversal Symmetric Momenta (TRIM) points of the BZ. In the one-dimensional case, this implies that $\mQ, \mQp \in \{0,\pi\}$.

The above analysis results in two distinct cases (i) when $\mQ=\mQp$ such that the CBS and the wire can hybridize to gap each other out as schematically shown in \Fig{fig_idea}(b), (ii) when $\mQ \neq \mQp$ and doesn't share a neighborhood, the CBS can gap out, but the criticality of the wire is symmetry-protected (see \Fig{fig_idea}(c)).  This symmetry-constrained embedding of one lower dimensional theory within a higher dimensional theory is what we term {\it symmetry-protected embedding}.

\begin{figure}
	\centering
	\includegraphics[width=0.95\columnwidth]{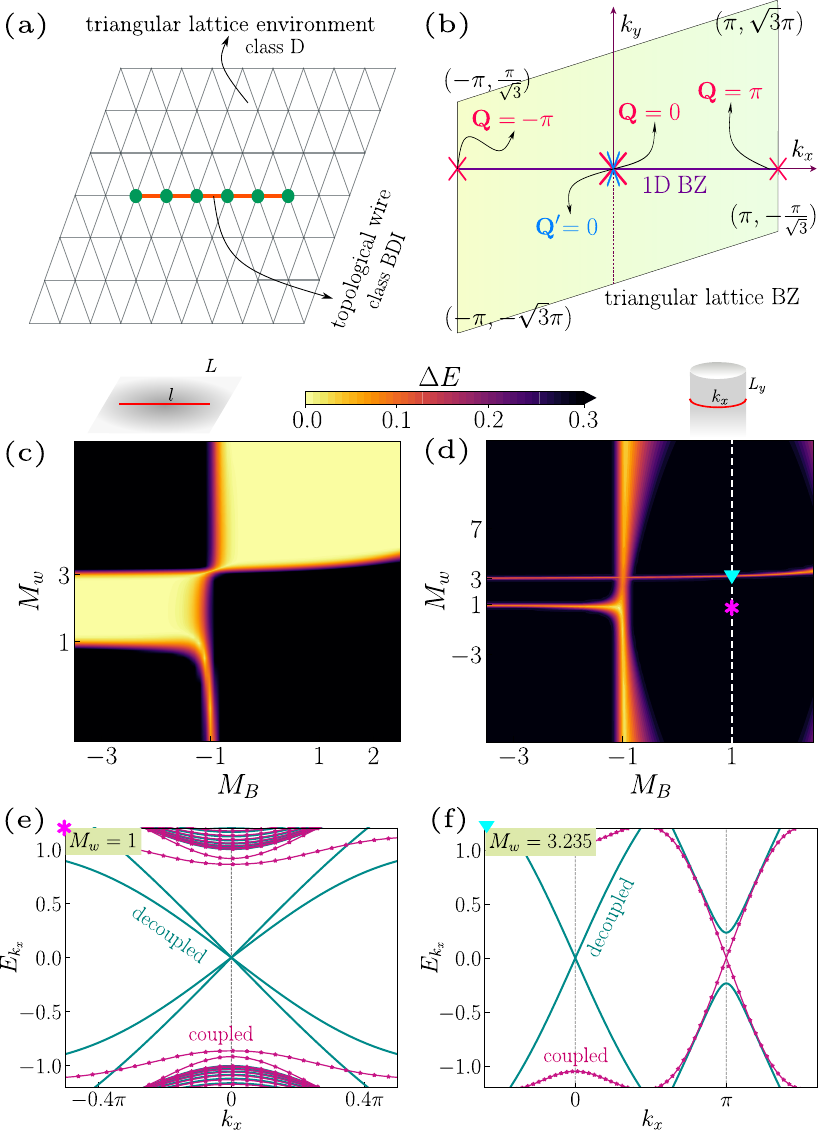}
	\caption{(a) Topological wire in a triangular lattice CI environment. (b) Special momenta of the wire ($\mQ$) and CBS ($\mQp$) projected on a 1D BZ from 2D BZ. The spectral gap in the ($M_B$ - $M_w$) space for (c) finite geometry and (d) ribbon geometry. Along the Cut $M_B=1$, we analyze the low energy bands at $M_w=1$ and $3.235$ in (e) and (f), respectively, for both coupled and isolated cases. (e) Both the low energy modes which appear at $k_x=0$ gap-out. (f) CBS at $k_x=0$ also gaps out, but subsystem metal shifts the critical $M_w$ value.}
	\label{fig_tribhz}
\end{figure}

Given the Hamiltonian respects the symmetry structure as discussed above, indeed, at $M_B=1, M_w=1$ when $\mQ=0, \mQp= 0$, the system gaps out with $V = \kappa_s/2$ (see \eqn{eq_lingap} and \eqn{eq_hyb1}). However, when  $M_B=1, M_w=3$ $\mQ = \pi, \mQp= 0$, and due to the momentum mismatch, the gapless point of the wire remains protected - thus leading to the existence of sub-system metal. The situation flips when $M_B=3$, i.e., the environment is in $\mathcal{C}=+1$ phase. At $M_B=3$, $\mQp= \pi$ and thus $M_w=1$ $(\mQ=0)$ now remains protected even though $M_w=3$ $(\mQp=\pi))$ gets gapped out. This explains the numerical observation of the different phases and their character in the previous section. It is interesting to point out that the same symmetry structure only allows the renormalization of the $M_w$ near the sub-system metallic phase so that, in general, the sub-system metal can be realized away from $M_w=1, 3$. For instance in \eqn{eq_wirecorrection} when $\kappa_1 \neq \kappa_2$, although it seems the $M_w=3$ TQCP is gapped for $M_B=1$, actually the critical line shifts to $M'_w \sim (3 - \mu \kappa_a \kappa_s)$ where $\mu$ is constant factor (see Appendix~\ref{apndx_soft3}).

In order to further verify the generality of our results, we now study another system where the wire is now part of a triangular lattice environment (see \Fig{fig_tribhz}(a)). The triangular lattice Hamiltonian is now given by,
\begin{align}
	\mathcal{H}_{\text{T-QWZ}}(\vec{k}) = &~   \bigg(2-M_B-\cos k_x - \cos \frac{k_x}{2} \cos \frac{\sqrt{3}k_y}{2} \bigg)\sigma_z  \nonumber \\ & - \bigg(\sin k_x + \sin \frac{k_x}{2} \cos \frac{\sqrt{3}k_y}{2}\bigg) \sigma_x \nonumber \\  & - \bigg( \sqrt{3}\cos \frac{k_x}{2} \sin \frac{\sqrt{3}k_y}{2} \bigg) \sigma_y.\label{eq_tribhz}
\end{align}
This again has a $\mathcal{C}=-1$ phase between $-1<M_B<3$. The environment contains the same symmetries as the square lattice case. Here again $\mQp=0$ for the CBS, while for the wire, $\mQ=0$ at $M_w=1$ and $\mQ=\pi$ for $M_w=3$ (see \Fig{fig_tribhz}(b)). As discussed above, the $M_w=1$ line is expected to gap out while the criticality at $M_w=3$ line is expected to remain stable. As seen in \Fig{fig_tribhz}(c,d), while in the trivial phase ($M_B<-1$), the wire shows edge modes between $1<M_w<3$, in the strip geometry the gapless point only arises near $M_w=3$ (specifically at $M_w=3.235$ for $M_B =1$) for $M_B>-1$ showing the existence of sub-system metal. The low energy dispersion at the two cases $M_w=1$ and $M_w=3.235$ mirrors the results shown in \Fig{fig_edge}(b,d) and the general case discussed in \Fig{fig_idea}(b,c). Interestingly, unlike the square lattice case here in \Fig{fig_tribhz}(e), one notices that the effective velocities of the CBS and the wire are not the same. 

Having discussed the crucial role of the parity and the CGS symmetry in the physics of the composite system discussed above, in the next section we study the effect of breaking these protecting symmetries of the environment.

\section{SYMMETRY TUNABILITY OF ENVIRONMENT}\label{sec_symmetry}

\begin{figure}
	\centering
	\includegraphics[width=1\columnwidth]{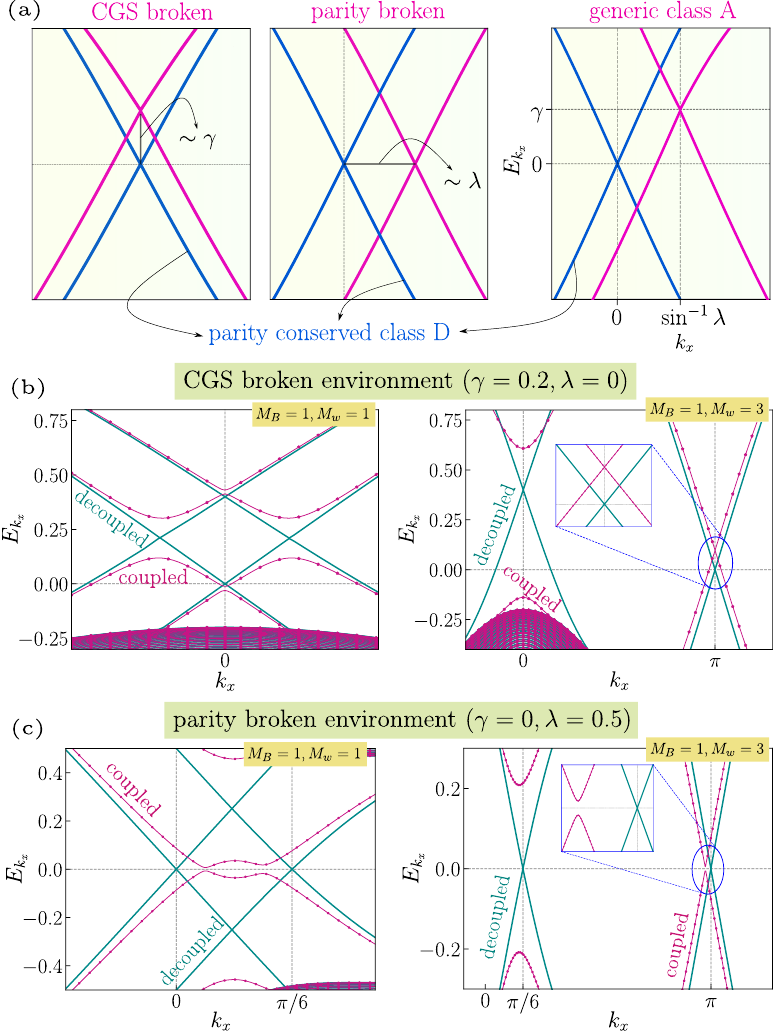}
	\caption{(a) The effect of symmetry breaking terms (see \eqn{eq_clsAHam}) on the CBS mode of the parity conserved class D QWZ model. (b) For the CGS broken topological environment ($M_B=1$), at $M_w=1$, both low energy modes (wire and CBS) appearing at $k_x=0$ open up a gap when coupled. For $M_w=3$, only subsystem metal at $k_x=\pi$ remains stable (zoomed-in data in inset). (c) Same as (b), but now the parity is broken in the environment, so low energy modes never appear at the same $k_x$, and both $M_w=1$ and $M_w=3$ subsystem metal gap-out.}
	\label{fig_clsA}
\end{figure}

The Chern insulating character of the environment is generically stable to breaking of either the CGS or the parity symmetry since a Chern insulator can exist even in Class A of the ten-fold symmetry classification \cite{Altland_PRB_1997,  Kitaevtenfold, chiu_topoclass_2016, Ludwig_topo_2016, Agarwala_AOP_2017}. Till now, we had limited our discussions to the case where the wire was in BDI symmetry class while the environment was in parity-symmetric class D. In this section, we show how to retain the symmetries of the wire, and what is the role of selectively breaking the environmental symmetries.

Systematically, the CGS and the parity can be broken in the parent Hamiltonian of the environment $\mathcal{H}_{\text{QWZ}}(\vec{k})$ (see \eqn{eq_clsDHam}), by introducing the two terms $\gamma$ and $\lambda$ as, 
\begin{align}
	\mathcal{H}_{\text{classA}}(\vec{k}) = &~ \mathcal{H}_{\text{QWZ}}(\vec{k})  -\lambda \sigma_x -\lambda \sigma_y \nonumber \\
	& + \gamma\left(\cos k_x + \cos k_y\right)\1.
	\label{eq_clsAHam}
\end{align}
The symmetry breaking in the presence of these specific terms can be seen using operations given in \eqn{eq_cgsp}. Importantly, even at a finite $\lambda$, and $\gamma$ the system retains finite parameter regimes where the system is Chern insulating (see Appendix~\ref{apndx_clsA}).

Since CGS and parity together relate $H(k) \leftrightarrow -H^*(k)$, at any momentum point, the spectrum is symmetric about $E=0$. This crucial symmetry governed the nature of hybridization between the wire and the CBS, as seen earlier. However, when $\gamma \neq 0$, breaking of CGS can result in a finite energy shift of the CBS manifold compared to the wire theory (see schematic in \Fig{fig_clsA}(a)). On the other hand - breaking of parity in the environment with $\lambda\neq 0$, allows the gap closing points of the CBS to not be restricted to TRIM points - thus, in the low energy theory now, the CBS manifold can shift in the one-dimensional BZ (see schematic in \Fig{fig_clsA}(a)). In the presence of both $\lambda$, and $\gamma$ (i.e., in generic class A), the low energy theory of the CBS can be shifted from that of the wire both in energy and momentum as seen in \Fig{fig_clsA}(a)).

As in the previous section, considering $M_B=1$ (i.e. the environment is in $\mathcal{C}=-1$ phase), we focus our study at $M_w=1$ and $M_w=3$, where the wire is critical at $\mQ = 0$ and $\mQ = \pi$ respectively. In the decoupled limit, for $M_B=M_w=1$, $\mQp = 0$, which continues to remain so, given $\lambda = 0$ (parity preserving). The introduction of $\gamma$ (CGS breaking), however, shifts the CBS energetically, as can be seen in \Fig{fig_clsA}(b). The gapless point of the wire, which earlier opened up a gap to linear order in $\kappa$ (see \eqn{eq_lingap}) due to the existence of an exact degeneracy, now opens up a gap as second order. The gap at $\mQ=0$ for the wire, in fact, opens up a perturbative gap due to the presence of $\gamma$. Thus, the selective breaking of CGS for the environment can allow for a tunable gap for the wire. For $M_w=3$, the existence of parity still ensures that $\mQp=0$, and, therefore, the wire theory at $\mQ=\pi$ continues to remain protected. So, the introduction of $\gamma$ continues to stabilize the sub-system metal phase and, consequently, the proximity-induced topology (see Appendix~\ref{apndx_clsA}). 

The introduction of $\lambda$ (parity-breaking), however, is more drastic. At $\lambda \neq 0$, it is interesting to see that the CBS manifold is shifted from $\mQp =0$ or $\mQp=\pi$ (as seen in \Fig{fig_clsA}(c)). However, now existence of $\kappa$ thus opens up the gap for the wire generically.  Therefore, in such a case, the signatures of both the sub-system metal, as well as proximity topology, are expected to vanish (see Appendix~\ref{apndx_clsA}). This is also what happens in class A, where at a function of $M_B$, the wire has no critical phenomena, and the wire becomes trivial for all values of $M_w$ once in the topological environment. Therefore, a general class A topological environment (vacuum), completely {\it trivializes} a topological wire without any features of proximity-induced topology.

\begin{table}
	\centering{
		\begin{tabular}{ c c c c}
			\hline
			\hline
			~~Symmetry~ & wire & \multicolumn{2}{c}{environment}  \\
			
			Class & ~$d=1$~  & ~$D=2$~ & ~$D=3$~ \\
			\hline
			A    & $0$          & \mycircs{$\mathbb{Z}$}        & $0$ \\
			AIII & $\mathbb{Z}$ & $0$                 & $\mathbb{Z}$ \\
			AI   & $0$          & $0$                 & $0$ \\
			BDI  &  \mycirc{$\mathbb{Z}$} & $0$                 & $0$ \\
			D    & $\mathbb{Z}_2$ & \mycirct{$\mathbb{Z}$}     & $0$ \\
			DIII & $\mathbb{Z}_2$ & $\mathbb{Z}_2$    & $\mathbb{Z}$ \\
			AII  & $0$          & $\mathbb{Z}_2$      & $\mathbb{Z}_2$ \\
			CII  & $2\mathbb{Z}$ & $0$                & $\mathbb{Z}_2$ \\
			C    & $0$          & $2\mathbb{Z}$       & $0$ \\
			CI   & $0$          & $0$                 & $2\mathbb{Z}$ \\
			\hline
			\hline
	\end{tabular}}
	\caption{Symmetry classification of topological phases where the wire resides in dimension $d$ and the environment in dimension $D$ \cite{Altland_PRB_1997,  Kitaevtenfold, chiu_topoclass_2016, Ludwig_topo_2016, Agarwala_AOP_2017}. The circled cases, paired with lines, are the ones discussed in this work.}
	\label{tab_classification}

    \begin{tikzpicture}[remember picture, overlay]
    \draw[ultra thick, black] (a1) -- (a2);
    \draw[ultra thick, red] (a1) -- (a3);
    \end{tikzpicture}
\end{table}

\section{DISCUSSION}\label{sec_dis}

In our study, we have focused on a class BDI topological wire in one dimension embedded in a two-dimensional system belonging to class D and consequently in class A, interplaying with parity symmetry. However, in both 1D and 2D, multiple symmetry classes are known to be topological (see Table~\ref{tab_classification}) by Kitaev's tenfold classification table \cite{Altland_PRB_1997,  Kitaevtenfold, chiu_topoclass_2016, Ludwig_topo_2016, Agarwala_AOP_2017}. 

While we expect that equivalent phenomena of subsystem metal, proximity-induced topology, and symmetry embedding must continue in other parity symmetric classes - it will be interesting to investigate the specificities of each of the cases. For instance, one-dimensional topological systems in class AIII and class D can be realized by the following Hamiltonians
\begin{align}
\mathcal{H}^{\text{AIII}}_{\text{wire}}(k_x) & = \left(2- M_w -\sin k_x\right)\sigma_z + \cos k_x \sigma_x, \label{eq_wireaiii} \\
\mathcal{H}^{\text{D}}_{\text{wire}}(k_x) & = \left(2- M_w -\cos k_x\right)\sigma_z + \sin k_x \left(\frac{\sigma_x + \sigma_y}{\sqrt{2}}\right),
\label{eq_wired}
\end{align}
Given $H^{\text{D}}_{\text{wire}}$ still retains parity, the exotic features and physics discussed here will hold true, but for $H^{\text{AIII}}_{\text{wire}}$ which breaks parity, we expect a complete trivialization of the wire when embedded in a class D CI. In another case for the class CII wire - where the topological wire has a $2\mathbb{Z}$ classification, it will be interesting to pose if the proximal topological aspects also retain $2\mathbb{Z}$ or can become $\mathbb{Z}$?

Similarly, the environment we studied is exclusively in two dimensions. It might be interesting to ask what happens to the wire in a topological three-dimensional vacuum. While the standard AII class of the 3D topological insulators may lead to a mismatch in the types of degrees of freedom, classes such as AIII/CI, which have time-reversal broken, may immediately gap out a BDI, which has a time-reversal symmetry protecting the edge state. Similarly, the nature of protection of the subsystem metal (vis-a-vis symmetry embedding) can be interpreted as a form of impedance mismatch between inter-dimensional topological phases, and one can pose its generalization to other classes and dimensions.

In our study, the effective physics of the embedded wire can still be understood within a one-dimensional theory with effective parameters changed perturbatively due to coupling with the environment. Such a description may, however, fail when the environment length scales strongly compete with the system length scales, leading to strong dimensional renormalization \cite{carrol_I_2021, carrol_II_2021}. Here, one often finds effective long-range couplings between the sites of the embedded system, leading to qualitative change in the relevant gap size, such as studied in the context of the role of magnetic textures on one-dimensional impurities in superconducting substrates \cite{carrol_I_2021, carrol_II_2021}. Given that, in our problem, the electronic wavefunctions on any impurity site are strongly localized (see details in Appendix~\ref{apndx_singleimpurity}), we expect no long-range couplings within sites of the wire, thus resulting in an effective one-dimensional nearest-neighbor tight-binding description. However, in the presence of long-wavelength magnetic textures, it might be interesting to study the role of dimensional renormalization in the context of this system.

While these theoretical studies are themselves open and interesting as future works, the experimental realizations of such exotic phenomena are also crucial to advancing the study of topological qubits and their application in quantum technology. Cousins of the microscopic Hamiltonian of the class D Chern insulating environment (see \eqn{eq_clsDHam}) have been realized both in material systems such as Hg-Te, Cd-Te quantum wells \cite{BHZ_Sci_2006} and in spin-orbit-coupled ultracold fermions \cite{CIM_ultracoldfermion_2023} based on the optical Raman lattice technique \cite{Wang_ultracold_2018, Sun_ultracold_2018, CIM_coldatoms_2014}. In these systems, introducing an engineered {\it mass} inhomogeneity will create a composite structure, enabling the experimental realization of a topological wire in a topological vacuum. For instance, in the ultracold fermion setup, such inhomogeneity in the optical lattice can be created by spatially modulated tuning of laser frequencies or local Zeeman fields. Also, in material-based systems,  the inhomogeneity engineering using impurity, dislocation, and fault stacking in three-dimensional topological insulators has been proposed recently \cite{shoman2015topological, schindler_zddefectin3d_2022, hu_dislocation_2024}. 

In addition to these specific systems, there are other materials that exhibit Chern insulating properties. In general, a Chern insulating material requires both strong spin-orbit coupling and spontaneous magnetic order, making it challenging to realize with a significant bulk gap. Most of the candidate materials are magnetic topological insulators such as Cr- and V-doped (Bi,Sb)$_2$Te$_3$ \cite{CIM_qah_2013, CIM_CrBST_2015, CIM_crbst2_2015}, thin film layers of MnBi$_2$Te$_4$ \cite{CIM_MBT_2020}, Moiré heterostructures like graphene-hBN \cite{CIM_moire_2020}, twisted bilayer transition metal dichalcogenide \cite{CIM_abMote2, CIM_TMD_2024} and Bi$_x$Sb$_{1-x}$ alloys \cite{CIM_bisb_2024}. Few high-temperature CIs with large charge gaps have been proposed recently in stanene \cite{CIM_stanene_2018}  and monolayers of colinear antiferromagnets such as CrO and MoO \cite{CIM_CrO_2022, CIM_MoO_2023, CIM_afci_2024}. Given these candidate materials, it might be interesting to conduct first-principles studies to explore the stability of such composite structures and determine which of them are experimentally viable. Interestingly, our results, derived for an electronic system, may have direct implications for superconducting systems where Majorana physics has been extensively pursued. In fact, similar composite systems have been experimentally achieved in superconductor-semiconductor heterostructure \cite{das_zbp_2012, finck_zbp_2013, mourik_nanowire_2012} and magnetic adatoms-superconductor hybrid \cite{nadj-perge_observation_2014, pawlak_probing_2016, liebhaber_quantum_2022}.

\section{CONCLUSION}\label{sec_summ}
The idea of gapless boundary modes in free-fermionic SPT phases is one of the widely investigated topics in the quantum condensed matter community owing to their potential applications in technological aspects. In the search for these exotic zero-modes, along with clean systems, the role of impurities such as defects, vortices, heterojunctions, dislocations, and disclinations has become crucial. The question of their presence in composite interfaces between subsystems of different dimensionality has recently drawn a lot of attention, both theoretically and experimentally. In this article, we investigated a composite system consisting of different-dimensional topological subsystems and studied the emerging phases and phase transitions. While the case of lower-dimensional topological systems coupled to a higher-dimensional trivial environment is well known as an embedded topological insulator, as mentioned in the Introduction (\sect{sec_intro}), we delve into the seemingly complicated limit of both being topological - a topological wire in a topological vacuum. Considering a lattice model of a one-dimensional topological system (class BDI) coupled with a class D Chern insulating (square lattice) environment (\Sect{sec_ham}), we numerically uncover the exotic features of such a heterostructural system (\Sect{sec_numerics}). We find that while the wire retains both its parent TQCPs in the trivial insulating environment, similar to an embedded topological insulator, in the topological environment, remarkably, one of these TQCPs selectively vanishes, rendering triviality in an otherwise topological phase of the wire. This intriguing selective nature is a direct consequence of the coupling between multi-dimensional low-energy modes appearing in our composite systems, which can be comprehended in the proposed analytical theory (\Sect{sec_edge}). Our theory also explains numerically observed features such as `sub-system metal' and the emergence of `proximity-induced topology' in the trivial regime of the wire. We further unravel the crucial role of parity symmetry in protecting these features of the embedded wire and explain their generality in an exemplary case where the vacuum of the wire is a triangular lattice class D Chern insulator (\Sect{sec_spe}). We also explore the effects of breaking symmetries of the topological environment selectively (\Sect{sec_symmetry}). Furthermore, we discuss the generalization of these rich phenomena in other topological classes and in higher dimensions (\Sect{sec_dis}). We also discuss the possible materials and techniques required to realize our composite system, which may have important implications in experimental research and application in quantum technology.

While our study deals with a class of heterostructural topological systems where the environment dimension ($D=2$) is higher than that of the embedded subsystem ($d=1$), another class of interesting composite structure would be where a two-dimensional higher-order topological insulator \cite{benalcazar_HOTI_2017, frank_hoti_2018, matsugai_hoti_2018, queiroz_hoti_2019, trifunovic_hoti_2019, roy_hoti_2021} is embedded within a CI. In such systems, even though $d=D=2$, the corresponding boundary theories that come into play will reside in zero (for higher-order topological insulator) and in one (for CI) dimension. In our work, we have focused on symmetry-protected topological phases; however, these can be extended to topologically ordered systems where the study of `topological defects' has been explored for its implications on quantum qubits \cite{Bombin_twist_2010, Wen_anyoncondensation_2013, Barkeshli_defcts_2013, Barkeshli_defectsclass_2013, Andrej_topodefects_2013, Barkeshli_twist_2013, Diptarka_topoparamagnet_2016, Barkeshli_defects2_2019, Wang_topodefects_2024}. Presence of Floquet driving \cite{oka2019floquet, harper2020topology, mondal_fmbs_2023, mondal_fmbs2_2023}, Landau-like magnetic order \cite{Rachel_2018, tokura2019magnetic}, and their effect on such composite systems are some of the exciting 
future directions.

\section*{ACKNOWLEDGMENTS}
We acknowledge fruitful discussions with Amit Agarwal, Diptiman Sen, Ajit C. Balram, Saikat Mondal, Ritajit Kundu, Nirnoy Basak, and Soumya Sur. S.P. acknowledges funding from the IIT Kanpur Institute Fellowship. A.A. acknowledges support from the IITK Initiation Grant (IITK/PHY/2022010). Numerical calculations were performed on the workstations {\it Wigner} and {\it Syahi} at IITK.

\appendix

\section*{APPENDIX}

\section{QWZ model and QWZ wire}\label{apndx_bhz}
\subsection{QWZ model}\label{apndx_bhz1}
The topological phase diagram of the QWZ Hamiltonian (\eqn{eq_clsDHam}), characterized by Chern number ($\mathcal{C}$), is shown in \Fig{fig_bhzwire}(a). The transitions between different topological ($\mathcal{C} =\pm 1$) and trivial ($\mathcal{C} = 0$) phases are marked by gapless Dirac cones at special points on the Brillouin zone $(k_x, k_y)$. These points have also been shown in \Fig{fig_bhzwire}(a). In the topological regime, the system hosts chiral boundary modes. In \Fig{fig_bhzwire}(b), we compare the eigenspectrum when the lattice is kept under periodic or open boundary conditions (PBC/OBC). The spectrum for OBC shows the presence of mid-gap states at the Fermi energy ($E_F = 0$).  

To investigate the nature of these mid-gap states, we calculate their local density of states, LDOS, for a range of eigenstates between $E=E_F-\Delta$ to $E_F+\Delta$ defined as following
\begin{equation}
\text{LDOS}~(E_F, \Delta, \br) = \frac{1}{N_{E}}\sum_{\substack{i\\ |E_i -E_F| <\Delta}} \sum_{\alpha} 	{\big|{\psi_{i,\alpha}(\br)}\big|}^2,
\label{eq_ldos}
\end{equation}
where $\psi_{i,\alpha}(\br)$ specifies the $\alpha$ orbital-component ($\alpha\equiv A, B$) of the $i^{\text{th}}$ eigenstate with energy $E_i$ at lattice position $\br$ and $N_E$ is the total number of states within the energy range. For $\Delta = 0.1, E_F=0$ (OBC), the LDOS is shown in the inset of \Fig{fig_bhzwire}(b), which confirms the edge character of such states.

\subsection{QWZ wire}\label{apndx_bhz2}
\begin{figure}
\centering
\includegraphics[width=1\columnwidth]{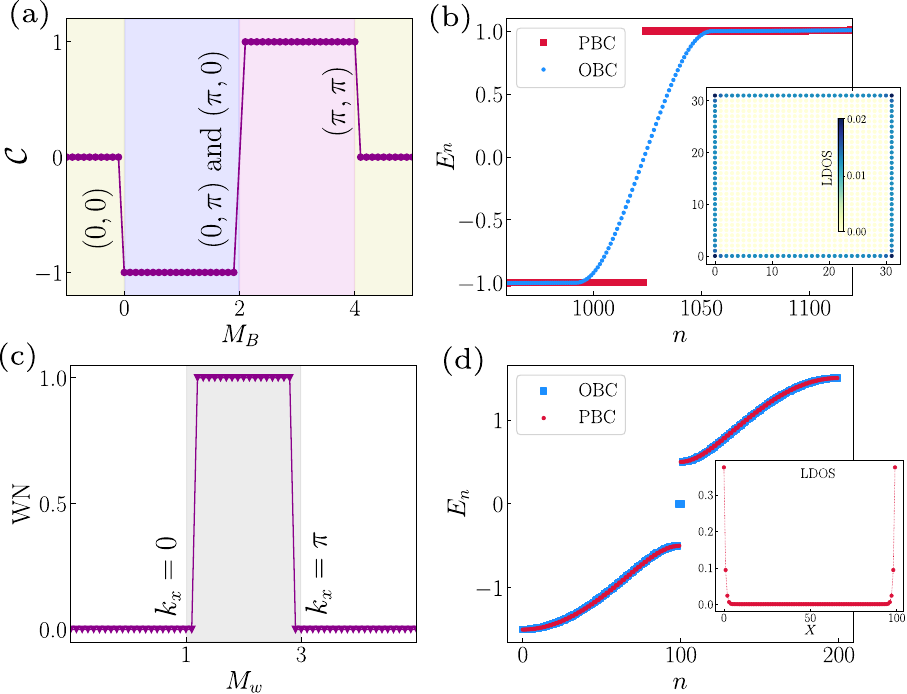}
\caption{(a) Topological phase diagram of QWZ model (\eqn{eq_clsDHam}) characterized by Chern number ($\mathcal{C}$) and TQCPs separating different topological phases. (b) Energy eigenvalues at $M_B = 1$. While the system with PBC shows a gap, mid-gap states appear with OBC. The system size is $32 \times 32$, but here, we zoom in close to the in-gap states. The inset shows LDOS of in-gap states localized at the boundary of the lattice, forming a CBS. (c) The QWZ wire Hamiltonian (\eqn{eq_wire}) shows a topological phase with a unit quantized winding number (WN). Corresponding TQCPs in one-dimensional Brillouin zone are also shown. (d) Energy eigenvalues at $M_w=2$ with both PBC and OBC for a wire of length $100$ suggest the presence of two in-gap modes with open boundaries, and the LDOS (inset) of those modes shows localization at the edges of the wire (EM).}
\label{fig_bhzwire}
\end{figure}

The QWZ wire Hamiltonian (\eqn{eq_wire}) has a topological phase characterized by winding number (WN) shown in \Fig{fig_bhzwire}(c), similar to paradigmatic Su-Schrieffer-Heeger (SSH) model \cite{SSH_PRB_1980}. In fact, the microscopic Hamiltonian in \eqn{eq_wire} can be mapped from the SSH model $\mathcal{H}_{\text{SSH}}(k) = \big[v-w\cos k\big]\sigma_x - w\sin k \sigma_y$, via unitary rotation as follows: $\mathcal{H}_{\text{SSH}}(k) \xrightarrow{U_R} \mathcal{H}_{\text{wire}}(k)$ where,
\beq
U_R = \exp{\left(-i\frac{\pi}{4}\left(\sigma_z+\sigma_y\right)\right)}.
\eeq
along with a parameter relabeling, $\frac{v}{w}\rightarrow (M_w -2)$. In the topological phase, an open QWZ wire consists of two zero-energy modes in its spectrum, shown in \Fig{fig_bhzwire}(d). Their corresponding LDOS (inset of \Fig{fig_bhzwire}(d)) shows they live on the open edges of the wire (EM).

\section{Bott index formulation}\label{apndx_bott}
In general, the Chern number ($\mathcal{C}$) of a two-dimensional system is computed from momentum-space eigenfunctions defined over the Brillouin zone. Such a momentum-space description becomes ill-defined in the absence of translational invariance. 
In disordered systems \cite{Agarwala_PRL_2017, Sahlberg_PRR_2020}, therefore, a topological system can be characterized using Bott index \cite{Loring_EPL_2010, Hastings_AOP_2011} - a real space analog of the Chern number.

Bott index is given by,
\begin{equation}
\mathcal{B} = \frac{1}{2\pi} \text{Im} \left[\text{Tr}\left(\log\big\{WUW^{\dagger}U^{\dagger}\big\}\right)\right], \label{eq_bott} 
\end{equation} 
where the matrices $U$ and $W$ can be obtained as follows: first, all the lattice points ($x_i, y_i$) are compactified on a torus using two position operators 
$\exp(i\Theta), \exp(i\Phi)$ where
$\Theta = \text{diag}\left[2\pi x_i/L_x\right]$ and $\Phi = \text{diag}\left[2\pi y_i/L_y\right]$. These are further projected on all the occupied states below the Fermi energy $E_F$ to get the matrices, 
\begin{align}
U &= \hat{P}\exp(i\Theta)\hat{P},\label{eq_U} \\
W &= \hat{P}\exp(i\Phi)\hat{P}.\label{eq_W} 
\end{align}
Here, $\hat{P} = \sum_{n, E_n\leq E_F} \ket{\psi_n}\bra{\psi_n}$ is the ground state projection operator,  $\ket{\psi_n}$ is the single-particle eigenstate with energy $E_n$.

\section{CBS solution and its orbital content}\label{apndx_cbs}
In order to obtain the low-energy theory of the CBS, we use the junction setup (see \Fig{fig_edge}(a) in the main text)  where two semi-infinite QWZ square lattices couple to a QWZ wire at $y=0$. The upper (lower) QWZ model has CBS, which decays $y>0$ ($y<0$). To obtain the dispersion of such CBS, we expand the bulk Hamiltonian near the gap-closing points. For instance, for 
$\mathcal{C}=+1$ phase, given the  
TQCP occurs at $(k_x, k_y) = (\pi, \pi)$
($M_B=4$), a low energy expansion of  $\mathcal{H}_{\text{QWZ}}(\vec{k})$ gives
\begin{align}
h_b = m_b \sigma_z - k_x \sigma_x - (-i\partial_y) \sigma_{y}, 
\end{align}
where $k_y=-i\partial_y$ and $M_B=(4-m_b)$. Considering the following trial wavefunctions \cite{shen_tibook_2012, bernevig_topobook_2013} for these CBSs,
\begin{align}
\ket{k_x, +} & \sim \begin{pmatrix}
	a_+\\
	b_+
\end{pmatrix} e^{-y/\xi} e^{ik_x x} ~\text{for $y>0$},\nonumber \\
\ket{k_x, -} & \sim \begin{pmatrix}
	a_-\\
	b_-
\end{pmatrix} e^{y/\xi} e^{ik_x x} ~~~\text{for $y<0$}, 
\end{align}
where $a_{\pm}, b_{\pm }$ are the orbital components. With $\xi \sim 1/m_b$, one finds $a_\pm=\pm b_\pm$ and the effective low energy Hamiltonian is given by $h_b \ket{k_x, \pm} = \mp k_x \ket{k_x, \pm}$. Thus defining a basis comprising of $c^{\dagger}_{k_x, \pm}|\Omega\rangle \equiv \ket{k_x, \pm}$, $h_{\text{CBS}}=-k_x \sigma_z$. A similar calculation  for $\mathcal C=-1$ requires an expansion of $M_B$ near $M_B=0$ leading to $h_{\text{CBS}}=k_x \sigma_z$.

\begin{widetext}

\section{Low-energy perturbation analysis of the junction Hamiltonian}\label{apndx_soft}
Let us consider the junction Hamiltonian of our composite system of study in the ribbon geometry $H_{\text{junction}} = H_{\text{wire}} + H_{\text{QWZ}} + H_{\text{coupling}}$, as pointed out in \eqn{eq_bhzcoupledwire} in the main text. Since $k_x$ is a good quantum number, we can rewrite the wire and the environment Hamiltonian in the one-dimensional Brillouin zone as, 
\begin{align}
	H_{\text{wire}} (k_x) & = \big[\sin k_x \sigma_x + \left(2- M_w -\cos k_x\right)\sigma_z \big]\psi_{k_x}^{\dagger} \psi_{k_x}, \nonumber \\
	H_{\text{QWZ}} (k_x) & = \sum_{n=1}^{L_y/2} \big[\sin k_x \sigma_x + \left(2- M_B -\cos k_x\right)\sigma_z\big] \psi_{\pm n\hat{y},k_x}^{\dagger} \psi_{\pm n\hat{y},k_x} \nonumber \\
	& ~~~~~+ \sum_{n=1}^{(L_y/2) - 1} T_{y} \Big(\psi_{+ n\hat{y},k_x}^{\dagger} \psi_{+ (n+1)\hat{y},k_x} + \psi_{- (n+1)\hat{y},k_x}^{\dagger} \psi_{ - n\hat{y},k_x}\Big) + \text{h.c.}
\end{align} 
with $\psi_{k_x}^{\dagger} = \begin{pmatrix}
	c^{\dagger}_{k_x, A} & c^{\dagger}_{k_x, B}
\end{pmatrix}$ represents the basis of the wire at $y=0$, $\psi_{\pm n\hat{y},k_x}^{\dagger} = \begin{pmatrix}
d^{\dagger}_{\pm n\hat{y},k_x, A} & d^{\dagger}_{\pm n\hat{y}, k_x, B}
\end{pmatrix}$ are the basis of environment above ($+n\hat{y}$) and below ($-n\hat{y}$) the wire and $T_y$ is $y$-hopping of the QWZ model given in \eqn{eq_rham}. The overall system has width $L_y$ in the $y$-direction with periodic boundary conditions. Now the microscopic coupling between the wire and the environment (see \eqn{eq_junction}) can be expressed for every $k_x$ as follows, 
\begin{align}
	H_{\text{coupling}} = \tilde{T}_y \Big(\psi_{k_x}^{\dagger}\psi_{+\hat{y},k_x} + \psi_{-\hat{y},k_x}^{\dagger}\psi_{k_x}\Big) + \text{h.c.} \label{eq_apndxcoup}
\end{align}
where $\tilde{T}_y$ given in \eqn{eq_kcoupmat} has two coupling parameter $\kappa_1$ and $\kappa_2$
which characterizes the strength of the coupling. 

\subsection{Vanishing of topological criticality of the wire}\label{apndx_soft1}
Following the main text, we first focus on the $\mathcal{C}=-1$ phase of the environment and $M_w=1$ (expanded as $M_w=1+m_w$) TQCP of the wire. Considering $H_{\text{coupling}}=0$ we perform a low energy expansion of the theory $H_{\text{junction}}$, and express it in the basis of CBS: $\ket{k_x, \pm}  =c^{\dagger}_{k_x, \pm}|\Omega\rangle$ and modes of the embedded wire: $|k_x, A/B \rangle =  c^{\dagger}_{k_x, A/B}|\Omega \rangle$, given by \eqn{cbs}, (\ref{eq_w1}) and (\ref{decou}),

\begin{align}
	\hat{H}_{\text{eff}} (\kappa_s=\kappa_a=0) = \begin{pmatrix}
		c^{\dagger}_{k_x, +} & c^{\dagger}_{k_x, -} & c^{\dagger}_{k_x, A} & c^{\dagger}_{k_x, B}
	\end{pmatrix} \begin{pmatrix}
		k_x & 0 & 0 & 0\\
		0 & -k_x & 0 & 0\\
		0 & 0 & -m_w & k_x\\
		0 & 0 & k_x & m_w
	\end{pmatrix} \begin{pmatrix}
	c_{k_x, +} \\ c_{k_x, -} \\ c_{k_x, A} \\ c_{k_x, B}
	\end{pmatrix}
\end{align}
Now, we turn on the coupling between the edge of the environment and the wire. The coupling terms in \eqn{eq_apndxcoup} are expanded in real space to give, 
\begin{align}
	\hat{H}_{\text{coupling}} = & \sum_{I,\epsilon=\pm1} \frac{\kappa_1}{2} \left( - c^{\dagger}_{I,A} d_{I+\epsilon \hat{y},A} + c^{\dagger}_{I,B} d_{I+\epsilon \hat{y},B} + \text{h.c.} \right) \nonumber \\
	&~~~~~~~~ + \sum_{I} \frac{\kappa_2}{2}  \bigg[\left(- c^{\dagger}_{I,A} d_{I+\hat{y},B} + c^{\dagger}_{I,B} d_{I+\hat{y},A} + \text{h.c.} \right) + \left(c^{\dagger}_{I,A} d_{I-\hat{y},B} -c^{\dagger}_{I,B} d_{I-\hat{y},A} + \text{h.c.} \right)\bigg]. \label{eq_microHy}
\end{align}
where the site indices $I$ represent $x$-positions along the wire. To project the coupling into the low-energy sector, we use the {\it soft-mode} expansion - which separates the {\it slow} and {\it rapid} spatial variations of the fermionic operators \cite{Houghton_bosonization_1993, shankar_RG_1994}. Since the low energy theory of the wire is near $k_x=0$, the soft-mode expansion will be, $c^{\dagger}_{I,A/B} \sim e^{i.0. I}\phi ^{\dagger}_{I,A/B}$ and in the $k_x$ space the operator support at each position will be $\frac{1}{\sqrt{L}} c^{\dagger}_{k_x,A/B}$, where $L$ is the length of the wire. As the CBS is also at the same momentum, we calculate the soft modes of the environment near $k_x=0$ and project them in the low energy CBS basis, using their solutions from Appendix~\ref{apndx_cbs} (also see \eqn{eq_cbssolutions} in the main text). The projection protocol is carried out in the following way,
\begin{align}
	d^{\dagger}_{I+\hat{y},A} & \rightarrow e^{i.0.I} \phi^{\dagger}_{I+\hat{y},A} \rightarrow \frac{1}{\sqrt{L}} d^{\dagger}_{k_x,A} \rightarrow \frac{1}{2 \sqrt{L}} \Big[ \Big(d^{\dagger}_{k_x,A} + d^{\dagger}_{k_x,B} \Big) + \Big(d^{\dagger}_{k_x,A} - d^{\dagger}_{k_x,B} \Big)\Big] \sim \frac{1}{\sqrt{L}} c^{\dagger}_{k_x,+}.
\end{align}
Table~\ref{tbl_soft} shows the soft-mode expansion of all such modes and their low-energy projection. Using these, we write the coupling Hamiltonian (\eqn{eq_microHy}) in the low-energy sector as follows, 
\begin{align}
	{\tilde{H}}_{\text{coupling}} = & - \frac{\kappa_1}{2} \bigg[\Big(c^{\dagger}_{k_x,A} c_{k_x,+} + c^{\dagger}_{k_x,A} c_{k_x,-} + \text{h.c.} \Big) -\Big(c^{\dagger}_{k_x,B} c_{k_x,+} - c^{\dagger}_{k_x,B} c_{k_x,-} +\text{h.c.} \Big)\bigg] \nonumber \\
	& ~~~~- \frac{\kappa_2}{2} \bigg[\Big(c^{\dagger}_{k_x,A} c_{k_x,+} + c^{\dagger}_{k_x,A} c_{k_x,-} + \text{h.c.} \Big) -\Big(c^{\dagger}_{k_x,B} c_{k_x,+} - c^{\dagger}_{k_x,B} c_{k_x,-} +\text{h.c.} \Big)\bigg].
\end{align}
$1/L$ factor vanishes after summing over all positions $I$. Now putting $\kappa_s = \frac{(\kappa_1 + \kappa_2)}{2}$, the low-energy effective Hamiltonian of the whole system in the presence of ${H}_{\text{coupling}}$ becomes, 
\begin{align}
	\hat{H}_{\text{eff}} =\hat{H}_{\text{eff}}(\kappa_s=\kappa_a=0)+ {\tilde{H}}_{\text{coupling}}= \begin{pmatrix}
		c^{\dagger}_{k_x, +} & c^{\dagger}_{k_x, -} & c^{\dagger}_{k_x, A} & c^{\dagger}_{k_x, B}
	\end{pmatrix} \begin{pmatrix}
		k_x & 0 & -\kappa_s & \kappa_s\\
		0 & -k_x & -\kappa_s & -\kappa_s\\
		-\kappa_s & -\kappa_s & -m_w & k_x\\
		\kappa_s & -\kappa_s & k_x & m_w
	\end{pmatrix} \begin{pmatrix}
		c_{k_x, +} \\ c_{k_x, -} \\ c_{k_x, A} \\ c_{k_x, B}
	\end{pmatrix},
\end{align}
which can be rewritten in the matrix form: $H_{\text{eff}} = H_{\text{eff}}(\kappa_s=\kappa_a=0) - \kappa_s \Big( \1 \otimes \tau_x + \sigma_y \otimes \tau_y \Big)$, thus both CBS and the topological criticality of the wire at $M_w=1$, gap out each other, where the gap is linearly proportional to $\kappa_s$. The numerical results obtained using the ribbon geometry depicted in \Fig{fig_gapbehave}(a) exhibit the same characteristics as those predicted by our analytical calculations.

\begin{table*}
	\centering
	\begin{tabular}{c c c} 
		\hline
		\hline
		
		~~~~~~Operator & Soft-mode & {Projection on CBS~~~~} \\
		\midrule
		$d^{\dagger}_{I+\hat{y},A}$ &~~~~ $e^{i.0. I} \phi^{\dagger}_{I+\hat{y},A} \rightarrow \frac{1}{\sqrt{L}} d^{\dagger}_{k_x,A}$ ~~~~& $\frac{1}{\sqrt{L}}  c^{\dagger}_{k_x,+}$\\[0.15cm]
		$d^{\dagger}_{I+\hat{y},B}$ &~~ $e^{i.0.I} \phi^{\dagger}_{I+\hat{y},B} \rightarrow \frac{1}{\sqrt{L}} d^{\dagger}_{k_x,B}$ ~~& $\frac{1}{\sqrt{L}}  c^{\dagger}_{k_x,+}$\\ 
		[0.15cm]
		$d^{\dagger}_{I-\hat{y},A}$ & $e^{i.0. I} \phi^{\dagger}_{I-\hat{y},A} \rightarrow \frac{1}{\sqrt{L}} d^{\dagger}_{k_x,A}$ & $\frac{1}{\sqrt{L}}  c^{\dagger}_{k_x,-}$\\ 
		[0.15cm]
		$d^{\dagger}_{I-\hat{y},B}$ & $e^{i.0.I} \phi^{\dagger}_{I-\hat{y},B} \rightarrow \frac{1}{\sqrt{L}} d^{\dagger}_{k_x,B}$ & $-\frac{1}{\sqrt{L}} c^{\dagger}_{k_x,-}$\\[0.1cm] 
		\hline
		\hline
	\end{tabular}
	\caption{Soft-mode expansion of fermions near $k_x=0$ living on the edges of the environment with $\mathcal{C}=-1$ and their projection on CBS.}
	\label{tbl_soft}
\end{table*} 

\subsection{Virtual coupling between CBS above and below the wire}\label{apndx_soft2}
Since the low energy theory governing the CBS is at $k_x=0$, it gets dominantly affected by the wire Hamiltonian near $k_x=0$; thus, to capture the latter's effect, we model the wire Hamiltonian up to zeroth order in $k_x$ as $H_{\text{wire}} = \left(1- M_w \right)\sigma_z$.  The correction to the CBS sector (\eqn{cbs}) under coupling with these higher energy sites will be of the form $\tilde{h}_{\text{CBS}} = H_{\text{above}} + H_{\text{below}} + H_{\text{above-below}}$, where $H_{\text{above}}$ and $H_{\text{below}}$ are the diagonal correction and $H_{\text{above-below}}$ is the cross term representing effective couplings between the upper and lower CBS via wire in the middle. We now use the second-order perturbation theory treating $\hat{H}_{\text{coupling}}$ from \eqn{eq_microHy} as perturbation to get, 
\begin{align}
	H_{\text{above/below}} & =  \sum_{I}\left[\frac{\kappa_{1}^2-\kappa_{2}^2  }{4(1-M_w)}\right] \left(d^{\dagger}_{I\pm \hat{y},A} d_{I\pm \hat{y},A} - d^{\dagger}_{I\pm \hat{y},B} d_{I\pm \hat{y},B}\right) \nonumber \\
	H_{\text{above-below}} & = \frac{1}{4} \sum_{I}\left[\frac{\kappa_{1}^2 +\kappa_{2}^2 }{1-M_w}\right] \left\{ \left(d^{\dagger}_{I+ \hat{y},A} d_{I - \hat{y},A} - d^{\dagger}_{I+\hat{y},B} d_{I-\hat{y},B} \right) + \text{h.c.} \right\} \nonumber\\
	& ~~~~~~- \frac{1}{4}  \sum_{I}\left[\frac{2\kappa_{1} \kappa_{2}}{1-M_w}\right] \left\{ \left(d^{\dagger}_{I+ \hat{y},A} d_{I - \hat{y},B} - d^{\dagger}_{I+\hat{y},B} d_{I-\hat{y},A} \right) + \text{h.c.} \right\},
\end{align}
After projecting these into the CBS basis using soft modes given in TABLE~\ref{tbl_soft}, $H_{\text{above/below}}$ vanishes, and the overall correction becomes,
\begin{align}
	\tilde{h}_{\text{CBS}} = \half \left[\frac{\left(\kappa_{1} + \kappa_{2}\right)^2}{1-M_w}\right]\left(c^{\dagger}_{k_x,+} c_{k_x,-} + ~\text{h.c.}\right) = \left[\frac{2\kappa_{s}^2}{1-M_w}\right]\left(c^{\dagger}_{k_x,+} c_{k_x,-} + ~\text{h.c.}\right)\label{apndx_hb}
\end{align} 
This, in fact, holds true for any value of $M_w$ apart from $M_w=1$. However, at $M_w=1$, this second order correction is redundant since the CBS and the wire gaps out in the first order (see Appendix~\ref{apndx_soft1}). Numerical data shown in \Fig{fig_gapbehave}(b,c) confirms this. Note that, at very large $|M_w|$ the gap between CBS vanishes as $\sim \frac{1}{M_w}$.

\begin{figure}
	\centering
	\includegraphics[width=1\columnwidth]{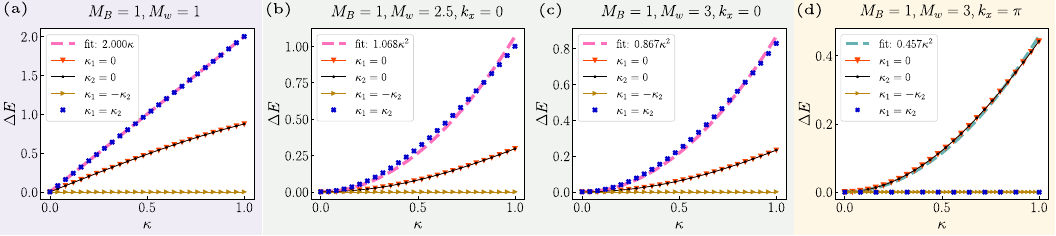}
	\caption{(a) Gap opening ($\Delta E$) for both the CBS (at $M_B=1 $) and $M_w=1$ critical point of the wire (both are at $k_x=0$) in presence of coupling parameter $\kappa$. Numerical results with ribbon geometry show $\Delta E \sim(\kappa_1 + \kappa_2)\sim \kappa_s$. (b) When the wire is gapped ($M_w=2.5$), CBS above and below the wire coupled with each other and gap out as $\Delta E \sim(\kappa_1 + \kappa_2)^2\sim \kappa_{s}^2$. (c) Similar phenomena happen to CBS at $M_w=3$, where although the wire is critical, it is at $k_x=\pi$. (d) The correction to the wire at $M_w=3$ critical point show gap opening $\Delta E \sim (\kappa_{1}^2 - \kappa_{2}^2)\sim \kappa_{s}\kappa_a$. Note that this correction can only shift the $M_w=3$ critical point as described in the main text.}
	\label{fig_gapbehave}
\end{figure}

\subsection{Survival of the subsystem metal}\label{apndx_soft3}
At $M_w=3$ where a TQCP for the wire occurs at $k_x=\pi$, the wire Hamiltonian can be expanded  as: $h_{\text{wire}} = -m_w\sigma_{z} -k_x \sigma_x$ with $m_w\rightarrow0$ ($M_w=3+m_w$) (see \eqn{eq_w3}). The CBS of the environment is still at $k_x=0$, thus soft modes in the environment shown in TABLE~\ref{tbl_soft} remain the same, but for the critical wire, it becomes, 
\begin{align}
	c^{\dagger}_{I,A/B} \sim e^{i\pi I}\phi ^{\dagger}_{I,A/B} \longrightarrow \frac{1}{\sqrt{L}}(-1)^I c^{\dagger}_{k_x,A/B}.
\end{align}
Consequently, the coupling $\hat{H}_{\text{coupling}}$ (\eqn{eq_microHy}) under projection now has a $(-1)^I$ factor at every site, which when summed leads to {\it zero}. Thus, in the first order, the coupling term cannot gap out the system. In the second order, while the CBS sector gets gapped (see Appendix~\ref{apndx_soft2}), the wire sector remains gapless, as we now show. At $k_x=\pi$, ${H}_{\text{QWZ}}(k_x=\pi) = \sin k_y \sigma_y +  \left(3-M_B-\cos k_y\right)\sigma_z$ which has an emergent chiral symmetry, $\sigma_{x}{H}_{\text{QWZ}}(k_x=\pi)\sigma_{x} = -{H}_{\text{QWZ}}(k_x=\pi)$ and time-reversal symmetry, $\sigma_{z}{H}^{*}_{\text{QWZ}}(k_x=\pi)\sigma_{z} = {H}_{\text{QWZ}}(k_x=\pi)$. This restricts at any $k_x$ the bulk eigenspectrum has the following form,
\begin{align}
	d^{\dagger}_{+\epsilon}\ket{\Omega} & = \begin{pmatrix}
		a\\
		i\sqrt{1-a^2}
	\end{pmatrix}\text{~with energy~} (+\epsilon) \text{~and~} d^{\dagger}_{-\epsilon}\ket{\Omega} = \begin{pmatrix}
		-\sqrt{1-a^2}\\
		ia
	\end{pmatrix}\text{~with energy~} (-\epsilon), 
\end{align}
where $0<a<1$. Thus the site just adjacent to the wire can be expanded into pairs of states with energies $\epsilon$ and $-\epsilon$ with weights such that   
\begin{align}
	d^{\dagger}_{\pm \hat{y},k_x, A} & \propto ad^{\dagger}_{+\epsilon} -\sqrt{1-a^2} d^{\dagger}_{-\epsilon},\nonumber\\ 
	d^{\dagger}_{\pm \hat{y},k_x, B} & \propto i\sqrt{1-a^2}d^{\dagger}_{+\epsilon} +ia d^{\dagger}_{-\epsilon}
\end{align}
Using this, we expand the coupling  Hamiltonian (\eqn{eq_apndxcoup}) in the basis of $d^{\dagger}_{\pm\epsilon}\ket{\Omega}$ and then carry out the second order perturbation correction in the low energy sector of the wire, which thus has the form, 
\begin{align}
	\tilde{h}_{\text{wire}} & \propto (2a^2-1) \left[\frac{\kappa_{1}^2-\kappa_{2}^2}{\epsilon} \right]\left( c^{\dagger}_{k_x,A} c_{k_x,A} - c^{\dagger}_{k_x,B} c_{k_x,B} \right)\nonumber \\ & \propto (2a^2-1) \left[\frac{4\kappa_{s}\kappa_{a}}{\epsilon} \right]\left( c^{\dagger}_{k_x,A} c_{k_x,A} - c^{\dagger}_{k_x,B} c_{k_x,B} \right).
\end{align}
Overall theory of the wire will now become $h_{\text{wire}} = -(m_w - \kappa_{s}\kappa_{a} \mu/\Delta)\sigma_{z} -k_x \sigma_x$ where $\mu$ is a constant and $\Delta$ is the bulk gap scale. 
In \Fig{fig_gapbehave}(d), we show the numerically obtained gap of the $k_x=\pi$ mode at $M_w=3$, following the same behavior. Interestingly, this perturbation does not open up any gap; instead, it shifts the criticality of the wire to a different $M_w$ value, $M^{'}_w=3-4\kappa_{s}\kappa_{a}\mu/\Delta$, thus making the subsystem metal robust. In atomic limit of the environment where, $\Delta = (2-M_B)$, the shift will be $M^{'}_w=3 + \mu \kappa_{s}\kappa_{a}/(M_B-2)$.
		
\end{widetext}

\section{Polarization of proximity-induced edge-modes}\label{apndx_pol}

Polarization is the real space representation of the topological invariant in one-dimensional systems such as winding number \cite{resta_1998, Bianco_PRB_2011}. It is defined as
\begin{equation}
	\mathcal{P} = \text{Im} \Big\{\text{Tr}\big[\ln(D)\big]\Big\}~ \text{mod $2\pi$}, \label{eq_pol}
\end{equation}
where $D \equiv U$ in order to measure $x$-polarization $\mathcal{P}_x$ and $D \equiv W$ for measuring the $y$-polarization $\mathcal{P}_y$ ($U$ and $W$ is given in \eqn{eq_U} and \eqn{eq_W} respectively). For one-dimensional SPT phases, in the presence of two topological EM, the polarization will take a quantized value of $1$ in the unit of $\pi$. Now, due to the complex nature of our composite system, we take the following protocol in order to numerically deduce the polarization.

\begin{figure}
	\centering
	\includegraphics[width=1\columnwidth]{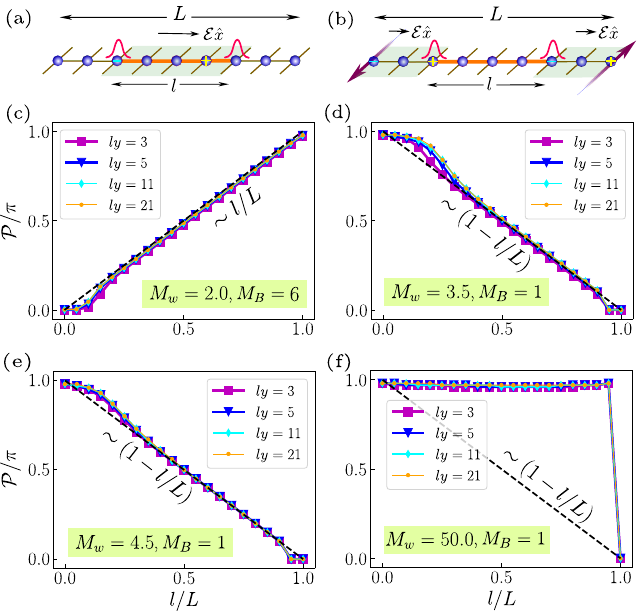}
	\caption{(a) In the trivial environment of linear size $L$, edge modes in a topological wire of length $l$ create two charge centers when a small electric field $\mathcal{E} \hat{x}$ is applied in the wire (shaded region). (b) For the topological environment, proximity-induced EMs, along with CBS at global boundaries, create four charge centers. (c) Polarization shows $\mathcal{P}/\pi = l/L$ behavior in the topological regime of the wire, i.e., $M_w=2$ when the environment is trivial ($M_B=6$). For $M_w=3.5$ (d) and $M_w=4.5$ (e), in the topological environment ($M_B=1$), $\mathcal{P}/\pi = (1-l/L)$. (f) No such behavior is observed at large $M_w=50$, where EM disappears. For all the cases, $L=60$, and the electric field strength is $0.0001$. All the results are independent of lattice width along $y$-direction $l_y$.}
	\label{fig_pol}
\end{figure}

\subsection{Polarization calculation protocol} Since the embedding of the wire is along $x$-direction, we will always calculate the $\mathcal{P} = \mathcal{P}_x$ only. Our composite system is a $l$ size wire placed in the middle of a $L \times l_y$ lattice of topological environment. While in $y$-direction, the environment is periodic, $x$-direction is kept open such that it creates $l_y$ number of one-dimensional systems of length $L$ stacked along $y$-direction, and the CBS on the open global boundaries looks like $2l_y$ pseudo-EMs. Now, according to the definition, every pair of pseudo-EMs will contribute $\pi$ polarization, then for the whole system, $\mathcal{P} = l_y\pi$ \text{mod $2\pi$}. This restricts us from probing the CBS as it will give zero for even $l_y$. To counter that, we always choose odd $l_y$ such that $\mathcal{P}$ takes a non-trivial value in the presence of global CBS. In the trivial limit of the environment (absence of CBS), $\mathcal{P}$ will probe the topological wire hosting EM. This protocol allows us to calculate $\mathcal{P}$ of the one-dimensional region (we call it a pseudo-wire), along which embedding has taken place. 

Now, as the system intuitively reduces to a $l$ sized wire embedded in the middle of a $L$ sized pseudo-wire (see \Fig{fig_pol}(a,b) for the trivial and topological limit of the environment), to get the polarization response, we now apply $x$-directional infinitesimal electric fields locally of the EMs so that it creates virtual charge center at the EM sites. For instance, in the trivial environment, since there is no CBS, the pseudo-wire does not host any pseudo-EMs; only the embedded topological wire has EM, hence the electric field will be applied only on the sites of the embedded wire (shaded region \Fig{fig_pol}(a)). In the topological environment, the presence of CBS will create pseudo-EMs, and with certain parameter values, the embedded wire will host proximity-induced EMs, thus creating two pairs of EMs at the two far sides of the whole system (see \Fig{fig_pol}(b)). So, in this limit, we have to apply the electric field on the sites excluding the bulk of the embedded wire as shown in shaded regions in \Fig{fig_pol}(b). The applied electric field of infinitesimal strength $\mathcal{E}$ will result in an extra onsite potential energy term $\hat{V}=-\mathcal{E}\hat{X}$ in the Hamiltonian, where the $x$-position operator $\hat{X}$ is defined only on the local regions depending on the environment as described just above. Note that to maintain the Fermi level at zero energy, we choose $\hat{X}$ to be centered (i.e., $x=0$ site) in the middle of a shaded region. Such protocols to calculate polarization numerically by applying an external electric field have been studied very recently \cite{agarwala_ssh_2021, Cristiane_polE_2025}.

With these protocols, we now calculate $\text{Im} \left\{\text{Tr}[\ln(D)]\right\}$ from \eqn{eq_pol}. In general, $\text{Im} \left\{\ln(Z)\right\}$ calculate the phase of the complex number $Z$ within [$-\pi$, $\pi$), but because of the summation (trace), results may not be bounded, so using $2\pi$ periodicity we bring it back within that range and add $\pi$ to make it between $0$ to $2\pi$. After this, we deduce the polarization using \eqn{eq_pol}.

\subsection{Non-quantized polarization}
Using the above-mentioned protocol, we now calculate $\mathcal{P}$ for our composite system with $\mathcal{E} =0.0001$ (in the unit of hopping/distance). When in the trivial environment, a $l$ sized region containing EMs appears in the middle of a $L$ sized pseudo-wire (see \Fig{fig_pol}(a)), we expect non-quantized polarization, which grows linearly with $l/L$ \cite{Mondal_PRB_2023}. In \Fig{fig_pol}(c), we numerically show the non-quantized linear behavior of $\mathcal{P}$, which is independent of width $l_y$. In the topological environment, for a phase-space region where proximity-induced EMs appear along with pseudo-EMs, two regions of length $(L-l)/2$ are created at two opposite sides of the pseudo-wire. Since polarization is additive in nature \cite{Mondal_PRB_2023}, $\mathcal{P}/\pi$ for the whole system in \Fig{fig_pol}(d) and \Fig{fig_pol}(e) show $(1-l/L)$ behavior. At a very large value of $M_w$, proximity-induced EM vanishes, and only pseudo-EMs remain on the edges of the pseudo-wire rendering unit quantization of $\mathcal{P}/\pi$ shown in \Fig{fig_pol}(f) until where $l=L$, where pseudo-EMs can not appear.

\begin{figure}
	\centering
	\includegraphics[width=1\columnwidth]{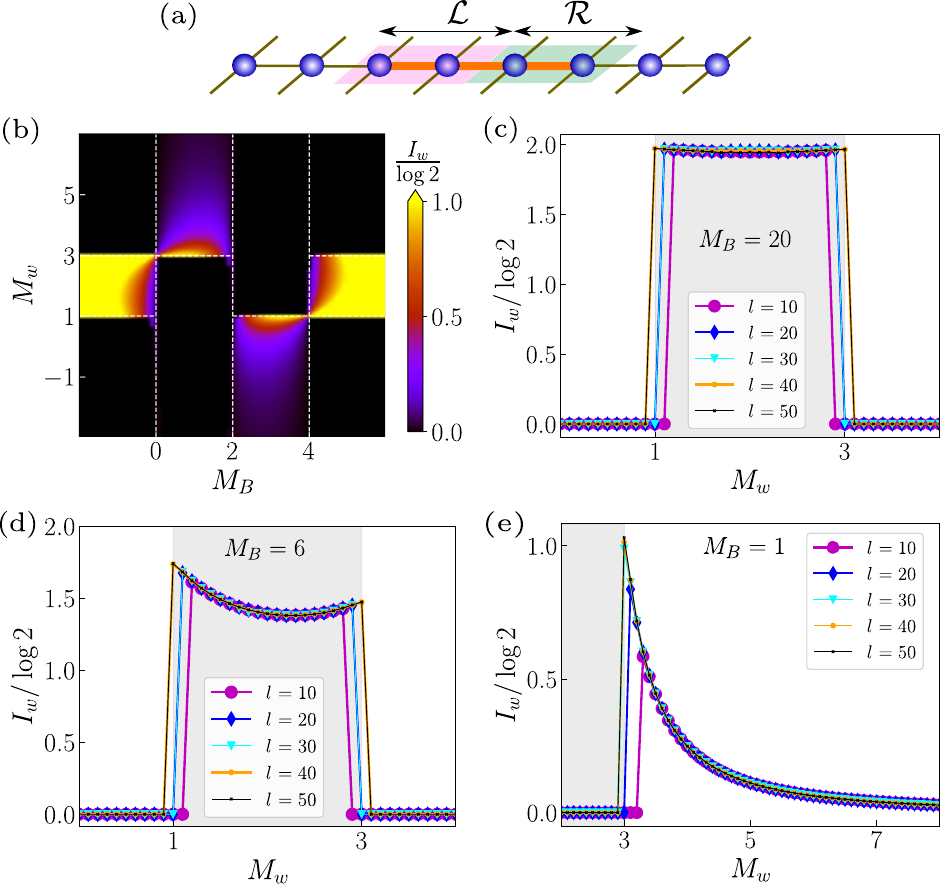}
	\caption{(a) Partition of the embedded wire into left ($\cal{L}$) and right ($\cal{R}$) half. Mutual information $I_w$ is calculated between subsystem $\cal{L}$ and $\cal{R}$. (b) $I_w$ phase diagram in the ($M_B$ - $M_w$) parameter space for the finite geometry case with $l=40$ sized wire embedded in the environment of size $60 \times 21$. (c) (d) and (e) show system size ($l$) independence of MI for both parent and proximity-induced edge modes along the Cut $M_B=20, 6$ (trivial environment; parent edge modes) and $M_B=1$ (topological environment; proximity-induced edge modes).}
	\label{fig_mi}
\end{figure}

\section{Mutual information of proximity-induced edge-modes}\label{apndx_mi}
Edge localized states of free fermionic topological systems are, in general, short-ranged entangled, leading to signatures in bipartite entanglement entropy and mutual information \cite{Fidkowski_PRL_2010, Turner_PRB_2011, agarwala_ssh_2021}. Mutual information (MI) captures entanglement between two parts of a system without any contribution from other remaining parts \cite{Adami_PRA_1997, Wolf_PRL_2008}, thus making it a good probe of edge-modes in composite systems such as ours.  Given two subsystems - the left ($\cal{L}$) and right ($\cal{R}$) half of the embedded wire (see \Fig{fig_mi}(a)), we first calculate the bipartite entanglement entropy of them $S_{\cal{L}}$ and $S_{\cal{R}}$, respectively, using correlation matrix formalism \cite{Peschel_JPA_2009}. Now, the MI between these two parts of the wire is, 
\beq
I_w = S_{\cal{L}} + S_{\cal{R}} - S_{w}, \label{eq_mi}
\eeq
where $S_w$ (where $w\equiv\cal{L} \cup \cal{R}$) is the bipartite entanglement entropy of the embedded wire with the environment. In the presence of gapless topological EM, we expect MI will give rise to a finite value in the unit of $\log 2$. \Fig{fig_mi}(b) shows the MI phase diagram evaluated in the unit of $\log 2$ for states within the energy range $\Delta E \sim 0.1$ (to capture only edge contribution) when the system comprises a $l=40$ sized wire is embedded on a $60 \times 21$ environment. We find not only does EM appearing in the trivial environment show a signature in MI, but the EMs induced by the topological environment are also finitely entangled. Since these proximity-induced EMs vanish at large $M_w$, MI goes to zero smoothly. In \Fig{fig_mi}(c,d,e) we show $I_w$ behavior with $M_w$ in the trivial environment ($M_B=20, 6$) and in the topological environment ($M_B=1$) for different sizes of wire $l$. These results suggest that MI for proximity-induced EMs is independent of system size, similar to the parent edge modes of the one-dimensional edge modes.

\begin{figure*}
	\centering
	\includegraphics[width=0.85\linewidth]{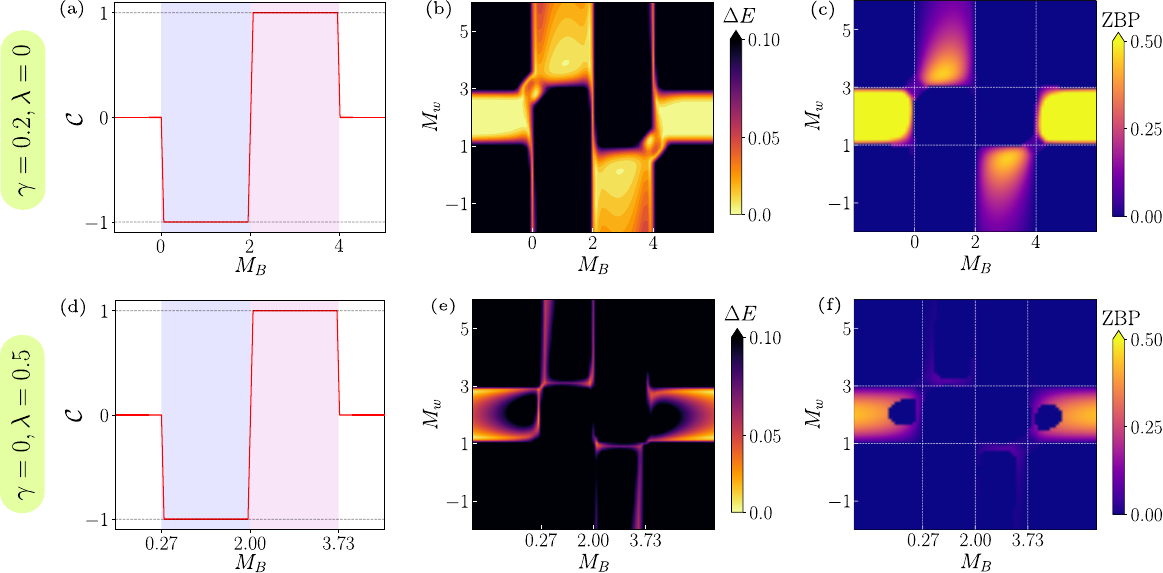}
	\caption{(a) Global Chern insulating phases of the class A Hamiltonian in \eqn{eq_clsAk} with $\gamma=0.2$ and $\lambda=0$. In this class A topological environment of linear size $L=40$, we embed a $l=24$ sized wire and calculate (b) spectral gap ($\Delta E$) and (c) zero-bias peak (ZBP) phase diagram in the ($M_B$ - $M_w$) parameter space. (d), (e) and (f) are the same as (a), (b), and (c), respectively, but for $\gamma=0$ and $\lambda=0.5$. For all cases, the global outer boundary of the environment is periodic.}
	\label{fig_clsAwire}
\end{figure*}

\section{Symmetry structure of the coupling Hamiltonian}\label{apndx_symmetry}
In the main text in \eqn{eq_cgsp} and \eqn{eq_tcsp}, we show that the environment and the embedded wire both have charge-conjugation symmetry (CGS) with $\sigma_{x}$ operator and parity with $\sigma_z$, as their common symmetries. Under the CGS, fermionic degrees of freedom (dof) in real space transform as follows,
\begin{align}
	\begin{pmatrix}
		c_{A} \\ c_{B}
	\end{pmatrix}\rightarrow \sigma_{x}
	\begin{pmatrix}
		c^{\dagger}_{A} \\ c^{\dagger}_{B}
	\end{pmatrix}=
	\begin{pmatrix}
		c^{\dagger}_{B} \\ c^{\dagger}_{A}
	\end{pmatrix}, \label{eq_symtrans1}
\end{align}
(see \eqn{eq_sxpsi}).  So, in the momentum space, they will follow, $c^\dagger_{k,\alpha} \rightarrow c_{-k,\overline{\alpha}}$,
where if $\alpha =A$, then $\overline{\alpha}=B$ and vice-versa. Now we use parity operation $k \rightarrow -k$, on top of this such that,
\beq
c^\dagger_{k,\alpha} \xrightarrow{\text{~CGS $\otimes$ Parity~}} c_{k,\overline{\alpha}}, \label{eq_symtrans}
\eeq
implement the symmetry of our composite system. Consequently, the relabeled dof of the wire $f^\dagger_{k_x,\alpha}$ ($\equiv c^\dagger_{k_x,\alpha}$ with $\alpha=A, B$), used in \sect{sec_spe}, will transform as: $f^\dagger_{k_x,\alpha} \rightarrow f_{k_x,\overline{\alpha}}$, following \eqn{eq_symtrans}. For the CBS, the relabeled dof are given by $d^\dagger_{k_x, \beta} \equiv c^\dagger_{k_x,\beta} \sim (c^\dagger_{k_x,A} + \beta c^\dagger_{k_x,B})$ with $\beta=\pm 1$. The symmetry implementation on these operators will be, 
\begin{align}
	d^\dagger_{k_x, \pm} \sim \left(c^\dagger_{k_x,A} \pm c^\dagger_{x,B}\right) & \xrightarrow{\text{~CGS $\otimes$ parity~}} \left(c_{k_x,B} \pm c_{k_x,A}\right) \nonumber \\
	&~~~~~~~~~~~ \longrightarrow \pm d_{k_x, \pm}
\end{align}
thus, $d^\dagger_{k_x, \beta} \rightarrow \beta d_{k_x, \beta}$. 

A general form of the hybridization $H_{\text{hyb}}$, between the wire and the CBS, is given in \eqn{eq_hyb1}. This coupling Hamiltonian $H_{\text{hyb}}$, under the symmetry of the composite system, will transform as,
\begin{align}
	 & \sum_{\alpha \beta} \Big( V_{\alpha \beta} f^\dagger_{k_x, \alpha} d_{k_x, \beta} + V^{*}_{\alpha \beta} d^\dagger_{k_x, \beta} f_{k_x, \alpha} \Big) \nonumber\\
	 \longrightarrow &  \sum_{\alpha \beta} \Big( \beta V_{\alpha \beta} f_{k_x, \overline{\alpha}} d^\dagger_{k_x, \beta} + \beta V^{*}_{\alpha \beta} d_{k_x, \beta} f^\dagger_{k_x, \overline{\alpha}} \Big)
	  \nonumber\\
	 \longrightarrow &  \sum_{\alpha \beta} \Big(-\beta V_{\overline{\alpha} \beta} d^\dagger_{k_x, \beta} f_{k_x, \alpha} - \beta V^{*}_{\overline{\alpha} \beta} f^\dagger_{k_x, \alpha} d_{k_x, \beta} \Big)
\end{align}
for every $k_x$. Hence, the symmetry allowed hybridization will have the condition $V_{\alpha \beta} = -\beta V^{*}_{\overline{\alpha} \beta}$, which gives the following restrictions in the coupling terms: $V_{A,+} = -V^{*}_{B, +}$ and $V_{A,-} = V^{*}_{B, -}$. 

Similarly, symmetry restrictions on the hybridization via higher-order processes $H^{\text{wire}}_{\text{hyb}}$ and $H^{\text{CBS}}_{\text{hyb}}$ (see \eqn{eq_hyb2} and (\ref{eq_hyb3})), we get the nature of the allowed coupling terms as discussed in the main text.

\section{Topological wire in class A topological vacuum}\label{apndx_clsA}
The class A Chern insulator model (see \eqn{eq_clsAHam}) is described by the momentum space Hamiltonian,
\begin{align}
	\mathcal{H}_{\text{classA}}(\vec{k}) = & ~\gamma\big(\cos k_x + \cos k_y\big)\1 \nonumber \\
	& + \big(\sin k_x - \lambda\big) \sigma_x + \big(\sin k_y - \lambda \big) \sigma_y \nonumber \\
	& + \big(2 - M_B - \cos k_x-\cos k_y\big)\sigma_z.
	\label{eq_clsAk}
\end{align}
In the presence of $\gamma$, the system breaks charge-conjugation symmetry, and a finite $\lambda$ breaks parity, as discussed in the main text. The system has both $\mathcal{C}=\pm1$ phases separated by the critical points shown in Table~\ref{tab_clsAtqcps}. In particular the parameter region $(2-2\sqrt{1-\lambda^2}) <M_B<2$ is $\mathcal{C}=-1$ phase and $(2-2\sqrt{1-\lambda^2}) <M_B<2$ has $\mathcal{C}=+1$. Thus a non-zero $\gamma$ cannot change the TQCPs if $\lambda=0$. On a square lattice, the hopping structure of this model is given by, 
\begin{align}
	H & = \sum_i \begin{pmatrix}
		c^{\dagger}_{iA} & c^{\dagger}_{iB}
	\end{pmatrix} 
	\begin{pmatrix}
		\frac{(\gamma-1)}{2} & -\frac{i}{2} \\
		-\frac{i}{2} & \frac{(\gamma+1)}{2}
	\end{pmatrix} 
	\begin{pmatrix}
		c_{i+\hat{x} A} \\ c_{i+\hat{x} B}
	\end{pmatrix} + \text{h.c.} \nonumber \\
	& + \sum_i 
	\begin{pmatrix}
		c^{\dagger}_{iA} & 	c^{\dagger}_{iB}
	\end{pmatrix} 
	\begin{pmatrix}
		\frac{(\gamma-1)}{2} & 	-\frac{1}{2} \\
		\frac{1}{2} & 	\frac{(\gamma+1)}{2}
	\end{pmatrix} 
	\begin{pmatrix}
		c_{i+\hat{y} A} \\ 	c_{i+\hat{y} B}
	\end{pmatrix} + \text{h.c.} \nonumber \\ 
	& + \sum_i 
	\begin{pmatrix}
			c^{\dagger}_{iA} & 	c^{\dagger}_{iB}
	\end{pmatrix} 
	\begin{pmatrix}
		2-M_B & \lambda(i-1) \\
		\lambda(-i-1) & M_B-2
	\end{pmatrix} 
	\begin{pmatrix}
		c_{iA} \\ c_{iB}
	\end{pmatrix}
	\label{eq_rhamclsA}
\end{align}

\begin{table}[H]
	\centering
	\begin{tabular}{c c} 
		\hline
		\hline
		~Gap closing $\vec{k}$ points~ & ~critical $M_B$ value~\\
		\midrule
		$\left(\sin^{-1}\lambda, \sin^{-1}\lambda\right)$ & $2-2\sqrt{1-\lambda^2}$\\ 
		[0.1cm]
		~$\left(\pi - \sin^{-1}\lambda, \sin^{-1}\lambda\right)$~ & $2$\\ [0.1cm]
		~$\left( \sin^{-1}\lambda, \pi - \sin^{-1}\lambda\right)$~ & $2$\\ [0.1cm]
		$\left(\pi - \sin^{-1}\lambda, \pi- \sin^{-1}\lambda\right)$ & ~$2+2\sqrt{1-\lambda^2}$~ \\ [0.1cm]
		\hline
		\hline
	\end{tabular}
	\caption{Gap closing in the Brillouin zone and corresponding TQCPs of the class A CI model given in \eqn{eq_clsAk}.}
	\label{tab_clsAtqcps}
\end{table}

In this class A topological environment (of size $L \times L$), we now create an $l$ sized topological wire having the same Hamiltonian as given in \eqn{eq_wire}, to study symmetry tunability of the vacuum. We consider the following cases: CGS broken environment with $\gamma=0.2, \lambda=0$ and parity broken environment with $\gamma=0, \lambda=0.5$. For these two cases, we first show the global phases of the environment in $M_B$ parameter and then illustrate numerically obtained ($M_B$ - $M_w$) phase diagrams in terms of spectral gap ($\Delta E$) and zero-bias peak (ZBP) for the finite geometry ($l<L$) in \Fig{fig_clsAwire}. As seen from \Fig{fig_clsAwire}(b-c), when parity is conserved even in the CGS broken topological environment, the embedded wire shows features like gapping-out of subsystem metals and proximity-induced topology similar to the case of charge-conjugation symmetric class D environment (see \Fig{fig3}(A, D)). While for the parity broken case (see \Fig{fig_clsAwire}(e-f)), these novel features are absent, trivializing the wire all through in the topological vacuum.

\begin{figure*}
\centering
\includegraphics[width=1\linewidth]{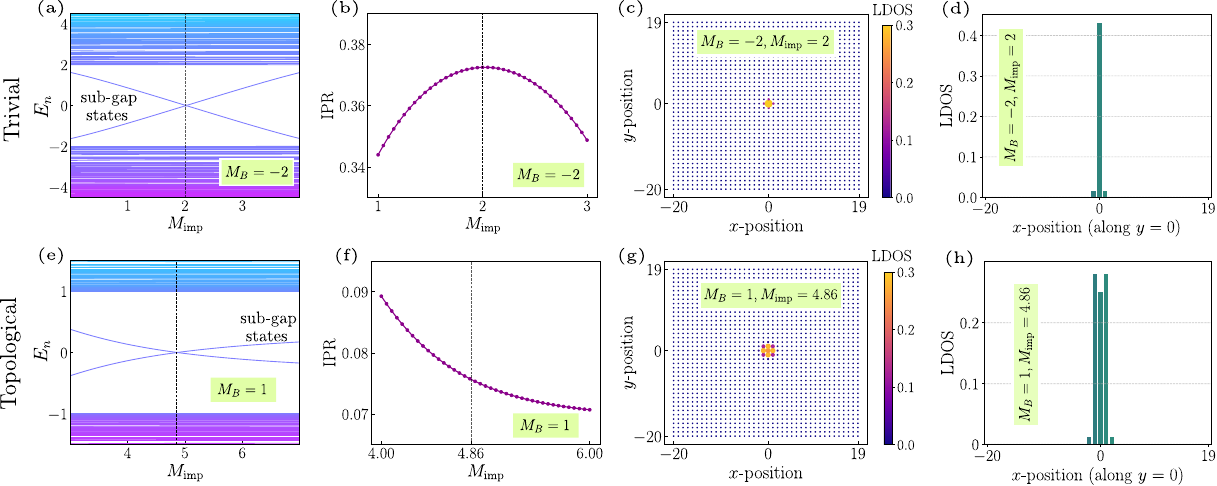}
\caption{{(a) Sub-gap states in the energy spectrum of a single-site impurity of {\it mass} $M_{\text{imp}}$ embedded on a trivial insulator characterized by $M_B=-2$. (b) The average inverse participation ratio (IPR) of the two sub-gap states as a function of $M_{\text{imp}}$. (c) The LDOS of these sub-gap states at $M_{\text{imp}}=2$ shows they are localized only near the position of the impurity ($x=0, y=0$). (f) LDOS along the effective one-dimensional wire at $y=0$ confirms strong localization. Equivalent to (a-d), (e-h) are the results for the topological environment with $M_B=1$. The average IPR (f) and LDOS (g, h) of the sub-gap states in this regime also show localization near the impurity. For all the calculations, the linear system size is $L=40$, and periodic boundary conditions are used.}}
\label{fig_bhzimpurity}
\end{figure*}

 \section{Single site impurity on the Chern insulator}\label{apndx_singleimpurity}

The wire Hamiltonian can also be interpreted as a one-dimensional array of impurities. In order to study the nature of overlaps between such impurities, we investigate the physics of a single impurity site within the Chern insulator environment. The real space hopping Hamiltonian of such a composite system is given by, 
\begin{align}
    H_{\text{QWZ-imp}} = & \sum_{i, \vec{\delta}}\left( \Psi^\dagger_{i} T_{\vec{\delta}}\Psi_{i+\vec{\delta}} + \text{h.c.} \right) \notag \\ +& \sum_{i \notin \text{impurity site}} \Psi^\dagger_i (2-M_B)\sigma_z \Psi_i   \notag \\ +& \sum_{i \in \text{impurity site}} \Psi^\dagger_i (2-M_{\text{imp}})\sigma_z \Psi_i,\label{eq_impurityham}
\end{align}
where $\Psi_i$, $T_{\vec{\delta}}$ and $\vec{\delta}$ are the same as previously defined below \eqn{eq_rham}. $M_B$ and $M_{\text{imp}}$ are the {\it mass} term for the environment and the impurity respectively. We numerically analyze this composite system for both trivial ($M_B=-2$) and topological ($M_B=1$) environments. In the energy spectrum, we find two sub-gap states induced by the impurity for both trivial and topological environments, as shown in Fig.~\ref{fig_bhzimpurity}(a) and Fig.~\ref{fig_bhzimpurity}(e), respectively. To understand these two low-energy impurity wavefunctions (namely $\psi_{0_\pm}$) and their localization properties, we calculate their average inverse participation ratio (IPR) defined as, 
\begin{equation}
\text{IPR} = \frac{1}{2}\sum_{i, \alpha} \Big(|\psi_{0_-}(i, \alpha)|^4 + |\psi_{0_+}(i,\alpha)|^4 \Big), \label{eq_ipr}  
\end{equation}
where, $\psi_{0_\pm}(i,\alpha)$ is the impurity wavefunction at site $i$ for the orbital $\alpha=A/B$. The IPR data shown in Fig.~\ref{fig_bhzimpurity}(b, f), along with LDOS (see Fig.~\ref{fig_bhzimpurity}(c, g)), suggest that the sub-gap states are localized severely over a few sites near the impurity at $(x, y)\equiv (0,0)$. Such a strong localization results in a very small spreading of these impurity-induced states ($3$ - $5$ sites in any direction) upon reduction to an effective one-dimensional structure along $y=0$, as shown in Fig.~\ref{fig_bhzimpurity}(d, h). 

Now, when there are consecutive such embedded impurities placed along $y=0$ as in a wire, given the exponentially falling overlap of the sub-gap states due to strong localization,  we expect no long-range coupling between them. Thus, in our study, the nearest-neighbor tight-binding description of the embedded wire effectively captures the low-energy physics.

\bibliography{topo_ref.bib}

\begin{thebibliography}{174}%
\makeatletter
\providecommand \@ifxundefined [1]{%
 \@ifx{#1\undefined}
}%
\providecommand \@ifnum [1]{%
 \ifnum #1\expandafter \@firstoftwo
 \else \expandafter \@secondoftwo
 \fi
}%
\providecommand \@ifx [1]{%
 \ifx #1\expandafter \@firstoftwo
 \else \expandafter \@secondoftwo
 \fi
}%
\providecommand \natexlab [1]{#1}%
\providecommand \enquote  [1]{``#1''}%
\providecommand \bibnamefont  [1]{#1}%
\providecommand \bibfnamefont [1]{#1}%
\providecommand \citenamefont [1]{#1}%
\providecommand \href@noop [0]{\@secondoftwo}%
\providecommand \href [0]{\begingroup \@sanitize@url \@href}%
\providecommand \@href[1]{\@@startlink{#1}\@@href}%
\providecommand \@@href[1]{\endgroup#1\@@endlink}%
\providecommand \@sanitize@url [0]{\catcode `\\12\catcode `\$12\catcode
  `\&12\catcode `\#12\catcode `\^12\catcode `\_12\catcode `\%12\relax}%
\providecommand \@@startlink[1]{}%
\providecommand \@@endlink[0]{}%
\providecommand \url  [0]{\begingroup\@sanitize@url \@url }%
\providecommand \@url [1]{\endgroup\@href {#1}{\urlprefix }}%
\providecommand \urlprefix  [0]{URL }%
\providecommand \Eprint [0]{\href }%
\providecommand \doibase [0]{https://doi.org/}%
\providecommand \selectlanguage [0]{\@gobble}%
\providecommand \bibinfo  [0]{\@secondoftwo}%
\providecommand \bibfield  [0]{\@secondoftwo}%
\providecommand \translation [1]{[#1]}%
\providecommand \BibitemOpen [0]{}%
\providecommand \bibitemStop [0]{}%
\providecommand \bibitemNoStop [0]{.\EOS\space}%
\providecommand \EOS [0]{\spacefactor3000\relax}%
\providecommand \BibitemShut  [1]{\csname bibitem#1\endcsname}%
\let\auto@bib@innerbib\@empty
\bibitem [{\citenamefont {Moore}(2010)}]{moore_tibirth_2010}%
  \BibitemOpen
  \bibfield  {author} {\bibinfo {author} {\bibfnamefont {J.~E.}\ \bibnamefont
  {Moore}},\ }\bibfield  {title} {\bibinfo {title} {The birth of topological
  insulators},\ }\href {https://www.nature.com/articles/nature08916} {\bibfield
   {journal} {\bibinfo  {journal} {Nature}\ }\textbf {\bibinfo {volume}
  {464}},\ \bibinfo {pages} {194} (\bibinfo {year} {2010})}\BibitemShut
  {NoStop}%
\bibitem [{\citenamefont {Hasan}\ and\ \citenamefont
  {Kane}(2010)}]{hasan_topo_2010}%
  \BibitemOpen
  \bibfield  {author} {\bibinfo {author} {\bibfnamefont {M.~Z.}\ \bibnamefont
  {Hasan}}\ and\ \bibinfo {author} {\bibfnamefont {C.~L.}\ \bibnamefont
  {Kane}},\ }\bibfield  {title} {\bibinfo {title} {Colloquium: Topological
  insulators},\ }\href {https://doi.org/10.1103/RevModPhys.82.3045} {\bibfield
  {journal} {\bibinfo  {journal} {Rev. Mod. Phys.}\ }\textbf {\bibinfo {volume}
  {82}},\ \bibinfo {pages} {3045} (\bibinfo {year} {2010})}\BibitemShut
  {NoStop}%
\bibitem [{\citenamefont {Ryu}\ \emph {et~al.}(2010)\citenamefont {Ryu},
  \citenamefont {Schnyder}, \citenamefont {Furusaki},\ and\ \citenamefont
  {Ludwig}}]{Ryu_topocls_2010}%
  \BibitemOpen
  \bibfield  {author} {\bibinfo {author} {\bibfnamefont {S.}~\bibnamefont
  {Ryu}}, \bibinfo {author} {\bibfnamefont {A.~P.}\ \bibnamefont {Schnyder}},
  \bibinfo {author} {\bibfnamefont {A.}~\bibnamefont {Furusaki}},\ and\
  \bibinfo {author} {\bibfnamefont {A.~W.~W.}\ \bibnamefont {Ludwig}},\
  }\bibfield  {title} {\bibinfo {title} {Topological insulators and
  superconductors: tenfold way and dimensional hierarchy},\ }\href
  {https://doi.org/10.1088/1367-2630/12/6/065010} {\bibfield  {journal}
  {\bibinfo  {journal} {New Journal of Physics}\ }\textbf {\bibinfo {volume}
  {12}},\ \bibinfo {pages} {065010} (\bibinfo {year} {2010})}\BibitemShut
  {NoStop}%
\bibitem [{\citenamefont {Qi}\ and\ \citenamefont
  {Zhang}(2011)}]{qi_toposc_2011}%
  \BibitemOpen
  \bibfield  {author} {\bibinfo {author} {\bibfnamefont {X.-L.}\ \bibnamefont
  {Qi}}\ and\ \bibinfo {author} {\bibfnamefont {S.-C.}\ \bibnamefont {Zhang}},\
  }\bibfield  {title} {\bibinfo {title} {Topological insulators and
  superconductors},\ }\href {https://doi.org/10.1103/RevModPhys.83.1057}
  {\bibfield  {journal} {\bibinfo  {journal} {Rev. Mod. Phys.}\ }\textbf
  {\bibinfo {volume} {83}},\ \bibinfo {pages} {1057} (\bibinfo {year}
  {2011})}\BibitemShut {NoStop}%
\bibitem [{\citenamefont {Shen}(2012)}]{shen_tibook_2012}%
  \BibitemOpen
  \bibfield  {author} {\bibinfo {author} {\bibfnamefont {S.-Q.}\ \bibnamefont
  {Shen}},\ }\href {https://link.springer.com/book/10.1007/978-981-10-4606-3}
  {\emph {\bibinfo {title} {Topological insulators}}},\ Vol.\ \bibinfo {volume}
  {174}\ (\bibinfo  {publisher} {Springer},\ \bibinfo {year}
  {2012})\BibitemShut {NoStop}%
\bibitem [{\citenamefont {Bernevig}(2013)}]{bernevig_topobook_2013}%
  \BibitemOpen
  \bibfield  {author} {\bibinfo {author} {\bibfnamefont {B.~A.}\ \bibnamefont
  {Bernevig}},\ }\href
  {https://press.princeton.edu/books/hardcover/9780691151755/topological-insulators-and-topological-superconductors?srsltid=AfmBOopPBNQcu3v6qSC6wwkEYvPHj-OiIPlSlfhabAXLHklhXO0XrJiS}
  {\emph {\bibinfo {title} {Topological insulators and topological
  superconductors}}}\ (\bibinfo  {publisher} {Princeton university press},\
  \bibinfo {year} {2013})\BibitemShut {NoStop}%
\bibitem [{\citenamefont {Altland}\ and\ \citenamefont
  {Zirnbauer}(1997)}]{Altland_PRB_1997}%
  \BibitemOpen
  \bibfield  {author} {\bibinfo {author} {\bibfnamefont {A.}~\bibnamefont
  {Altland}}\ and\ \bibinfo {author} {\bibfnamefont {M.~R.}\ \bibnamefont
  {Zirnbauer}},\ }\bibfield  {title} {\bibinfo {title} {Nonstandard symmetry
  classes in mesoscopic normal-superconducting hybrid structures},\ }\href
  {https://doi.org/10.1103/PhysRevB.55.1142} {\bibfield  {journal} {\bibinfo
  {journal} {Phys. Rev. B}\ }\textbf {\bibinfo {volume} {55}},\ \bibinfo
  {pages} {1142} (\bibinfo {year} {1997})}\BibitemShut {NoStop}%
\bibitem [{\citenamefont {Kitaev}(2009)}]{Kitaevtenfold}%
  \BibitemOpen
  \bibfield  {author} {\bibinfo {author} {\bibfnamefont {A.}~\bibnamefont
  {Kitaev}},\ }\bibfield  {title} {\bibinfo {title} {Periodic table for
  topological insulators and superconductors},\ }\href
  {https://doi.org/10.1063/1.3149495} {\bibfield  {journal} {\bibinfo
  {journal} {AIP Conference Proceedings}\ }\textbf {\bibinfo {volume} {1134}},\
  \bibinfo {pages} {22} (\bibinfo {year} {2009})}\BibitemShut {NoStop}%
\bibitem [{\citenamefont {Chiu}\ \emph {et~al.}(2016)\citenamefont {Chiu},
  \citenamefont {Teo}, \citenamefont {Schnyder},\ and\ \citenamefont
  {Ryu}}]{chiu_topoclass_2016}%
  \BibitemOpen
  \bibfield  {author} {\bibinfo {author} {\bibfnamefont {C.-K.}\ \bibnamefont
  {Chiu}}, \bibinfo {author} {\bibfnamefont {J.~C.}\ \bibnamefont {Teo}},
  \bibinfo {author} {\bibfnamefont {A.~P.}\ \bibnamefont {Schnyder}},\ and\
  \bibinfo {author} {\bibfnamefont {S.}~\bibnamefont {Ryu}},\ }\bibfield
  {title} {\bibinfo {title} {Classification of topological quantum matter with
  symmetries},\ }\href {https://doi.org/10.1103/RevModPhys.88.035005}
  {\bibfield  {journal} {\bibinfo  {journal} {Reviews of Modern Physics}\
  }\textbf {\bibinfo {volume} {88}},\ \bibinfo {pages} {035005} (\bibinfo
  {year} {2016})}\BibitemShut {NoStop}%
\bibitem [{\citenamefont {Ludwig}(2015)}]{Ludwig_topo_2016}%
  \BibitemOpen
  \bibfield  {author} {\bibinfo {author} {\bibfnamefont {A.~W.~W.}\
  \bibnamefont {Ludwig}},\ }\bibfield  {title} {\bibinfo {title} {Topological
  phases: classification of topological insulators and superconductors of
  non-interacting fermions, and beyond},\ }\href
  {https://doi.org/10.1088/0031-8949/2015/T168/014001} {\bibfield  {journal}
  {\bibinfo  {journal} {Physica Scripta}\ }\textbf {\bibinfo {volume} {2016}},\
  \bibinfo {pages} {014001} (\bibinfo {year} {2015})}\BibitemShut {NoStop}%
\bibitem [{\citenamefont {Agarwala}\ \emph {et~al.}(2017)\citenamefont
  {Agarwala}, \citenamefont {Haldar},\ and\ \citenamefont
  {Shenoy}}]{Agarwala_AOP_2017}%
  \BibitemOpen
  \bibfield  {author} {\bibinfo {author} {\bibfnamefont {A.}~\bibnamefont
  {Agarwala}}, \bibinfo {author} {\bibfnamefont {A.}~\bibnamefont {Haldar}},\
  and\ \bibinfo {author} {\bibfnamefont {V.~B.}\ \bibnamefont {Shenoy}},\
  }\bibfield  {title} {\bibinfo {title} {The tenfold way redux: Fermionic
  systems with n-body interactions},\ }\href
  {https://doi.org/https://doi.org/10.1016/j.aop.2017.07.016} {\bibfield
  {journal} {\bibinfo  {journal} {Annals of Physics}\ }\textbf {\bibinfo
  {volume} {385}},\ \bibinfo {pages} {469} (\bibinfo {year}
  {2017})}\BibitemShut {NoStop}%
\bibitem [{\citenamefont {Kruthoff}\ \emph {et~al.}(2017)\citenamefont
  {Kruthoff}, \citenamefont {de~Boer}, \citenamefont {van Wezel}, \citenamefont
  {Kane},\ and\ \citenamefont {Slager}}]{Jorrit_topoclass_2017}%
  \BibitemOpen
  \bibfield  {author} {\bibinfo {author} {\bibfnamefont {J.}~\bibnamefont
  {Kruthoff}}, \bibinfo {author} {\bibfnamefont {J.}~\bibnamefont {de~Boer}},
  \bibinfo {author} {\bibfnamefont {J.}~\bibnamefont {van Wezel}}, \bibinfo
  {author} {\bibfnamefont {C.~L.}\ \bibnamefont {Kane}},\ and\ \bibinfo
  {author} {\bibfnamefont {R.-J.}\ \bibnamefont {Slager}},\ }\bibfield  {title}
  {\bibinfo {title} {Topological classification of crystalline insulators
  through band structure combinatorics},\ }\href
  {https://doi.org/10.1103/PhysRevX.7.041069} {\bibfield  {journal} {\bibinfo
  {journal} {Phys. Rev. X}\ }\textbf {\bibinfo {volume} {7}},\ \bibinfo {pages}
  {041069} (\bibinfo {year} {2017})}\BibitemShut {NoStop}%
\bibitem [{\citenamefont {Halperin}(1982)}]{halperin_roedge_1982}%
  \BibitemOpen
  \bibfield  {author} {\bibinfo {author} {\bibfnamefont {B.~I.}\ \bibnamefont
  {Halperin}},\ }\bibfield  {title} {\bibinfo {title} {Quantized hall
  conductance, current-carrying edge states, and the existence of extended
  states in a two-dimensional disordered potential},\ }\href
  {https://doi.org/10.1103/PhysRevB.25.2185} {\bibfield  {journal} {\bibinfo
  {journal} {Phys. Rev. B}\ }\textbf {\bibinfo {volume} {25}},\ \bibinfo
  {pages} {2185} (\bibinfo {year} {1982})}\BibitemShut {NoStop}%
\bibitem [{\citenamefont {Xu}\ and\ \citenamefont
  {Moore}(2006)}]{xu_robust_2006}%
  \BibitemOpen
  \bibfield  {author} {\bibinfo {author} {\bibfnamefont {C.}~\bibnamefont
  {Xu}}\ and\ \bibinfo {author} {\bibfnamefont {J.~E.}\ \bibnamefont {Moore}},\
  }\bibfield  {title} {\bibinfo {title} {Stability of the quantum spin hall
  effect: Effects of interactions, disorder, and $\mathbb{Z}_{2}$ topology},\
  }\href {https://doi.org/10.1103/PhysRevB.73.045322} {\bibfield  {journal}
  {\bibinfo  {journal} {Phys. Rev. B}\ }\textbf {\bibinfo {volume} {73}},\
  \bibinfo {pages} {045322} (\bibinfo {year} {2006})}\BibitemShut {NoStop}%
\bibitem [{\citenamefont {Kitaev}(2003)}]{kitaev_fault-tolerant_2003}%
  \BibitemOpen
  \bibfield  {author} {\bibinfo {author} {\bibfnamefont {A.~Y.}\ \bibnamefont
  {Kitaev}},\ }\bibfield  {title} {\bibinfo {title} {Fault-tolerant quantum
  computation by anyons},\ }\href
  {https://doi.org/10.1016/S0003-4916(02)00018-0} {\bibfield  {journal}
  {\bibinfo  {journal} {Annals of Physics}\ }\textbf {\bibinfo {volume}
  {303}},\ \bibinfo {pages} {2} (\bibinfo {year} {2003})}\BibitemShut {NoStop}%
\bibitem [{\citenamefont {Nayak}\ \emph {et~al.}(2008)\citenamefont {Nayak},
  \citenamefont {Simon}, \citenamefont {Stern}, \citenamefont {Freedman},\ and\
  \citenamefont {Das~Sarma}}]{nyak_topocomp_2008}%
  \BibitemOpen
  \bibfield  {author} {\bibinfo {author} {\bibfnamefont {C.}~\bibnamefont
  {Nayak}}, \bibinfo {author} {\bibfnamefont {S.~H.}\ \bibnamefont {Simon}},
  \bibinfo {author} {\bibfnamefont {A.}~\bibnamefont {Stern}}, \bibinfo
  {author} {\bibfnamefont {M.}~\bibnamefont {Freedman}},\ and\ \bibinfo
  {author} {\bibfnamefont {S.}~\bibnamefont {Das~Sarma}},\ }\bibfield  {title}
  {\bibinfo {title} {Non-abelian anyons and topological quantum computation},\
  }\href {https://doi.org/10.1103/RevModPhys.80.1083} {\bibfield  {journal}
  {\bibinfo  {journal} {Rev. Mod. Phys.}\ }\textbf {\bibinfo {volume} {80}},\
  \bibinfo {pages} {1083} (\bibinfo {year} {2008})}\BibitemShut {NoStop}%
\bibitem [{\citenamefont {Pachos}(2012)}]{pachos_introduction_2012}%
  \BibitemOpen
  \bibfield  {author} {\bibinfo {author} {\bibfnamefont {J.~K.}\ \bibnamefont
  {Pachos}},\ }\href
  {https://www.cambridge.org/core/books/introduction-to-topological-quantum-computation/F6C4B2C9F83E434E9BF3F73E492231F0}
  {\emph {\bibinfo {title} {Introduction to Topological Quantum Computation}}}\
  (\bibinfo  {publisher} {Cambridge University Press},\ \bibinfo {year}
  {2012})\BibitemShut {NoStop}%
\bibitem [{\citenamefont {Wu}\ \emph {et~al.}(2006)\citenamefont {Wu},
  \citenamefont {Bernevig},\ and\ \citenamefont {Zhang}}]{wu_em_2006}%
  \BibitemOpen
  \bibfield  {author} {\bibinfo {author} {\bibfnamefont {C.}~\bibnamefont
  {Wu}}, \bibinfo {author} {\bibfnamefont {B.~A.}\ \bibnamefont {Bernevig}},\
  and\ \bibinfo {author} {\bibfnamefont {S.-C.}\ \bibnamefont {Zhang}},\
  }\bibfield  {title} {\bibinfo {title} {Helical liquid and the edge of quantum
  spin hall systems},\ }\href {https://doi.org/10.1103/PhysRevLett.96.106401}
  {\bibfield  {journal} {\bibinfo  {journal} {Phys. Rev. Lett.}\ }\textbf
  {\bibinfo {volume} {96}},\ \bibinfo {pages} {106401} (\bibinfo {year}
  {2006})}\BibitemShut {NoStop}%
\bibitem [{\citenamefont {Gurarie}\ and\ \citenamefont
  {Radzihovsky}(2007)}]{Gurarie_PRB_2007}%
  \BibitemOpen
  \bibfield  {author} {\bibinfo {author} {\bibfnamefont {V.}~\bibnamefont
  {Gurarie}}\ and\ \bibinfo {author} {\bibfnamefont {L.}~\bibnamefont
  {Radzihovsky}},\ }\bibfield  {title} {\bibinfo {title} {Zero modes of
  two-dimensional chiral $p$-wave superconductors},\ }\href
  {https://doi.org/10.1103/PhysRevB.75.212509} {\bibfield  {journal} {\bibinfo
  {journal} {Phys. Rev. B}\ }\textbf {\bibinfo {volume} {75}},\ \bibinfo
  {pages} {212509} (\bibinfo {year} {2007})}\BibitemShut {NoStop}%
\bibitem [{\citenamefont {Fu}\ and\ \citenamefont {Kane}(2008)}]{fu_mbs_2008}%
  \BibitemOpen
  \bibfield  {author} {\bibinfo {author} {\bibfnamefont {L.}~\bibnamefont
  {Fu}}\ and\ \bibinfo {author} {\bibfnamefont {C.~L.}\ \bibnamefont {Kane}},\
  }\bibfield  {title} {\bibinfo {title} {Superconducting proximity effect and
  majorana fermions at the surface of a topological insulator},\ }\href
  {https://doi.org/10.1103/PhysRevLett.100.096407} {\bibfield  {journal}
  {\bibinfo  {journal} {Phys. Rev. Lett.}\ }\textbf {\bibinfo {volume} {100}},\
  \bibinfo {pages} {096407} (\bibinfo {year} {2008})}\BibitemShut {NoStop}%
\bibitem [{\citenamefont {Teo}\ \emph {et~al.}(2008)\citenamefont {Teo},
  \citenamefont {Fu},\ and\ \citenamefont {Kane}}]{teo_sm_2008}%
  \BibitemOpen
  \bibfield  {author} {\bibinfo {author} {\bibfnamefont {J.~C.~Y.}\
  \bibnamefont {Teo}}, \bibinfo {author} {\bibfnamefont {L.}~\bibnamefont
  {Fu}},\ and\ \bibinfo {author} {\bibfnamefont {C.~L.}\ \bibnamefont {Kane}},\
  }\bibfield  {title} {\bibinfo {title} {Surface states and topological
  invariants in three-dimensional topological insulators: Application to
  $\text{Bi}_{1\ensuremath{-}x}\text{Sb}_{x}$},\ }\href
  {https://doi.org/10.1103/PhysRevB.78.045426} {\bibfield  {journal} {\bibinfo
  {journal} {Phys. Rev. B}\ }\textbf {\bibinfo {volume} {78}},\ \bibinfo
  {pages} {045426} (\bibinfo {year} {2008})}\BibitemShut {NoStop}%
\bibitem [{\citenamefont {Roy}(2010)}]{Roy_PRL_2010}%
  \BibitemOpen
  \bibfield  {author} {\bibinfo {author} {\bibfnamefont {R.}~\bibnamefont
  {Roy}},\ }\bibfield  {title} {\bibinfo {title} {Topological majorana and
  dirac zero modes in superconducting vortex cores},\ }\href
  {https://doi.org/10.1103/PhysRevLett.105.186401} {\bibfield  {journal}
  {\bibinfo  {journal} {Phys. Rev. Lett.}\ }\textbf {\bibinfo {volume} {105}},\
  \bibinfo {pages} {186401} (\bibinfo {year} {2010})}\BibitemShut {NoStop}%
\bibitem [{\citenamefont {Teo}\ and\ \citenamefont
  {Kane}(2010{\natexlab{a}})}]{teo_mbs_2010}%
  \BibitemOpen
  \bibfield  {author} {\bibinfo {author} {\bibfnamefont {J.~C.~Y.}\
  \bibnamefont {Teo}}\ and\ \bibinfo {author} {\bibfnamefont {C.~L.}\
  \bibnamefont {Kane}},\ }\bibfield  {title} {\bibinfo {title} {Majorana
  fermions and non-abelian statistics in three dimensions},\ }\href
  {https://doi.org/10.1103/PhysRevLett.104.046401} {\bibfield  {journal}
  {\bibinfo  {journal} {Phys. Rev. Lett.}\ }\textbf {\bibinfo {volume} {104}},\
  \bibinfo {pages} {046401} (\bibinfo {year} {2010}{\natexlab{a}})}\BibitemShut
  {NoStop}%
\bibitem [{\citenamefont {Potter}\ and\ \citenamefont
  {Lee}(2010)}]{Potter_multichannel_2010}%
  \BibitemOpen
  \bibfield  {author} {\bibinfo {author} {\bibfnamefont {A.~C.}\ \bibnamefont
  {Potter}}\ and\ \bibinfo {author} {\bibfnamefont {P.~A.}\ \bibnamefont
  {Lee}},\ }\bibfield  {title} {\bibinfo {title} {Multichannel generalization
  of kitaev's majorana end states and a practical route to realize them in thin
  films},\ }\href {https://doi.org/10.1103/PhysRevLett.105.227003} {\bibfield
  {journal} {\bibinfo  {journal} {Phys. Rev. Lett.}\ }\textbf {\bibinfo
  {volume} {105}},\ \bibinfo {pages} {227003} (\bibinfo {year}
  {2010})}\BibitemShut {NoStop}%
\bibitem [{\citenamefont {Beenakker}(2013)}]{beenakker_mbs_2013}%
  \BibitemOpen
  \bibfield  {author} {\bibinfo {author} {\bibfnamefont {C.}~\bibnamefont
  {Beenakker}},\ }\bibfield  {title} {\bibinfo {title} {Search for majorana
  fermions in superconductors},\ }\href
  {https://www.annualreviews.org/content/journals/10.1146/annurev-conmatphys-030212-184337}
  {\bibfield  {journal} {\bibinfo  {journal} {Annu. Rev. Condens. Matter
  Phys.}\ }\textbf {\bibinfo {volume} {4}},\ \bibinfo {pages} {113} (\bibinfo
  {year} {2013})}\BibitemShut {NoStop}%
\bibitem [{\citenamefont {Thakurathi}\ \emph {et~al.}(2014)\citenamefont
  {Thakurathi}, \citenamefont {Sengupta},\ and\ \citenamefont
  {Sen}}]{Thakurathi_MBS_2014}%
  \BibitemOpen
  \bibfield  {author} {\bibinfo {author} {\bibfnamefont {M.}~\bibnamefont
  {Thakurathi}}, \bibinfo {author} {\bibfnamefont {K.}~\bibnamefont
  {Sengupta}},\ and\ \bibinfo {author} {\bibfnamefont {D.}~\bibnamefont
  {Sen}},\ }\bibfield  {title} {\bibinfo {title} {Majorana edge modes in the
  kitaev model},\ }\href {https://doi.org/10.1103/PhysRevB.89.235434}
  {\bibfield  {journal} {\bibinfo  {journal} {Phys. Rev. B}\ }\textbf {\bibinfo
  {volume} {89}},\ \bibinfo {pages} {235434} (\bibinfo {year}
  {2014})}\BibitemShut {NoStop}%
\bibitem [{\citenamefont {Thakurathi}\ \emph {et~al.}(2015)\citenamefont
  {Thakurathi}, \citenamefont {Deb},\ and\ \citenamefont
  {Sen}}]{Thakurathi_mbs_2015}%
  \BibitemOpen
  \bibfield  {author} {\bibinfo {author} {\bibfnamefont {M.}~\bibnamefont
  {Thakurathi}}, \bibinfo {author} {\bibfnamefont {O.}~\bibnamefont {Deb}},\
  and\ \bibinfo {author} {\bibfnamefont {D.}~\bibnamefont {Sen}},\ }\bibfield
  {title} {\bibinfo {title} {Majorana modes and transport across junctions of
  superconductors and normal metals},\ }\href
  {https://doi.org/10.1088/0953-8984/27/27/275702} {\bibfield  {journal}
  {\bibinfo  {journal} {Journal of Physics: Condensed Matter}\ }\textbf
  {\bibinfo {volume} {27}},\ \bibinfo {pages} {275702} (\bibinfo {year}
  {2015})}\BibitemShut {NoStop}%
\bibitem [{\citenamefont {Yazdani}\ \emph {et~al.}(2023)\citenamefont
  {Yazdani}, \citenamefont {von Oppen}, \citenamefont {Halperin},\ and\
  \citenamefont {Yacoby}}]{yazdani_mzmhunt_2023}%
  \BibitemOpen
  \bibfield  {author} {\bibinfo {author} {\bibfnamefont {A.}~\bibnamefont
  {Yazdani}}, \bibinfo {author} {\bibfnamefont {F.}~\bibnamefont {von Oppen}},
  \bibinfo {author} {\bibfnamefont {B.~I.}\ \bibnamefont {Halperin}},\ and\
  \bibinfo {author} {\bibfnamefont {A.}~\bibnamefont {Yacoby}},\ }\bibfield
  {title} {\bibinfo {title} {Hunting for majoranas},\ }\href
  {https://doi.org/10.1126/science.ade0850} {\bibfield  {journal} {\bibinfo
  {journal} {Science}\ }\textbf {\bibinfo {volume} {380}},\ \bibinfo {pages}
  {eade0850} (\bibinfo {year} {2023})}\BibitemShut {NoStop}%
\bibitem [{\citenamefont {Misra}\ \emph {et~al.}(2002)\citenamefont {Misra},
  \citenamefont {Oh}, \citenamefont {Hornbaker}, \citenamefont {DiLuccio},
  \citenamefont {Eckstein},\ and\ \citenamefont {Yazdani}}]{misra_em_2002}%
  \BibitemOpen
  \bibfield  {author} {\bibinfo {author} {\bibfnamefont {S.}~\bibnamefont
  {Misra}}, \bibinfo {author} {\bibfnamefont {S.}~\bibnamefont {Oh}}, \bibinfo
  {author} {\bibfnamefont {D.~J.}\ \bibnamefont {Hornbaker}}, \bibinfo {author}
  {\bibfnamefont {T.}~\bibnamefont {DiLuccio}}, \bibinfo {author}
  {\bibfnamefont {J.~N.}\ \bibnamefont {Eckstein}},\ and\ \bibinfo {author}
  {\bibfnamefont {A.}~\bibnamefont {Yazdani}},\ }\bibfield  {title} {\bibinfo
  {title} {Formation of an andreev bound state at the step edges of
  $\mathrm{Bi}_{2}\mathrm{Sr}_{2}\mathrm{CaCu}_{2}\mathrm{O}_{8+\ensuremath{\delta}}$
  surface},\ }\href {https://doi.org/10.1103/PhysRevB.66.100510} {\bibfield
  {journal} {\bibinfo  {journal} {Phys. Rev. B}\ }\textbf {\bibinfo {volume}
  {66}},\ \bibinfo {pages} {100510} (\bibinfo {year} {2002})}\BibitemShut
  {NoStop}%
\bibitem [{\citenamefont {Hsieh}\ \emph {et~al.}(2008)\citenamefont {Hsieh},
  \citenamefont {Qian}, \citenamefont {Wray}, \citenamefont {Xia},
  \citenamefont {Hor}, \citenamefont {Cava},\ and\ \citenamefont
  {Hasan}}]{hsieh_topological_2008}%
  \BibitemOpen
  \bibfield  {author} {\bibinfo {author} {\bibfnamefont {D.}~\bibnamefont
  {Hsieh}}, \bibinfo {author} {\bibfnamefont {D.}~\bibnamefont {Qian}},
  \bibinfo {author} {\bibfnamefont {L.}~\bibnamefont {Wray}}, \bibinfo {author}
  {\bibfnamefont {Y.}~\bibnamefont {Xia}}, \bibinfo {author} {\bibfnamefont
  {Y.~S.}\ \bibnamefont {Hor}}, \bibinfo {author} {\bibfnamefont {R.~J.}\
  \bibnamefont {Cava}},\ and\ \bibinfo {author} {\bibfnamefont {M.~Z.}\
  \bibnamefont {Hasan}},\ }\bibfield  {title} {\bibinfo {title} {A topological
  dirac insulator in a quantum spin hall phase},\ }\href
  {https://www.nature.com/articles/nature06843} {\bibfield  {journal} {\bibinfo
   {journal} {Nature}\ }\textbf {\bibinfo {volume} {452}},\ \bibinfo {pages}
  {970} (\bibinfo {year} {2008})}\BibitemShut {NoStop}%
\bibitem [{\citenamefont {Bradlyn}\ \emph {et~al.}(2017)\citenamefont
  {Bradlyn}, \citenamefont {Elcoro}, \citenamefont {Cano}, \citenamefont
  {Vergniory}, \citenamefont {Wang}, \citenamefont {Felser}, \citenamefont
  {Aroyo},\ and\ \citenamefont {Bernevig}}]{bernevig_nature_2017}%
  \BibitemOpen
  \bibfield  {author} {\bibinfo {author} {\bibfnamefont {B.}~\bibnamefont
  {Bradlyn}}, \bibinfo {author} {\bibfnamefont {L.}~\bibnamefont {Elcoro}},
  \bibinfo {author} {\bibfnamefont {J.}~\bibnamefont {Cano}}, \bibinfo {author}
  {\bibfnamefont {M.~G.}\ \bibnamefont {Vergniory}}, \bibinfo {author}
  {\bibfnamefont {Z.}~\bibnamefont {Wang}}, \bibinfo {author} {\bibfnamefont
  {C.}~\bibnamefont {Felser}}, \bibinfo {author} {\bibfnamefont {M.~I.}\
  \bibnamefont {Aroyo}},\ and\ \bibinfo {author} {\bibfnamefont {B.~A.}\
  \bibnamefont {Bernevig}},\ }\bibfield  {title} {\bibinfo {title} {Topological
  quantum chemistry},\ }\href {https://doi.org/10.1038/nature23268} {\bibfield
  {journal} {\bibinfo  {journal} {Nature}\ }\textbf {\bibinfo {volume} {547}},\
  \bibinfo {pages} {298–305} (\bibinfo {year} {2017})}\BibitemShut {NoStop}%
\bibitem [{ber()}]{bernevig_topmat_database}%
  \BibitemOpen
  \href@noop {} {\bibinfo {title} {Topological materials database}},\ \bibinfo
  {howpublished} {\url{https://topologicalquantumchemistry.com}}\BibitemShut
  {NoStop}%
\bibitem [{\citenamefont {Kanungo}\ \emph {et~al.}(2022)\citenamefont
  {Kanungo}, \citenamefont {Whalen}, \citenamefont {Lu}, \citenamefont {Yuan},
  \citenamefont {Dasgupta}, \citenamefont {Dunning}, \citenamefont {Hazzard},\
  and\ \citenamefont {Killian}}]{kanungo2022realizing}%
  \BibitemOpen
  \bibfield  {author} {\bibinfo {author} {\bibfnamefont {S.}~\bibnamefont
  {Kanungo}}, \bibinfo {author} {\bibfnamefont {J.}~\bibnamefont {Whalen}},
  \bibinfo {author} {\bibfnamefont {Y.}~\bibnamefont {Lu}}, \bibinfo {author}
  {\bibfnamefont {M.}~\bibnamefont {Yuan}}, \bibinfo {author} {\bibfnamefont
  {S.}~\bibnamefont {Dasgupta}}, \bibinfo {author} {\bibfnamefont
  {F.}~\bibnamefont {Dunning}}, \bibinfo {author} {\bibfnamefont
  {K.}~\bibnamefont {Hazzard}},\ and\ \bibinfo {author} {\bibfnamefont
  {T.}~\bibnamefont {Killian}},\ }\bibfield  {title} {\bibinfo {title}
  {Realizing topological edge states with rydberg-atom synthetic dimensions},\
  }\href@noop {} {\bibfield  {journal} {\bibinfo  {journal} {Nature
  communications}\ }\textbf {\bibinfo {volume} {13}},\ \bibinfo {pages} {972}
  (\bibinfo {year} {2022})}\BibitemShut {NoStop}%
\bibitem [{\citenamefont {Meier}\ \emph {et~al.}(2016)\citenamefont {Meier},
  \citenamefont {An},\ and\ \citenamefont {Gadway}}]{meier_2016}%
  \BibitemOpen
  \bibfield  {author} {\bibinfo {author} {\bibfnamefont {E.~J.}\ \bibnamefont
  {Meier}}, \bibinfo {author} {\bibfnamefont {F.~A.}\ \bibnamefont {An}},\ and\
  \bibinfo {author} {\bibfnamefont {B.}~\bibnamefont {Gadway}},\ }\bibfield
  {title} {\bibinfo {title} {Observation of the topological soliton state in
  the su–schrieffer–heeger model},\ }\href
  {https://doi.org/10.1038/ncomms13986} {\bibfield  {journal} {\bibinfo
  {journal} {Nature Communications}\ }\textbf {\bibinfo {volume} {7}},\
  \bibinfo {pages} {13986} (\bibinfo {year} {2016})}\BibitemShut {NoStop}%
\bibitem [{\citenamefont {König}\ \emph {et~al.}(2007)\citenamefont {König},
  \citenamefont {Wiedmann}, \citenamefont {Brüne}, \citenamefont {Roth},
  \citenamefont {Buhmann}, \citenamefont {Molenkamp}, \citenamefont {Qi},\ and\
  \citenamefont {Zhang}}]{markus_hgte_2007}%
  \BibitemOpen
  \bibfield  {author} {\bibinfo {author} {\bibfnamefont {M.}~\bibnamefont
  {König}}, \bibinfo {author} {\bibfnamefont {S.}~\bibnamefont {Wiedmann}},
  \bibinfo {author} {\bibfnamefont {C.}~\bibnamefont {Brüne}}, \bibinfo
  {author} {\bibfnamefont {A.}~\bibnamefont {Roth}}, \bibinfo {author}
  {\bibfnamefont {H.}~\bibnamefont {Buhmann}}, \bibinfo {author} {\bibfnamefont
  {L.~W.}\ \bibnamefont {Molenkamp}}, \bibinfo {author} {\bibfnamefont {X.-L.}\
  \bibnamefont {Qi}},\ and\ \bibinfo {author} {\bibfnamefont {S.-C.}\
  \bibnamefont {Zhang}},\ }\bibfield  {title} {\bibinfo {title} {Quantum spin
  hall insulator state in hgte quantum wells},\ }\href
  {https://doi.org/10.1126/science.1148047} {\bibfield  {journal} {\bibinfo
  {journal} {Science}\ }\textbf {\bibinfo {volume} {318}},\ \bibinfo {pages}
  {766} (\bibinfo {year} {2007})}\BibitemShut {NoStop}%
\bibitem [{\citenamefont {Alpichshev}\ \emph {et~al.}(2010)\citenamefont
  {Alpichshev}, \citenamefont {Analytis}, \citenamefont {Chu}, \citenamefont
  {Fisher}, \citenamefont {Chen}, \citenamefont {Shen}, \citenamefont {Fang},\
  and\ \citenamefont {Kapitulnik}}]{Alpichshev_bite_2010}%
  \BibitemOpen
  \bibfield  {author} {\bibinfo {author} {\bibfnamefont {Z.}~\bibnamefont
  {Alpichshev}}, \bibinfo {author} {\bibfnamefont {J.~G.}\ \bibnamefont
  {Analytis}}, \bibinfo {author} {\bibfnamefont {J.-H.}\ \bibnamefont {Chu}},
  \bibinfo {author} {\bibfnamefont {I.~R.}\ \bibnamefont {Fisher}}, \bibinfo
  {author} {\bibfnamefont {Y.~L.}\ \bibnamefont {Chen}}, \bibinfo {author}
  {\bibfnamefont {Z.~X.}\ \bibnamefont {Shen}}, \bibinfo {author}
  {\bibfnamefont {A.}~\bibnamefont {Fang}},\ and\ \bibinfo {author}
  {\bibfnamefont {A.}~\bibnamefont {Kapitulnik}},\ }\bibfield  {title}
  {\bibinfo {title} {Stm imaging of electronic waves on the surface of
  $\mathrm{Bi}_{2}\mathrm{Te}_{3}$: Topologically protected surface states and
  hexagonal warping effects},\ }\href
  {https://doi.org/10.1103/PhysRevLett.104.016401} {\bibfield  {journal}
  {\bibinfo  {journal} {Phys. Rev. Lett.}\ }\textbf {\bibinfo {volume} {104}},\
  \bibinfo {pages} {016401} (\bibinfo {year} {2010})}\BibitemShut {NoStop}%
\bibitem [{\citenamefont {Roushan}\ \emph {et~al.}(2009)\citenamefont
  {Roushan}, \citenamefont {Seo}, \citenamefont {Parker}, \citenamefont {Hor},
  \citenamefont {Hsieh}, \citenamefont {Qian}, \citenamefont {Richardella},
  \citenamefont {Hasan}, \citenamefont {Cava},\ and\ \citenamefont
  {Yazdani}}]{roushan2009topological}%
  \BibitemOpen
  \bibfield  {author} {\bibinfo {author} {\bibfnamefont {P.}~\bibnamefont
  {Roushan}}, \bibinfo {author} {\bibfnamefont {J.}~\bibnamefont {Seo}},
  \bibinfo {author} {\bibfnamefont {C.~V.}\ \bibnamefont {Parker}}, \bibinfo
  {author} {\bibfnamefont {Y.~S.}\ \bibnamefont {Hor}}, \bibinfo {author}
  {\bibfnamefont {D.}~\bibnamefont {Hsieh}}, \bibinfo {author} {\bibfnamefont
  {D.}~\bibnamefont {Qian}}, \bibinfo {author} {\bibfnamefont {A.}~\bibnamefont
  {Richardella}}, \bibinfo {author} {\bibfnamefont {M.~Z.}\ \bibnamefont
  {Hasan}}, \bibinfo {author} {\bibfnamefont {R.~J.}\ \bibnamefont {Cava}},\
  and\ \bibinfo {author} {\bibfnamefont {A.}~\bibnamefont {Yazdani}},\
  }\bibfield  {title} {\bibinfo {title} {Topological surface states protected
  from backscattering by chiral spin texture},\ }\href
  {https://www.nature.com/articles/nature08308} {\bibfield  {journal} {\bibinfo
   {journal} {Nature}\ }\textbf {\bibinfo {volume} {460}},\ \bibinfo {pages}
  {1106} (\bibinfo {year} {2009})}\BibitemShut {NoStop}%
\bibitem [{\citenamefont {Teo}\ and\ \citenamefont
  {Kane}(2010{\natexlab{b}})}]{teo_topodefect_2010}%
  \BibitemOpen
  \bibfield  {author} {\bibinfo {author} {\bibfnamefont {J.~C.~Y.}\
  \bibnamefont {Teo}}\ and\ \bibinfo {author} {\bibfnamefont {C.~L.}\
  \bibnamefont {Kane}},\ }\bibfield  {title} {\bibinfo {title} {Topological
  defects and gapless modes in insulators and superconductors},\ }\href
  {https://doi.org/10.1103/PhysRevB.82.115120} {\bibfield  {journal} {\bibinfo
  {journal} {Phys. Rev. B}\ }\textbf {\bibinfo {volume} {82}},\ \bibinfo
  {pages} {115120} (\bibinfo {year} {2010}{\natexlab{b}})}\BibitemShut
  {NoStop}%
\bibitem [{\citenamefont {Choy}\ \emph {et~al.}(2011)\citenamefont {Choy},
  \citenamefont {Edge}, \citenamefont {Akhmerov},\ and\ \citenamefont
  {Beenakker}}]{choy_majorana_2011}%
  \BibitemOpen
  \bibfield  {author} {\bibinfo {author} {\bibfnamefont {T.-P.}\ \bibnamefont
  {Choy}}, \bibinfo {author} {\bibfnamefont {J.~M.}\ \bibnamefont {Edge}},
  \bibinfo {author} {\bibfnamefont {A.~R.}\ \bibnamefont {Akhmerov}},\ and\
  \bibinfo {author} {\bibfnamefont {C.~W.~J.}\ \bibnamefont {Beenakker}},\
  }\bibfield  {title} {\bibinfo {title} {Majorana fermions emerging from
  magnetic nanoparticles on a superconductor without spin-orbit coupling},\
  }\href {https://doi.org/10.1103/PhysRevB.84.195442} {\bibfield  {journal}
  {\bibinfo  {journal} {Physical Review B}\ }\textbf {\bibinfo {volume} {84}},\
  \bibinfo {pages} {195442} (\bibinfo {year} {2011})}\BibitemShut {NoStop}%
\bibitem [{\citenamefont {Nadj-Perge}\ \emph {et~al.}(2013)\citenamefont
  {Nadj-Perge}, \citenamefont {Drozdov}, \citenamefont {Bernevig},\ and\
  \citenamefont {Yazdani}}]{nadj-perge_proposal_2013}%
  \BibitemOpen
  \bibfield  {author} {\bibinfo {author} {\bibfnamefont {S.}~\bibnamefont
  {Nadj-Perge}}, \bibinfo {author} {\bibfnamefont {I.~K.}\ \bibnamefont
  {Drozdov}}, \bibinfo {author} {\bibfnamefont {B.~A.}\ \bibnamefont
  {Bernevig}},\ and\ \bibinfo {author} {\bibfnamefont {A.}~\bibnamefont
  {Yazdani}},\ }\bibfield  {title} {\bibinfo {title} {Proposal for realizing
  majorana fermions in chains of magnetic atoms on a superconductor},\ }\href
  {https://doi.org/10.1103/PhysRevB.88.020407} {\bibfield  {journal} {\bibinfo
  {journal} {Physical Review B}\ }\textbf {\bibinfo {volume} {88}},\ \bibinfo
  {pages} {020407} (\bibinfo {year} {2013})}\BibitemShut {NoStop}%
\bibitem [{\citenamefont {Soori}\ \emph {et~al.}(2013)\citenamefont {Soori},
  \citenamefont {Deb}, \citenamefont {Sengupta},\ and\ \citenamefont
  {Sen}}]{Soori_TISC_2013}%
  \BibitemOpen
  \bibfield  {author} {\bibinfo {author} {\bibfnamefont {A.}~\bibnamefont
  {Soori}}, \bibinfo {author} {\bibfnamefont {O.}~\bibnamefont {Deb}}, \bibinfo
  {author} {\bibfnamefont {K.}~\bibnamefont {Sengupta}},\ and\ \bibinfo
  {author} {\bibfnamefont {D.}~\bibnamefont {Sen}},\ }\bibfield  {title}
  {\bibinfo {title} {Transport across a junction of topological insulators and
  a superconductor},\ }\href {https://doi.org/10.1103/PhysRevB.87.245435}
  {\bibfield  {journal} {\bibinfo  {journal} {Phys. Rev. B}\ }\textbf {\bibinfo
  {volume} {87}},\ \bibinfo {pages} {245435} (\bibinfo {year}
  {2013})}\BibitemShut {NoStop}%
\bibitem [{\citenamefont {Shiozaki}\ and\ \citenamefont
  {Sato}(2014)}]{sgiozaki_topo_2014}%
  \BibitemOpen
  \bibfield  {author} {\bibinfo {author} {\bibfnamefont {K.}~\bibnamefont
  {Shiozaki}}\ and\ \bibinfo {author} {\bibfnamefont {M.}~\bibnamefont
  {Sato}},\ }\bibfield  {title} {\bibinfo {title} {Topology of crystalline
  insulators and superconductors},\ }\href
  {https://doi.org/10.1103/PhysRevB.90.165114} {\bibfield  {journal} {\bibinfo
  {journal} {Phys. Rev. B}\ }\textbf {\bibinfo {volume} {90}},\ \bibinfo
  {pages} {165114} (\bibinfo {year} {2014})}\BibitemShut {NoStop}%
\bibitem [{\citenamefont {Heimes}\ \emph {et~al.}(2014)\citenamefont {Heimes},
  \citenamefont {Kotetes},\ and\ \citenamefont {Sch\"on}}]{heimis_shiba_2014}%
  \BibitemOpen
  \bibfield  {author} {\bibinfo {author} {\bibfnamefont {A.}~\bibnamefont
  {Heimes}}, \bibinfo {author} {\bibfnamefont {P.}~\bibnamefont {Kotetes}},\
  and\ \bibinfo {author} {\bibfnamefont {G.}~\bibnamefont {Sch\"on}},\
  }\bibfield  {title} {\bibinfo {title} {Majorana fermions from shiba states in
  an antiferromagnetic chain on top of a superconductor},\ }\href
  {https://doi.org/10.1103/PhysRevB.90.060507} {\bibfield  {journal} {\bibinfo
  {journal} {Phys. Rev. B}\ }\textbf {\bibinfo {volume} {90}},\ \bibinfo
  {pages} {060507} (\bibinfo {year} {2014})}\BibitemShut {NoStop}%
\bibitem [{\citenamefont {Rainis}\ \emph {et~al.}(2014)\citenamefont {Rainis},
  \citenamefont {Saha}, \citenamefont {Klinovaja}, \citenamefont {Trifunovic},\
  and\ \citenamefont {Loss}}]{rainis_nano_2014}%
  \BibitemOpen
  \bibfield  {author} {\bibinfo {author} {\bibfnamefont {D.}~\bibnamefont
  {Rainis}}, \bibinfo {author} {\bibfnamefont {A.}~\bibnamefont {Saha}},
  \bibinfo {author} {\bibfnamefont {J.}~\bibnamefont {Klinovaja}}, \bibinfo
  {author} {\bibfnamefont {L.}~\bibnamefont {Trifunovic}},\ and\ \bibinfo
  {author} {\bibfnamefont {D.}~\bibnamefont {Loss}},\ }\bibfield  {title}
  {\bibinfo {title} {Transport signatures of fractional fermions in rashba
  nanowires},\ }\href {https://doi.org/10.1103/PhysRevLett.112.196803}
  {\bibfield  {journal} {\bibinfo  {journal} {Phys. Rev. Lett.}\ }\textbf
  {\bibinfo {volume} {112}},\ \bibinfo {pages} {196803} (\bibinfo {year}
  {2014})}\BibitemShut {NoStop}%
\bibitem [{\citenamefont {Jana}\ \emph {et~al.}(2019)\citenamefont {Jana},
  \citenamefont {Saha},\ and\ \citenamefont {Das}}]{jana_nano_2019}%
  \BibitemOpen
  \bibfield  {author} {\bibinfo {author} {\bibfnamefont {S.}~\bibnamefont
  {Jana}}, \bibinfo {author} {\bibfnamefont {A.}~\bibnamefont {Saha}},\ and\
  \bibinfo {author} {\bibfnamefont {S.}~\bibnamefont {Das}},\ }\bibfield
  {title} {\bibinfo {title} {Jackiw-rebbi zero modes in non-uniform topological
  insulator nanowire},\ }\href {https://doi.org/10.1103/PhysRevB.100.085428}
  {\bibfield  {journal} {\bibinfo  {journal} {Phys. Rev. B}\ }\textbf {\bibinfo
  {volume} {100}},\ \bibinfo {pages} {085428} (\bibinfo {year}
  {2019})}\BibitemShut {NoStop}%
\bibitem [{\citenamefont {Tuegel}\ \emph {et~al.}(2019)\citenamefont {Tuegel},
  \citenamefont {Chua},\ and\ \citenamefont {Hughes}}]{tuegel_embedded_2019}%
  \BibitemOpen
  \bibfield  {author} {\bibinfo {author} {\bibfnamefont {T.~I.}\ \bibnamefont
  {Tuegel}}, \bibinfo {author} {\bibfnamefont {V.}~\bibnamefont {Chua}},\ and\
  \bibinfo {author} {\bibfnamefont {T.~L.}\ \bibnamefont {Hughes}},\ }\bibfield
   {title} {\bibinfo {title} {Embedded topological insulators},\ }\href
  {https://doi.org/10.1103/PhysRevB.100.115126} {\bibfield  {journal} {\bibinfo
   {journal} {Physical Review B}\ }\textbf {\bibinfo {volume} {100}},\ \bibinfo
  {pages} {115126} (\bibinfo {year} {2019})}\BibitemShut {NoStop}%
\bibitem [{\citenamefont {Slager}(2019)}]{SLAGER_tbi_2019}%
  \BibitemOpen
  \bibfield  {author} {\bibinfo {author} {\bibfnamefont {R.-J.}\ \bibnamefont
  {Slager}},\ }\bibfield  {title} {\bibinfo {title} {The translational side of
  topological band insulators},\ }\href
  {https://doi.org/https://doi.org/10.1016/j.jpcs.2018.01.023} {\bibfield
  {journal} {\bibinfo  {journal} {Journal of Physics and Chemistry of Solids}\
  }\textbf {\bibinfo {volume} {128}},\ \bibinfo {pages} {24} (\bibinfo {year}
  {2019})},\ \bibinfo {note} {spin-Orbit Coupled Materials}\BibitemShut
  {NoStop}%
\bibitem [{\citenamefont {Deb}\ \emph {et~al.}(2021)\citenamefont {Deb},
  \citenamefont {Hoffman}, \citenamefont {Loss},\ and\ \citenamefont
  {Klinovaja}}]{Deb_nano_2021}%
  \BibitemOpen
  \bibfield  {author} {\bibinfo {author} {\bibfnamefont {O.}~\bibnamefont
  {Deb}}, \bibinfo {author} {\bibfnamefont {S.}~\bibnamefont {Hoffman}},
  \bibinfo {author} {\bibfnamefont {D.}~\bibnamefont {Loss}},\ and\ \bibinfo
  {author} {\bibfnamefont {J.}~\bibnamefont {Klinovaja}},\ }\bibfield  {title}
  {\bibinfo {title} {Yu-shiba-rusinov states and ordering of magnetic
  impurities near the boundary of a superconducting nanowire},\ }\href
  {https://doi.org/10.1103/PhysRevB.103.165403} {\bibfield  {journal} {\bibinfo
   {journal} {Phys. Rev. B}\ }\textbf {\bibinfo {volume} {103}},\ \bibinfo
  {pages} {165403} (\bibinfo {year} {2021})}\BibitemShut {NoStop}%
\bibitem [{\citenamefont {Velury}\ and\ \citenamefont
  {Hughes}(2022)}]{velury_ts_2022}%
  \BibitemOpen
  \bibfield  {author} {\bibinfo {author} {\bibfnamefont {S.}~\bibnamefont
  {Velury}}\ and\ \bibinfo {author} {\bibfnamefont {T.~L.}\ \bibnamefont
  {Hughes}},\ }\bibfield  {title} {\bibinfo {title} {Embedded topological
  semimetals},\ }\href {https://doi.org/10.1103/PhysRevB.105.184105} {\bibfield
   {journal} {\bibinfo  {journal} {Phys. Rev. B}\ }\textbf {\bibinfo {volume}
  {105}},\ \bibinfo {pages} {184105} (\bibinfo {year} {2022})}\BibitemShut
  {NoStop}%
\bibitem [{\citenamefont {Adak}\ \emph {et~al.}(2022)\citenamefont {Adak},
  \citenamefont {Mukhopadhyay}, \citenamefont {De}, \citenamefont {Khanna},
  \citenamefont {Rao},\ and\ \citenamefont {Das}}]{adak_tisc_2022}%
  \BibitemOpen
  \bibfield  {author} {\bibinfo {author} {\bibfnamefont {V.}~\bibnamefont
  {Adak}}, \bibinfo {author} {\bibfnamefont {A.}~\bibnamefont {Mukhopadhyay}},
  \bibinfo {author} {\bibfnamefont {S.~J.}\ \bibnamefont {De}}, \bibinfo
  {author} {\bibfnamefont {U.}~\bibnamefont {Khanna}}, \bibinfo {author}
  {\bibfnamefont {S.}~\bibnamefont {Rao}},\ and\ \bibinfo {author}
  {\bibfnamefont {S.}~\bibnamefont {Das}},\ }\bibfield  {title} {\bibinfo
  {title} {Chiral detection of majorana bound states at the edge of a quantum
  spin hall insulator},\ }\href {https://doi.org/10.1103/PhysRevB.106.045422}
  {\bibfield  {journal} {\bibinfo  {journal} {Phys. Rev. B}\ }\textbf {\bibinfo
  {volume} {106}},\ \bibinfo {pages} {045422} (\bibinfo {year}
  {2022})}\BibitemShut {NoStop}%
\bibitem [{\citenamefont {Saxena}\ \emph {et~al.}(2022)\citenamefont {Saxena},
  \citenamefont {Grosfeld}, \citenamefont {E~de Graaf}, \citenamefont
  {Lindstrom}, \citenamefont {Lombardi}, \citenamefont {Deb},\ and\
  \citenamefont {Ginossar}}]{Saxena_nano_2022}%
  \BibitemOpen
  \bibfield  {author} {\bibinfo {author} {\bibfnamefont {R.}~\bibnamefont
  {Saxena}}, \bibinfo {author} {\bibfnamefont {E.}~\bibnamefont {Grosfeld}},
  \bibinfo {author} {\bibfnamefont {S.}~\bibnamefont {E~de Graaf}}, \bibinfo
  {author} {\bibfnamefont {T.}~\bibnamefont {Lindstrom}}, \bibinfo {author}
  {\bibfnamefont {F.}~\bibnamefont {Lombardi}}, \bibinfo {author}
  {\bibfnamefont {O.}~\bibnamefont {Deb}},\ and\ \bibinfo {author}
  {\bibfnamefont {E.}~\bibnamefont {Ginossar}},\ }\bibfield  {title} {\bibinfo
  {title} {Electronic confinement of surface states in a topological insulator
  nanowire},\ }\href {https://doi.org/10.1103/PhysRevB.106.035407} {\bibfield
  {journal} {\bibinfo  {journal} {Phys. Rev. B}\ }\textbf {\bibinfo {volume}
  {106}},\ \bibinfo {pages} {035407} (\bibinfo {year} {2022})}\BibitemShut
  {NoStop}%
\bibitem [{\citenamefont {Chatterjee}\ \emph {et~al.}(2023)\citenamefont
  {Chatterjee}, \citenamefont {Pradhan}, \citenamefont {Nandy},\ and\
  \citenamefont {Saha}}]{pritam_prb_2023}%
  \BibitemOpen
  \bibfield  {author} {\bibinfo {author} {\bibfnamefont {P.}~\bibnamefont
  {Chatterjee}}, \bibinfo {author} {\bibfnamefont {S.}~\bibnamefont {Pradhan}},
  \bibinfo {author} {\bibfnamefont {A.~K.}\ \bibnamefont {Nandy}},\ and\
  \bibinfo {author} {\bibfnamefont {A.}~\bibnamefont {Saha}},\ }\bibfield
  {title} {\bibinfo {title} {Tailoring the phase transition from topological
  superconductor to trivial superconductor induced by magnetic textures of a
  spin chain on a $p$-wave superconductor},\ }\href
  {https://doi.org/10.1103/PhysRevB.107.085423} {\bibfield  {journal} {\bibinfo
   {journal} {Phys. Rev. B}\ }\textbf {\bibinfo {volume} {107}},\ \bibinfo
  {pages} {085423} (\bibinfo {year} {2023})}\BibitemShut {NoStop}%
\bibitem [{\citenamefont {Nyári}\ \emph {et~al.}(2023)\citenamefont {Nyári},
  \citenamefont {Lászlóffy}, \citenamefont {Csire}, \citenamefont
  {Szunyogh},\ and\ \citenamefont {Újfalussy}}]{nyari_topological_2023}%
  \BibitemOpen
  \bibfield  {author} {\bibinfo {author} {\bibfnamefont {B.}~\bibnamefont
  {Nyári}}, \bibinfo {author} {\bibfnamefont {A.}~\bibnamefont {Lászlóffy}},
  \bibinfo {author} {\bibfnamefont {G.}~\bibnamefont {Csire}}, \bibinfo
  {author} {\bibfnamefont {L.}~\bibnamefont {Szunyogh}},\ and\ \bibinfo
  {author} {\bibfnamefont {B.}~\bibnamefont {Újfalussy}},\ }\bibfield  {title}
  {\bibinfo {title} {Topological superconductivity from first principles. i.
  shiba band structure and topological edge states of artificial spin chains},\
  }\href {https://doi.org/10.1103/PhysRevB.108.134512} {\bibfield  {journal}
  {\bibinfo  {journal} {Physical Review B}\ }\textbf {\bibinfo {volume}
  {108}},\ \bibinfo {pages} {134512} (\bibinfo {year} {2023})}\BibitemShut
  {NoStop}%
\bibitem [{\citenamefont {Lászlóffy}\ \emph {et~al.}(2023)\citenamefont
  {Lászlóffy}, \citenamefont {Nyári}, \citenamefont {Csire}, \citenamefont
  {Szunyogh},\ and\ \citenamefont {Újfalussy}}]{laszloffy_topological_2023}%
  \BibitemOpen
  \bibfield  {author} {\bibinfo {author} {\bibfnamefont {A.}~\bibnamefont
  {Lászlóffy}}, \bibinfo {author} {\bibfnamefont {B.}~\bibnamefont {Nyári}},
  \bibinfo {author} {\bibfnamefont {G.}~\bibnamefont {Csire}}, \bibinfo
  {author} {\bibfnamefont {L.}~\bibnamefont {Szunyogh}},\ and\ \bibinfo
  {author} {\bibfnamefont {B.}~\bibnamefont {Újfalussy}},\ }\bibfield  {title}
  {\bibinfo {title} {Topological superconductivity from first principles. ii.
  effects from manipulation of spin spirals: Topological fragmentation,
  braiding, and quasi-majorana bound states},\ }\href
  {https://doi.org/10.1103/PhysRevB.108.134513} {\bibfield  {journal} {\bibinfo
   {journal} {Physical Review B}\ }\textbf {\bibinfo {volume} {108}},\ \bibinfo
  {pages} {134513} (\bibinfo {year} {2023})}\BibitemShut {NoStop}%
\bibitem [{\citenamefont {MullineauxSanders}\ and\ \citenamefont
  {Braunecker}(2024)}]{harry_interface_2024}%
  \BibitemOpen
  \bibfield  {author} {\bibinfo {author} {\bibfnamefont {H.}~\bibnamefont
  {MullineauxSanders}}\ and\ \bibinfo {author} {\bibfnamefont {B.}~\bibnamefont
  {Braunecker}},\ }\bibfield  {title} {\bibinfo {title} {Topological
  classification of one-dimensional chiral symmetric interfaces},\ }\href
  {https://doi.org/10.1103/PhysRevB.110.L241409} {\bibfield  {journal}
  {\bibinfo  {journal} {Phys. Rev. B}\ }\textbf {\bibinfo {volume} {110}},\
  \bibinfo {pages} {L241409} (\bibinfo {year} {2024})}\BibitemShut {NoStop}%
\bibitem [{\citenamefont {Mondal}\ \emph {et~al.}(2025)\citenamefont {Mondal},
  \citenamefont {Pal}, \citenamefont {Saha},\ and\ \citenamefont
  {Nag}}]{mondal_amzm_2024}%
  \BibitemOpen
  \bibfield  {author} {\bibinfo {author} {\bibfnamefont {D.}~\bibnamefont
  {Mondal}}, \bibinfo {author} {\bibfnamefont {A.}~\bibnamefont {Pal}},
  \bibinfo {author} {\bibfnamefont {A.}~\bibnamefont {Saha}},\ and\ \bibinfo
  {author} {\bibfnamefont {T.}~\bibnamefont {Nag}},\ }\bibfield  {title}
  {\bibinfo {title} {Distinguishing between topological majorana and trivial
  zero modes via transport and shot noise study in an altermagnet
  heterostructure},\ }\href {https://doi.org/10.1103/PhysRevB.111.L121401}
  {\bibfield  {journal} {\bibinfo  {journal} {Phys. Rev. B}\ }\textbf {\bibinfo
  {volume} {111}},\ \bibinfo {pages} {L121401} (\bibinfo {year}
  {2025})}\BibitemShut {NoStop}%
\bibitem [{\citenamefont {Bhowmik}\ and\ \citenamefont
  {Saha}(2025)}]{bhowmik_fflo_2025}%
  \BibitemOpen
  \bibfield  {author} {\bibinfo {author} {\bibfnamefont {S.}~\bibnamefont
  {Bhowmik}}\ and\ \bibinfo {author} {\bibfnamefont {A.}~\bibnamefont {Saha}},\
  }\href@noop {} {\bibinfo {title} {Realizing fulde-ferrell-larkin-ovchinnikov
  pairing driven topological majorana zero modes and superconducting diode
  effect in helical shiba chain}} (\bibinfo {year} {2025}),\ \Eprint
  {https://arxiv.org/abs/2412.11784} {arXiv:2412.11784 [cond-mat.supr-con]}
  \BibitemShut {NoStop}%
\bibitem [{\citenamefont {Wang}\ and\ \citenamefont
  {Wang}(2004)}]{wnag_interface_2004}%
  \BibitemOpen
  \bibfield  {author} {\bibinfo {author} {\bibfnamefont {Q.-H.}\ \bibnamefont
  {Wang}}\ and\ \bibinfo {author} {\bibfnamefont {Z.~D.}\ \bibnamefont
  {Wang}},\ }\bibfield  {title} {\bibinfo {title} {Impurity and interface bound
  states in ${d}_{{x}^{2}\ensuremath{-}{y}^{2}}{+id}_{\mathrm{xy}}$ and
  ${p}_{x}{+ip}_{y}$ superconductors},\ }\href
  {https://doi.org/10.1103/PhysRevB.69.092502} {\bibfield  {journal} {\bibinfo
  {journal} {Phys. Rev. B}\ }\textbf {\bibinfo {volume} {69}},\ \bibinfo
  {pages} {092502} (\bibinfo {year} {2004})}\BibitemShut {NoStop}%
\bibitem [{\citenamefont {Ran}\ \emph {et~al.}(2009)\citenamefont {Ran},
  \citenamefont {Zhang},\ and\ \citenamefont {Vishwanath}}]{ran_em_2009}%
  \BibitemOpen
  \bibfield  {author} {\bibinfo {author} {\bibfnamefont {Y.}~\bibnamefont
  {Ran}}, \bibinfo {author} {\bibfnamefont {Y.}~\bibnamefont {Zhang}},\ and\
  \bibinfo {author} {\bibfnamefont {A.}~\bibnamefont {Vishwanath}},\ }\bibfield
   {title} {\bibinfo {title} {One-dimensional topologically protected modes in
  topological insulators with lattice dislocations},\ }\href
  {https://www.nature.com/articles/nphys1220} {\bibfield  {journal} {\bibinfo
  {journal} {Nature Physics}\ }\textbf {\bibinfo {volume} {5}},\ \bibinfo
  {pages} {298} (\bibinfo {year} {2009})}\BibitemShut {NoStop}%
\bibitem [{\citenamefont {Wimmer}\ \emph {et~al.}(2010)\citenamefont {Wimmer},
  \citenamefont {Akhmerov}, \citenamefont {Medvedyeva}, \citenamefont
  {Tworzyd\l{}o},\ and\ \citenamefont {Beenakker}}]{Wimmer_PRL_2010}%
  \BibitemOpen
  \bibfield  {author} {\bibinfo {author} {\bibfnamefont {M.}~\bibnamefont
  {Wimmer}}, \bibinfo {author} {\bibfnamefont {A.~R.}\ \bibnamefont
  {Akhmerov}}, \bibinfo {author} {\bibfnamefont {M.~V.}\ \bibnamefont
  {Medvedyeva}}, \bibinfo {author} {\bibfnamefont {J.}~\bibnamefont
  {Tworzyd\l{}o}},\ and\ \bibinfo {author} {\bibfnamefont {C.~W.~J.}\
  \bibnamefont {Beenakker}},\ }\bibfield  {title} {\bibinfo {title} {Majorana
  bound states without vortices in topological superconductors with
  electrostatic defects},\ }\href
  {https://doi.org/10.1103/PhysRevLett.105.046803} {\bibfield  {journal}
  {\bibinfo  {journal} {Phys. Rev. Lett.}\ }\textbf {\bibinfo {volume} {105}},\
  \bibinfo {pages} {046803} (\bibinfo {year} {2010})}\BibitemShut {NoStop}%
\bibitem [{\citenamefont {Shiozaki}\ and\ \citenamefont
  {Fujimoto}(2012)}]{shiozaki_gf_2012}%
  \BibitemOpen
  \bibfield  {author} {\bibinfo {author} {\bibfnamefont {K.}~\bibnamefont
  {Shiozaki}}\ and\ \bibinfo {author} {\bibfnamefont {S.}~\bibnamefont
  {Fujimoto}},\ }\bibfield  {title} {\bibinfo {title} {Green's function method
  for line defects and gapless modes in topological insulators: Beyond the
  semiclassical approach},\ }\href {https://doi.org/10.1103/PhysRevB.85.085409}
  {\bibfield  {journal} {\bibinfo  {journal} {Phys. Rev. B}\ }\textbf {\bibinfo
  {volume} {85}},\ \bibinfo {pages} {085409} (\bibinfo {year}
  {2012})}\BibitemShut {NoStop}%
\bibitem [{\citenamefont {Potter}\ and\ \citenamefont
  {Lee}(2012)}]{Potter_mzmsurface_2012}%
  \BibitemOpen
  \bibfield  {author} {\bibinfo {author} {\bibfnamefont {A.~C.}\ \bibnamefont
  {Potter}}\ and\ \bibinfo {author} {\bibfnamefont {P.~A.}\ \bibnamefont
  {Lee}},\ }\bibfield  {title} {\bibinfo {title} {Topological superconductivity
  and majorana fermions in metallic surface states},\ }\href
  {https://doi.org/10.1103/PhysRevB.85.094516} {\bibfield  {journal} {\bibinfo
  {journal} {Phys. Rev. B}\ }\textbf {\bibinfo {volume} {85}},\ \bibinfo
  {pages} {094516} (\bibinfo {year} {2012})}\BibitemShut {NoStop}%
\bibitem [{\citenamefont {Juri\ifmmode \check{c}\else
  \v{c}\fi{}i\ifmmode~\acute{c}\else \'{c}\fi{}}\ \emph
  {et~al.}(2012)\citenamefont {Juri\ifmmode \check{c}\else
  \v{c}\fi{}i\ifmmode~\acute{c}\else \'{c}\fi{}}, \citenamefont {Mesaros},
  \citenamefont {Slager},\ and\ \citenamefont {Zaanen}}]{Vladimir_piflux_2012}%
  \BibitemOpen
  \bibfield  {author} {\bibinfo {author} {\bibfnamefont {V.}~\bibnamefont
  {Juri\ifmmode \check{c}\else \v{c}\fi{}i\ifmmode~\acute{c}\else \'{c}\fi{}}},
  \bibinfo {author} {\bibfnamefont {A.}~\bibnamefont {Mesaros}}, \bibinfo
  {author} {\bibfnamefont {R.-J.}\ \bibnamefont {Slager}},\ and\ \bibinfo
  {author} {\bibfnamefont {J.}~\bibnamefont {Zaanen}},\ }\bibfield  {title}
  {\bibinfo {title} {Universal probes of two-dimensional topological
  insulators: Dislocation and $\ensuremath{\pi}$ flux},\ }\href
  {https://doi.org/10.1103/PhysRevLett.108.106403} {\bibfield  {journal}
  {\bibinfo  {journal} {Phys. Rev. Lett.}\ }\textbf {\bibinfo {volume} {108}},\
  \bibinfo {pages} {106403} (\bibinfo {year} {2012})}\BibitemShut {NoStop}%
\bibitem [{\citenamefont {Zhang}\ \emph {et~al.}(2013)\citenamefont {Zhang},
  \citenamefont {Kane},\ and\ \citenamefont {Mele}}]{Zhang_scnanowire_2013}%
  \BibitemOpen
  \bibfield  {author} {\bibinfo {author} {\bibfnamefont {F.}~\bibnamefont
  {Zhang}}, \bibinfo {author} {\bibfnamefont {C.~L.}\ \bibnamefont {Kane}},\
  and\ \bibinfo {author} {\bibfnamefont {E.~J.}\ \bibnamefont {Mele}},\
  }\bibfield  {title} {\bibinfo {title} {Time-reversal-invariant topological
  superconductivity and majorana kramers pairs},\ }\href
  {https://doi.org/10.1103/PhysRevLett.111.056402} {\bibfield  {journal}
  {\bibinfo  {journal} {Phys. Rev. Lett.}\ }\textbf {\bibinfo {volume} {111}},\
  \bibinfo {pages} {056402} (\bibinfo {year} {2013})}\BibitemShut {NoStop}%
\bibitem [{\citenamefont {Slager}\ \emph {et~al.}(2014)\citenamefont {Slager},
  \citenamefont {Mesaros}, \citenamefont {Juri\ifmmode \check{c}\else
  \v{c}\fi{}i\ifmmode~\acute{c}\else \'{c}\fi{}},\ and\ \citenamefont
  {Zaanen}}]{slager_dislocation_2014}%
  \BibitemOpen
  \bibfield  {author} {\bibinfo {author} {\bibfnamefont {R.-J.}\ \bibnamefont
  {Slager}}, \bibinfo {author} {\bibfnamefont {A.}~\bibnamefont {Mesaros}},
  \bibinfo {author} {\bibfnamefont {V.}~\bibnamefont {Juri\ifmmode
  \check{c}\else \v{c}\fi{}i\ifmmode~\acute{c}\else \'{c}\fi{}}},\ and\
  \bibinfo {author} {\bibfnamefont {J.}~\bibnamefont {Zaanen}},\ }\bibfield
  {title} {\bibinfo {title} {Interplay between electronic topology and crystal
  symmetry: Dislocation-line modes in topological band insulators},\ }\href
  {https://doi.org/10.1103/PhysRevB.90.241403} {\bibfield  {journal} {\bibinfo
  {journal} {Phys. Rev. B}\ }\textbf {\bibinfo {volume} {90}},\ \bibinfo
  {pages} {241403} (\bibinfo {year} {2014})}\BibitemShut {NoStop}%
\bibitem [{\citenamefont {Benalcazar}\ \emph {et~al.}(2014)\citenamefont
  {Benalcazar}, \citenamefont {Teo},\ and\ \citenamefont
  {Hughes}}]{waldimir_disc_2014}%
  \BibitemOpen
  \bibfield  {author} {\bibinfo {author} {\bibfnamefont {W.~A.}\ \bibnamefont
  {Benalcazar}}, \bibinfo {author} {\bibfnamefont {J.~C.~Y.}\ \bibnamefont
  {Teo}},\ and\ \bibinfo {author} {\bibfnamefont {T.~L.}\ \bibnamefont
  {Hughes}},\ }\bibfield  {title} {\bibinfo {title} {Classification of
  two-dimensional topological crystalline superconductors and majorana bound
  states at disclinations},\ }\href
  {https://doi.org/10.1103/PhysRevB.89.224503} {\bibfield  {journal} {\bibinfo
  {journal} {Phys. Rev. B}\ }\textbf {\bibinfo {volume} {89}},\ \bibinfo
  {pages} {224503} (\bibinfo {year} {2014})}\BibitemShut {NoStop}%
\bibitem [{\citenamefont {Slager}\ \emph {et~al.}(2015)\citenamefont {Slager},
  \citenamefont {Rademaker}, \citenamefont {Zaanen},\ and\ \citenamefont
  {Balents}}]{slager_gf_2015}%
  \BibitemOpen
  \bibfield  {author} {\bibinfo {author} {\bibfnamefont {R.-J.}\ \bibnamefont
  {Slager}}, \bibinfo {author} {\bibfnamefont {L.}~\bibnamefont {Rademaker}},
  \bibinfo {author} {\bibfnamefont {J.}~\bibnamefont {Zaanen}},\ and\ \bibinfo
  {author} {\bibfnamefont {L.}~\bibnamefont {Balents}},\ }\bibfield  {title}
  {\bibinfo {title} {Impurity-bound states and green's function zeros as local
  signatures of topology},\ }\href {https://doi.org/10.1103/PhysRevB.92.085126}
  {\bibfield  {journal} {\bibinfo  {journal} {Phys. Rev. B}\ }\textbf {\bibinfo
  {volume} {92}},\ \bibinfo {pages} {085126} (\bibinfo {year}
  {2015})}\BibitemShut {NoStop}%
\bibitem [{\citenamefont {Paulose}\ \emph {et~al.}(2015)\citenamefont
  {Paulose}, \citenamefont {Chen},\ and\ \citenamefont
  {Vitelli}}]{paulose_topo_2015}%
  \BibitemOpen
  \bibfield  {author} {\bibinfo {author} {\bibfnamefont {J.}~\bibnamefont
  {Paulose}}, \bibinfo {author} {\bibfnamefont {B.~G.}\ \bibnamefont {Chen}},\
  and\ \bibinfo {author} {\bibfnamefont {V.}~\bibnamefont {Vitelli}},\
  }\bibfield  {title} {\bibinfo {title} {Topological modes bound to
  dislocations in mechanical metamaterials},\ }\href
  {https://www.nature.com/articles/nphys3185} {\bibfield  {journal} {\bibinfo
  {journal} {Nature Physics}\ }\textbf {\bibinfo {volume} {11}},\ \bibinfo
  {pages} {153} (\bibinfo {year} {2015})}\BibitemShut {NoStop}%
\bibitem [{\citenamefont {Peng}\ \emph {et~al.}(2015)\citenamefont {Peng},
  \citenamefont {Pientka}, \citenamefont {Glazman},\ and\ \citenamefont {von
  Oppen}}]{peng_mbs_2015}%
  \BibitemOpen
  \bibfield  {author} {\bibinfo {author} {\bibfnamefont {Y.}~\bibnamefont
  {Peng}}, \bibinfo {author} {\bibfnamefont {F.}~\bibnamefont {Pientka}},
  \bibinfo {author} {\bibfnamefont {L.~I.}\ \bibnamefont {Glazman}},\ and\
  \bibinfo {author} {\bibfnamefont {F.}~\bibnamefont {von Oppen}},\ }\bibfield
  {title} {\bibinfo {title} {Strong localization of majorana end states in
  chains of magnetic adatoms},\ }\href
  {https://doi.org/10.1103/PhysRevLett.114.106801} {\bibfield  {journal}
  {\bibinfo  {journal} {Phys. Rev. Lett.}\ }\textbf {\bibinfo {volume} {114}},\
  \bibinfo {pages} {106801} (\bibinfo {year} {2015})}\BibitemShut {NoStop}%
\bibitem [{\citenamefont {Sablikov}\ and\ \citenamefont
  {Sukhanov}(2015)}]{Sablikov_nomag_2015}%
  \BibitemOpen
  \bibfield  {author} {\bibinfo {author} {\bibfnamefont {V.~A.}\ \bibnamefont
  {Sablikov}}\ and\ \bibinfo {author} {\bibfnamefont {A.~A.}\ \bibnamefont
  {Sukhanov}},\ }\bibfield  {title} {\bibinfo {title} {Electronic states
  induced by nonmagnetic defects in two-dimensional topological insulators},\
  }\href {https://doi.org/10.1103/PhysRevB.91.075412} {\bibfield  {journal}
  {\bibinfo  {journal} {Phys. Rev. B}\ }\textbf {\bibinfo {volume} {91}},\
  \bibinfo {pages} {075412} (\bibinfo {year} {2015})}\BibitemShut {NoStop}%
\bibitem [{\citenamefont {Neupert}\ \emph {et~al.}(2016)\citenamefont
  {Neupert}, \citenamefont {Yazdani},\ and\ \citenamefont
  {Bernevig}}]{neupart_shiba_2016}%
  \BibitemOpen
  \bibfield  {author} {\bibinfo {author} {\bibfnamefont {T.}~\bibnamefont
  {Neupert}}, \bibinfo {author} {\bibfnamefont {A.}~\bibnamefont {Yazdani}},\
  and\ \bibinfo {author} {\bibfnamefont {B.~A.}\ \bibnamefont {Bernevig}},\
  }\bibfield  {title} {\bibinfo {title} {Shiba chains of scalar impurities on
  unconventional superconductors},\ }\href
  {https://doi.org/10.1103/PhysRevB.93.094508} {\bibfield  {journal} {\bibinfo
  {journal} {Phys. Rev. B}\ }\textbf {\bibinfo {volume} {93}},\ \bibinfo
  {pages} {094508} (\bibinfo {year} {2016})}\BibitemShut {NoStop}%
\bibitem [{\citenamefont {P\"oyh\"onen}\ \emph {et~al.}(2016)\citenamefont
  {P\"oyh\"onen}, \citenamefont {Weststr\"om},\ and\ \citenamefont
  {Ojanen}}]{Poyhonen_topo_2016}%
  \BibitemOpen
  \bibfield  {author} {\bibinfo {author} {\bibfnamefont {K.}~\bibnamefont
  {P\"oyh\"onen}}, \bibinfo {author} {\bibfnamefont {A.}~\bibnamefont
  {Weststr\"om}},\ and\ \bibinfo {author} {\bibfnamefont {T.}~\bibnamefont
  {Ojanen}},\ }\bibfield  {title} {\bibinfo {title} {Topological
  superconductivity in ferromagnetic atom chains beyond the deep-impurity
  regime},\ }\href {https://doi.org/10.1103/PhysRevB.93.014517} {\bibfield
  {journal} {\bibinfo  {journal} {Phys. Rev. B}\ }\textbf {\bibinfo {volume}
  {93}},\ \bibinfo {pages} {014517} (\bibinfo {year} {2016})}\BibitemShut
  {NoStop}%
\bibitem [{\citenamefont {Sahlberg}\ \emph {et~al.}(2017)\citenamefont
  {Sahlberg}, \citenamefont {Weststr\"om}, \citenamefont {P\"oyh\"onen},\ and\
  \citenamefont {Ojanen}}]{sahlberg_1dtopo_2017}%
  \BibitemOpen
  \bibfield  {author} {\bibinfo {author} {\bibfnamefont {I.}~\bibnamefont
  {Sahlberg}}, \bibinfo {author} {\bibfnamefont {A.}~\bibnamefont
  {Weststr\"om}}, \bibinfo {author} {\bibfnamefont {K.}~\bibnamefont
  {P\"oyh\"onen}},\ and\ \bibinfo {author} {\bibfnamefont {T.}~\bibnamefont
  {Ojanen}},\ }\bibfield  {title} {\bibinfo {title} {Engineering
  one-dimensional topological phases on $p$-wave superconductors},\ }\href
  {https://doi.org/10.1103/PhysRevB.95.184512} {\bibfield  {journal} {\bibinfo
  {journal} {Phys. Rev. B}\ }\textbf {\bibinfo {volume} {95}},\ \bibinfo
  {pages} {184512} (\bibinfo {year} {2017})}\BibitemShut {NoStop}%
\bibitem [{\citenamefont {Zhang}\ \emph {et~al.}(2021)\citenamefont {Zhang},
  \citenamefont {Jiang}, \citenamefont {Zhang}, \citenamefont {Wang},\ and\
  \citenamefont {Wang}}]{Zhang_PRX_2021}%
  \BibitemOpen
  \bibfield  {author} {\bibinfo {author} {\bibfnamefont {Y.}~\bibnamefont
  {Zhang}}, \bibinfo {author} {\bibfnamefont {K.}~\bibnamefont {Jiang}},
  \bibinfo {author} {\bibfnamefont {F.}~\bibnamefont {Zhang}}, \bibinfo
  {author} {\bibfnamefont {J.}~\bibnamefont {Wang}},\ and\ \bibinfo {author}
  {\bibfnamefont {Z.}~\bibnamefont {Wang}},\ }\bibfield  {title} {\bibinfo
  {title} {Atomic line defects and topological superconductivity in
  unconventional superconductors},\ }\href
  {https://doi.org/10.1103/PhysRevX.11.011041} {\bibfield  {journal} {\bibinfo
  {journal} {Phys. Rev. X}\ }\textbf {\bibinfo {volume} {11}},\ \bibinfo
  {pages} {011041} (\bibinfo {year} {2021})}\BibitemShut {NoStop}%
\bibitem [{\citenamefont {Kaladzhyan}\ \emph {et~al.}(2018)\citenamefont
  {Kaladzhyan}, \citenamefont {Bena},\ and\ \citenamefont
  {Simon}}]{kaladzhyan_tft_2018}%
  \BibitemOpen
  \bibfield  {author} {\bibinfo {author} {\bibfnamefont {V.}~\bibnamefont
  {Kaladzhyan}}, \bibinfo {author} {\bibfnamefont {C.}~\bibnamefont {Bena}},\
  and\ \bibinfo {author} {\bibfnamefont {P.}~\bibnamefont {Simon}},\ }\bibfield
   {title} {\bibinfo {title} {Topology from triviality},\ }\href
  {https://doi.org/10.1103/PhysRevB.97.104512} {\bibfield  {journal} {\bibinfo
  {journal} {Phys. Rev. B}\ }\textbf {\bibinfo {volume} {97}},\ \bibinfo
  {pages} {104512} (\bibinfo {year} {2018})}\BibitemShut {NoStop}%
\bibitem [{\citenamefont {Sticlet}\ and\ \citenamefont
  {Morari}(2019)}]{sticlet_topological_2019}%
  \BibitemOpen
  \bibfield  {author} {\bibinfo {author} {\bibfnamefont {D.}~\bibnamefont
  {Sticlet}}\ and\ \bibinfo {author} {\bibfnamefont {C.}~\bibnamefont
  {Morari}},\ }\bibfield  {title} {\bibinfo {title} {Topological
  superconductivity from magnetic impurities on monolayer nbse$_2$},\ }\href
  {https://doi.org/10.1103/PhysRevB.100.075420} {\bibfield  {journal} {\bibinfo
   {journal} {Physical Review B}\ }\textbf {\bibinfo {volume} {100}},\ \bibinfo
  {pages} {075420} (\bibinfo {year} {2019})}\BibitemShut {NoStop}%
\bibitem [{\citenamefont {Kaladzhyan}\ and\ \citenamefont
  {Bena}(2019)}]{bena_imp_2019}%
  \BibitemOpen
  \bibfield  {author} {\bibinfo {author} {\bibfnamefont {V.}~\bibnamefont
  {Kaladzhyan}}\ and\ \bibinfo {author} {\bibfnamefont {C.}~\bibnamefont
  {Bena}},\ }\bibfield  {title} {\bibinfo {title} {Obtaining majorana and other
  boundary modes from the metamorphosis of impurity-induced states: Exact
  solutions via the t-matrix},\ }\href
  {https://doi.org/10.1103/PhysRevB.100.081106} {\bibfield  {journal} {\bibinfo
   {journal} {Phys. Rev. B}\ }\textbf {\bibinfo {volume} {100}},\ \bibinfo
  {pages} {081106} (\bibinfo {year} {2019})}\BibitemShut {NoStop}%
\bibitem [{\citenamefont {Li}\ \emph {et~al.}(2020)\citenamefont {Li},
  \citenamefont {Zhu}, \citenamefont {Benalcazar},\ and\ \citenamefont
  {Hughes}}]{Li_disc_2020}%
  \BibitemOpen
  \bibfield  {author} {\bibinfo {author} {\bibfnamefont {T.}~\bibnamefont
  {Li}}, \bibinfo {author} {\bibfnamefont {P.}~\bibnamefont {Zhu}}, \bibinfo
  {author} {\bibfnamefont {W.~A.}\ \bibnamefont {Benalcazar}},\ and\ \bibinfo
  {author} {\bibfnamefont {T.~L.}\ \bibnamefont {Hughes}},\ }\bibfield  {title}
  {\bibinfo {title} {Fractional disclination charge in two-dimensional
  ${C}_{n}$-symmetric topological crystalline insulators},\ }\href
  {https://doi.org/10.1103/PhysRevB.101.115115} {\bibfield  {journal} {\bibinfo
   {journal} {Phys. Rev. B}\ }\textbf {\bibinfo {volume} {101}},\ \bibinfo
  {pages} {115115} (\bibinfo {year} {2020})}\BibitemShut {NoStop}%
\bibitem [{\citenamefont {Sedlmayr}\ \emph {et~al.}(2021)\citenamefont
  {Sedlmayr}, \citenamefont {Kaladzhyan},\ and\ \citenamefont
  {Bena}}]{sedlmayr_shiba_2021}%
  \BibitemOpen
  \bibfield  {author} {\bibinfo {author} {\bibfnamefont {N.}~\bibnamefont
  {Sedlmayr}}, \bibinfo {author} {\bibfnamefont {V.}~\bibnamefont
  {Kaladzhyan}},\ and\ \bibinfo {author} {\bibfnamefont {C.}~\bibnamefont
  {Bena}},\ }\bibfield  {title} {\bibinfo {title} {Analytical and
  semianalytical tools to determine the topological character of shiba
  chains},\ }\href {https://doi.org/10.1103/PhysRevB.104.024508} {\bibfield
  {journal} {\bibinfo  {journal} {Phys. Rev. B}\ }\textbf {\bibinfo {volume}
  {104}},\ \bibinfo {pages} {024508} (\bibinfo {year} {2021})}\BibitemShut
  {NoStop}%
\bibitem [{\citenamefont {Carroll}\ and\ \citenamefont
  {Braunecker}(2021{\natexlab{a}})}]{carrol_I_2021}%
  \BibitemOpen
  \bibfield  {author} {\bibinfo {author} {\bibfnamefont {C.~J.~F.}\
  \bibnamefont {Carroll}}\ and\ \bibinfo {author} {\bibfnamefont
  {B.}~\bibnamefont {Braunecker}},\ }\bibfield  {title} {\bibinfo {title}
  {Subgap states at ferromagnetic and spiral-ordered magnetic chains in
  two-dimensional superconductors. i. continuum description},\ }\href
  {https://doi.org/10.1103/PhysRevB.104.245133} {\bibfield  {journal} {\bibinfo
   {journal} {Phys. Rev. B}\ }\textbf {\bibinfo {volume} {104}},\ \bibinfo
  {pages} {245133} (\bibinfo {year} {2021}{\natexlab{a}})}\BibitemShut
  {NoStop}%
\bibitem [{\citenamefont {Carroll}\ and\ \citenamefont
  {Braunecker}(2021{\natexlab{b}})}]{carrol_II_2021}%
  \BibitemOpen
  \bibfield  {author} {\bibinfo {author} {\bibfnamefont {C.~J.~F.}\
  \bibnamefont {Carroll}}\ and\ \bibinfo {author} {\bibfnamefont
  {B.}~\bibnamefont {Braunecker}},\ }\bibfield  {title} {\bibinfo {title}
  {Subgap states at ferromagnetic and spiral-ordered magnetic chains in
  two-dimensional superconductors. ii. topological classification},\ }\href
  {https://doi.org/10.1103/PhysRevB.104.245134} {\bibfield  {journal} {\bibinfo
   {journal} {Phys. Rev. B}\ }\textbf {\bibinfo {volume} {104}},\ \bibinfo
  {pages} {245134} (\bibinfo {year} {2021}{\natexlab{b}})}\BibitemShut
  {NoStop}%
\bibitem [{\citenamefont {Miao}\ \emph {et~al.}(2023)\citenamefont {Miao},
  \citenamefont {Wan}, \citenamefont {Sun},\ and\ \citenamefont
  {Zhang}}]{Miao_afm_2023}%
  \BibitemOpen
  \bibfield  {author} {\bibinfo {author} {\bibfnamefont {C.-M.}\ \bibnamefont
  {Miao}}, \bibinfo {author} {\bibfnamefont {Y.-H.}\ \bibnamefont {Wan}},
  \bibinfo {author} {\bibfnamefont {Q.-F.}\ \bibnamefont {Sun}},\ and\ \bibinfo
  {author} {\bibfnamefont {Y.-T.}\ \bibnamefont {Zhang}},\ }\bibfield  {title}
  {\bibinfo {title} {Engineering topologically protected zero-dimensional
  interface end states in antiferromagnetic heterojunction graphene
  nanoflakes},\ }\href {https://doi.org/10.1103/PhysRevB.108.075401} {\bibfield
   {journal} {\bibinfo  {journal} {Phys. Rev. B}\ }\textbf {\bibinfo {volume}
  {108}},\ \bibinfo {pages} {075401} (\bibinfo {year} {2023})}\BibitemShut
  {NoStop}%
\bibitem [{\citenamefont {Chakraborty}\ \emph {et~al.}(2024)\citenamefont
  {Chakraborty}, \citenamefont {Adak},\ and\ \citenamefont
  {Das}}]{chakraborty_mbs_2024}%
  \BibitemOpen
  \bibfield  {author} {\bibinfo {author} {\bibfnamefont {S.}~\bibnamefont
  {Chakraborty}}, \bibinfo {author} {\bibfnamefont {V.}~\bibnamefont {Adak}},\
  and\ \bibinfo {author} {\bibfnamefont {S.}~\bibnamefont {Das}},\ }\bibfield
  {title} {\bibinfo {title} {Robust majorana bound state in pseudospin domain
  wall of a two-dimensional topological insulator},\ }\href
  {https://doi.org/10.1103/PhysRevB.110.155424} {\bibfield  {journal} {\bibinfo
   {journal} {Phys. Rev. B}\ }\textbf {\bibinfo {volume} {110}},\ \bibinfo
  {pages} {155424} (\bibinfo {year} {2024})}\BibitemShut {NoStop}%
\bibitem [{\citenamefont {Pientka}\ \emph {et~al.}(2013)\citenamefont
  {Pientka}, \citenamefont {Glazman},\ and\ \citenamefont {von
  Oppen}}]{Felix_PRB_2013}%
  \BibitemOpen
  \bibfield  {author} {\bibinfo {author} {\bibfnamefont {F.}~\bibnamefont
  {Pientka}}, \bibinfo {author} {\bibfnamefont {L.~I.}\ \bibnamefont
  {Glazman}},\ and\ \bibinfo {author} {\bibfnamefont {F.}~\bibnamefont {von
  Oppen}},\ }\bibfield  {title} {\bibinfo {title} {Topological superconducting
  phase in helical shiba chains},\ }\href
  {https://doi.org/10.1103/PhysRevB.88.155420} {\bibfield  {journal} {\bibinfo
  {journal} {Phys. Rev. B}\ }\textbf {\bibinfo {volume} {88}},\ \bibinfo
  {pages} {155420} (\bibinfo {year} {2013})}\BibitemShut {NoStop}%
\bibitem [{\citenamefont {Pientka}\ \emph {et~al.}(2014)\citenamefont
  {Pientka}, \citenamefont {Glazman},\ and\ \citenamefont {von
  Oppen}}]{Felix_PRB_2014}%
  \BibitemOpen
  \bibfield  {author} {\bibinfo {author} {\bibfnamefont {F.}~\bibnamefont
  {Pientka}}, \bibinfo {author} {\bibfnamefont {L.~I.}\ \bibnamefont
  {Glazman}},\ and\ \bibinfo {author} {\bibfnamefont {F.}~\bibnamefont {von
  Oppen}},\ }\bibfield  {title} {\bibinfo {title} {Unconventional topological
  phase transitions in helical shiba chains},\ }\href
  {https://doi.org/10.1103/PhysRevB.89.180505} {\bibfield  {journal} {\bibinfo
  {journal} {Phys. Rev. B}\ }\textbf {\bibinfo {volume} {89}},\ \bibinfo
  {pages} {180505} (\bibinfo {year} {2014})}\BibitemShut {NoStop}%
\bibitem [{\citenamefont {Kaladzhyan}\ \emph {et~al.}(2017)\citenamefont
  {Kaladzhyan}, \citenamefont {Simon},\ and\ \citenamefont
  {Trif}}]{Vardan_PRB_2017}%
  \BibitemOpen
  \bibfield  {author} {\bibinfo {author} {\bibfnamefont {V.}~\bibnamefont
  {Kaladzhyan}}, \bibinfo {author} {\bibfnamefont {P.}~\bibnamefont {Simon}},\
  and\ \bibinfo {author} {\bibfnamefont {M.}~\bibnamefont {Trif}},\ }\bibfield
  {title} {\bibinfo {title} {Controlling topological superconductivity by
  magnetization dynamics},\ }\href {https://doi.org/10.1103/PhysRevB.96.020507}
  {\bibfield  {journal} {\bibinfo  {journal} {Phys. Rev. B}\ }\textbf {\bibinfo
  {volume} {96}},\ \bibinfo {pages} {020507} (\bibinfo {year}
  {2017})}\BibitemShut {NoStop}%
\bibitem [{\citenamefont {Braunecker}\ and\ \citenamefont
  {Simon}(2013)}]{braunecker_interplay_2013}%
  \BibitemOpen
  \bibfield  {author} {\bibinfo {author} {\bibfnamefont {B.}~\bibnamefont
  {Braunecker}}\ and\ \bibinfo {author} {\bibfnamefont {P.}~\bibnamefont
  {Simon}},\ }\bibfield  {title} {\bibinfo {title} {Interplay between classical
  magnetic moments and superconductivity in quantum one-dimensional conductors:
  Toward a self-sustained topological majorana phase},\ }\href
  {https://doi.org/10.1103/PhysRevLett.111.147202} {\bibfield  {journal}
  {\bibinfo  {journal} {Physical Review Letters}\ }\textbf {\bibinfo {volume}
  {111}},\ \bibinfo {pages} {147202} (\bibinfo {year} {2013})}\BibitemShut
  {NoStop}%
\bibitem [{\citenamefont {Vazifeh}\ and\ \citenamefont
  {Franz}(2013)}]{vazifeh_self-organized_2013}%
  \BibitemOpen
  \bibfield  {author} {\bibinfo {author} {\bibfnamefont {M.~M.}\ \bibnamefont
  {Vazifeh}}\ and\ \bibinfo {author} {\bibfnamefont {M.}~\bibnamefont
  {Franz}},\ }\bibfield  {title} {\bibinfo {title} {Self-organized topological
  state with majorana fermions},\ }\href
  {https://doi.org/10.1103/PhysRevLett.111.206802} {\bibfield  {journal}
  {\bibinfo  {journal} {Physical Review Letters}\ }\textbf {\bibinfo {volume}
  {111}},\ \bibinfo {pages} {206802} (\bibinfo {year} {2013})}\BibitemShut
  {NoStop}%
\bibitem [{\citenamefont {Klinovaja}\ \emph {et~al.}(2013)\citenamefont
  {Klinovaja}, \citenamefont {Stano}, \citenamefont {Yazdani},\ and\
  \citenamefont {Loss}}]{klinovaja_topological_2013}%
  \BibitemOpen
  \bibfield  {author} {\bibinfo {author} {\bibfnamefont {J.}~\bibnamefont
  {Klinovaja}}, \bibinfo {author} {\bibfnamefont {P.}~\bibnamefont {Stano}},
  \bibinfo {author} {\bibfnamefont {A.}~\bibnamefont {Yazdani}},\ and\ \bibinfo
  {author} {\bibfnamefont {D.}~\bibnamefont {Loss}},\ }\bibfield  {title}
  {\bibinfo {title} {Topological superconductivity and majorana fermions in
  rkky systems},\ }\href {https://doi.org/10.1103/PhysRevLett.111.186805}
  {\bibfield  {journal} {\bibinfo  {journal} {Physical Review Letters}\
  }\textbf {\bibinfo {volume} {111}},\ \bibinfo {pages} {186805} (\bibinfo
  {year} {2013})}\BibitemShut {NoStop}%
\bibitem [{\citenamefont {Das}\ \emph {et~al.}(2012)\citenamefont {Das},
  \citenamefont {Ronen}, \citenamefont {Most}, \citenamefont {Oreg},
  \citenamefont {Heiblum},\ and\ \citenamefont {Shtrikman}}]{das_zbp_2012}%
  \BibitemOpen
  \bibfield  {author} {\bibinfo {author} {\bibfnamefont {A.}~\bibnamefont
  {Das}}, \bibinfo {author} {\bibfnamefont {Y.}~\bibnamefont {Ronen}}, \bibinfo
  {author} {\bibfnamefont {Y.}~\bibnamefont {Most}}, \bibinfo {author}
  {\bibfnamefont {Y.}~\bibnamefont {Oreg}}, \bibinfo {author} {\bibfnamefont
  {M.}~\bibnamefont {Heiblum}},\ and\ \bibinfo {author} {\bibfnamefont
  {H.}~\bibnamefont {Shtrikman}},\ }\bibfield  {title} {\bibinfo {title}
  {Zero-bias peaks and splitting in an al--inas nanowire topological
  superconductor as a signature of majorana fermions},\ }\href
  {https://www.nature.com/articles/nphys2479} {\bibfield  {journal} {\bibinfo
  {journal} {Nature Physics}\ }\textbf {\bibinfo {volume} {8}},\ \bibinfo
  {pages} {887} (\bibinfo {year} {2012})}\BibitemShut {NoStop}%
\bibitem [{\citenamefont {Finck}\ \emph {et~al.}(2013)\citenamefont {Finck},
  \citenamefont {Van~Harlingen}, \citenamefont {Mohseni}, \citenamefont
  {Jung},\ and\ \citenamefont {Li}}]{finck_zbp_2013}%
  \BibitemOpen
  \bibfield  {author} {\bibinfo {author} {\bibfnamefont {A.~D.~K.}\
  \bibnamefont {Finck}}, \bibinfo {author} {\bibfnamefont {D.~J.}\ \bibnamefont
  {Van~Harlingen}}, \bibinfo {author} {\bibfnamefont {P.~K.}\ \bibnamefont
  {Mohseni}}, \bibinfo {author} {\bibfnamefont {K.}~\bibnamefont {Jung}},\ and\
  \bibinfo {author} {\bibfnamefont {X.}~\bibnamefont {Li}},\ }\bibfield
  {title} {\bibinfo {title} {Anomalous modulation of a zero-bias peak in a
  hybrid nanowire-superconductor device},\ }\href
  {https://doi.org/10.1103/PhysRevLett.110.126406} {\bibfield  {journal}
  {\bibinfo  {journal} {Phys. Rev. Lett.}\ }\textbf {\bibinfo {volume} {110}},\
  \bibinfo {pages} {126406} (\bibinfo {year} {2013})}\BibitemShut {NoStop}%
\bibitem [{\citenamefont {Mourik}\ \emph {et~al.}(2012)\citenamefont {Mourik},
  \citenamefont {Zuo}, \citenamefont {Frolov}, \citenamefont {Plissard},
  \citenamefont {Bakkers},\ and\ \citenamefont
  {Kouwenhoven}}]{mourik_nanowire_2012}%
  \BibitemOpen
  \bibfield  {author} {\bibinfo {author} {\bibfnamefont {V.}~\bibnamefont
  {Mourik}}, \bibinfo {author} {\bibfnamefont {K.}~\bibnamefont {Zuo}},
  \bibinfo {author} {\bibfnamefont {S.~M.}\ \bibnamefont {Frolov}}, \bibinfo
  {author} {\bibfnamefont {S.}~\bibnamefont {Plissard}}, \bibinfo {author}
  {\bibfnamefont {E.~P.}\ \bibnamefont {Bakkers}},\ and\ \bibinfo {author}
  {\bibfnamefont {L.~P.}\ \bibnamefont {Kouwenhoven}},\ }\bibfield  {title}
  {\bibinfo {title} {Signatures of majorana fermions in hybrid
  superconductor-semiconductor nanowire devices},\ }\href
  {https://www.science.org/doi/10.1126/science.1222360} {\bibfield  {journal}
  {\bibinfo  {journal} {Science}\ }\textbf {\bibinfo {volume} {336}},\ \bibinfo
  {pages} {1003} (\bibinfo {year} {2012})}\BibitemShut {NoStop}%
\bibitem [{\citenamefont {Nadj-Perge}\ \emph {et~al.}(2014)\citenamefont
  {Nadj-Perge}, \citenamefont {Drozdov}, \citenamefont {Li}, \citenamefont
  {Chen}, \citenamefont {Jeon}, \citenamefont {Seo}, \citenamefont {MacDonald},
  \citenamefont {Bernevig},\ and\ \citenamefont
  {Yazdani}}]{nadj-perge_observation_2014}%
  \BibitemOpen
  \bibfield  {author} {\bibinfo {author} {\bibfnamefont {S.}~\bibnamefont
  {Nadj-Perge}}, \bibinfo {author} {\bibfnamefont {I.~K.}\ \bibnamefont
  {Drozdov}}, \bibinfo {author} {\bibfnamefont {J.}~\bibnamefont {Li}},
  \bibinfo {author} {\bibfnamefont {H.}~\bibnamefont {Chen}}, \bibinfo {author}
  {\bibfnamefont {S.}~\bibnamefont {Jeon}}, \bibinfo {author} {\bibfnamefont
  {J.}~\bibnamefont {Seo}}, \bibinfo {author} {\bibfnamefont {A.~H.}\
  \bibnamefont {MacDonald}}, \bibinfo {author} {\bibfnamefont {B.~A.}\
  \bibnamefont {Bernevig}},\ and\ \bibinfo {author} {\bibfnamefont
  {A.}~\bibnamefont {Yazdani}},\ }\bibfield  {title} {\bibinfo {title}
  {Observation of {Majorana} fermions in ferromagnetic atomic chains on a
  superconductor},\ }\href {https://doi.org/10.1126/science.1259327} {\bibfield
   {journal} {\bibinfo  {journal} {Science}\ }\textbf {\bibinfo {volume}
  {346}},\ \bibinfo {pages} {602} (\bibinfo {year} {2014})}\BibitemShut
  {NoStop}%
\bibitem [{\citenamefont {Xu}\ \emph {et~al.}(2014)\citenamefont {Xu},
  \citenamefont {Liu}, \citenamefont {Wang}, \citenamefont {Ge}, \citenamefont
  {Liu}, \citenamefont {Yang}, \citenamefont {Chen}, \citenamefont {Liu},
  \citenamefont {Xu}, \citenamefont {Gao}, \citenamefont {Qian}, \citenamefont
  {Zhang},\ and\ \citenamefont {Jia}}]{Xu_ptsc_2014}%
  \BibitemOpen
  \bibfield  {author} {\bibinfo {author} {\bibfnamefont {J.-P.}\ \bibnamefont
  {Xu}}, \bibinfo {author} {\bibfnamefont {C.}~\bibnamefont {Liu}}, \bibinfo
  {author} {\bibfnamefont {M.-X.}\ \bibnamefont {Wang}}, \bibinfo {author}
  {\bibfnamefont {J.}~\bibnamefont {Ge}}, \bibinfo {author} {\bibfnamefont
  {Z.-L.}\ \bibnamefont {Liu}}, \bibinfo {author} {\bibfnamefont
  {X.}~\bibnamefont {Yang}}, \bibinfo {author} {\bibfnamefont {Y.}~\bibnamefont
  {Chen}}, \bibinfo {author} {\bibfnamefont {Y.}~\bibnamefont {Liu}}, \bibinfo
  {author} {\bibfnamefont {Z.-A.}\ \bibnamefont {Xu}}, \bibinfo {author}
  {\bibfnamefont {C.-L.}\ \bibnamefont {Gao}}, \bibinfo {author} {\bibfnamefont
  {D.}~\bibnamefont {Qian}}, \bibinfo {author} {\bibfnamefont {F.-C.}\
  \bibnamefont {Zhang}},\ and\ \bibinfo {author} {\bibfnamefont {J.-F.}\
  \bibnamefont {Jia}},\ }\bibfield  {title} {\bibinfo {title} {Artificial
  topological superconductor by the proximity effect},\ }\href
  {https://doi.org/10.1103/PhysRevLett.112.217001} {\bibfield  {journal}
  {\bibinfo  {journal} {Phys. Rev. Lett.}\ }\textbf {\bibinfo {volume} {112}},\
  \bibinfo {pages} {217001} (\bibinfo {year} {2014})}\BibitemShut {NoStop}%
\bibitem [{\citenamefont {Shoman}\ \emph {et~al.}(2015)\citenamefont {Shoman},
  \citenamefont {Takayama}, \citenamefont {Sato}, \citenamefont {Souma},
  \citenamefont {Takahashi}, \citenamefont {Oguchi}, \citenamefont {Segawa},\
  and\ \citenamefont {Ando}}]{shoman2015topological}%
  \BibitemOpen
  \bibfield  {author} {\bibinfo {author} {\bibfnamefont {T.}~\bibnamefont
  {Shoman}}, \bibinfo {author} {\bibfnamefont {A.}~\bibnamefont {Takayama}},
  \bibinfo {author} {\bibfnamefont {T.}~\bibnamefont {Sato}}, \bibinfo {author}
  {\bibfnamefont {S.}~\bibnamefont {Souma}}, \bibinfo {author} {\bibfnamefont
  {T.}~\bibnamefont {Takahashi}}, \bibinfo {author} {\bibfnamefont
  {T.}~\bibnamefont {Oguchi}}, \bibinfo {author} {\bibfnamefont
  {K.}~\bibnamefont {Segawa}},\ and\ \bibinfo {author} {\bibfnamefont
  {Y.}~\bibnamefont {Ando}},\ }\bibfield  {title} {\bibinfo {title}
  {Topological proximity effect in a topological insulator hybrid},\ }\href
  {https://www.nature.com/articles/ncomms7547} {\bibfield  {journal} {\bibinfo
  {journal} {Nature communications}\ }\textbf {\bibinfo {volume} {6}},\
  \bibinfo {pages} {6547} (\bibinfo {year} {2015})}\BibitemShut {NoStop}%
\bibitem [{\citenamefont {Ruby}\ \emph {et~al.}(2015)\citenamefont {Ruby},
  \citenamefont {Pientka}, \citenamefont {Peng}, \citenamefont {Von~Oppen},
  \citenamefont {Heinrich},\ and\ \citenamefont {Franke}}]{ruby_end_2015}%
  \BibitemOpen
  \bibfield  {author} {\bibinfo {author} {\bibfnamefont {M.}~\bibnamefont
  {Ruby}}, \bibinfo {author} {\bibfnamefont {F.}~\bibnamefont {Pientka}},
  \bibinfo {author} {\bibfnamefont {Y.}~\bibnamefont {Peng}}, \bibinfo {author}
  {\bibfnamefont {F.}~\bibnamefont {Von~Oppen}}, \bibinfo {author}
  {\bibfnamefont {B.~W.}\ \bibnamefont {Heinrich}},\ and\ \bibinfo {author}
  {\bibfnamefont {K.~J.}\ \bibnamefont {Franke}},\ }\bibfield  {title}
  {\bibinfo {title} {End states and subgap structure in proximity-coupled
  chains of magnetic adatoms},\ }\href
  {https://doi.org/10.1103/PhysRevLett.115.197204} {\bibfield  {journal}
  {\bibinfo  {journal} {Physical Review Letters}\ }\textbf {\bibinfo {volume}
  {115}},\ \bibinfo {pages} {197204} (\bibinfo {year} {2015})}\BibitemShut
  {NoStop}%
\bibitem [{\citenamefont {Ménard}\ \emph {et~al.}(2015)\citenamefont
  {Ménard}, \citenamefont {Guissart}, \citenamefont {Brun}, \citenamefont
  {Pons}, \citenamefont {Stolyarov}, \citenamefont {Debontridder},
  \citenamefont {Leclerc}, \citenamefont {Janod}, \citenamefont {Cario},
  \citenamefont {Roditchev}, \citenamefont {Simon},\ and\ \citenamefont
  {Cren}}]{menard_coherent_2015}%
  \BibitemOpen
  \bibfield  {author} {\bibinfo {author} {\bibfnamefont {G.~C.}\ \bibnamefont
  {Ménard}}, \bibinfo {author} {\bibfnamefont {S.}~\bibnamefont {Guissart}},
  \bibinfo {author} {\bibfnamefont {C.}~\bibnamefont {Brun}}, \bibinfo {author}
  {\bibfnamefont {S.}~\bibnamefont {Pons}}, \bibinfo {author} {\bibfnamefont
  {V.~S.}\ \bibnamefont {Stolyarov}}, \bibinfo {author} {\bibfnamefont
  {F.}~\bibnamefont {Debontridder}}, \bibinfo {author} {\bibfnamefont {M.~V.}\
  \bibnamefont {Leclerc}}, \bibinfo {author} {\bibfnamefont {E.}~\bibnamefont
  {Janod}}, \bibinfo {author} {\bibfnamefont {L.}~\bibnamefont {Cario}},
  \bibinfo {author} {\bibfnamefont {D.}~\bibnamefont {Roditchev}}, \bibinfo
  {author} {\bibfnamefont {P.}~\bibnamefont {Simon}},\ and\ \bibinfo {author}
  {\bibfnamefont {T.}~\bibnamefont {Cren}},\ }\bibfield  {title} {\bibinfo
  {title} {Coherent long-range magnetic bound states in a superconductor},\
  }\href {https://doi.org/10.1038/nphys3508} {\bibfield  {journal} {\bibinfo
  {journal} {Nature Physics}\ }\textbf {\bibinfo {volume} {11}},\ \bibinfo
  {pages} {1013} (\bibinfo {year} {2015})}\BibitemShut {NoStop}%
\bibitem [{\citenamefont {Pawlak}\ \emph {et~al.}(2016)\citenamefont {Pawlak},
  \citenamefont {Kisiel}, \citenamefont {Klinovaja}, \citenamefont {Meier},
  \citenamefont {Kawai}, \citenamefont {Glatzel}, \citenamefont {Loss},\ and\
  \citenamefont {Meyer}}]{pawlak_probing_2016}%
  \BibitemOpen
  \bibfield  {author} {\bibinfo {author} {\bibfnamefont {R.}~\bibnamefont
  {Pawlak}}, \bibinfo {author} {\bibfnamefont {M.}~\bibnamefont {Kisiel}},
  \bibinfo {author} {\bibfnamefont {J.}~\bibnamefont {Klinovaja}}, \bibinfo
  {author} {\bibfnamefont {T.}~\bibnamefont {Meier}}, \bibinfo {author}
  {\bibfnamefont {S.}~\bibnamefont {Kawai}}, \bibinfo {author} {\bibfnamefont
  {T.}~\bibnamefont {Glatzel}}, \bibinfo {author} {\bibfnamefont
  {D.}~\bibnamefont {Loss}},\ and\ \bibinfo {author} {\bibfnamefont
  {E.}~\bibnamefont {Meyer}},\ }\bibfield  {title} {\bibinfo {title} {Probing
  atomic structure and {Majorana} wavefunctions in mono-atomic {Fe} chains on
  superconducting {Pb} surface},\ }\href
  {https://doi.org/10.1038/npjqi.2016.35} {\bibfield  {journal} {\bibinfo
  {journal} {npj Quantum Information}\ }\textbf {\bibinfo {volume} {2}},\
  \bibinfo {pages} {1} (\bibinfo {year} {2016})}\BibitemShut {NoStop}%
\bibitem [{\citenamefont {Feldman}\ \emph {et~al.}(2017)\citenamefont
  {Feldman}, \citenamefont {Randeria}, \citenamefont {Li}, \citenamefont
  {Jeon}, \citenamefont {Xie}, \citenamefont {Wang}, \citenamefont {Drozdov},
  \citenamefont {Andrei~Bernevig},\ and\ \citenamefont
  {Yazdani}}]{feldman_high-resolution_2017}%
  \BibitemOpen
  \bibfield  {author} {\bibinfo {author} {\bibfnamefont {B.~E.}\ \bibnamefont
  {Feldman}}, \bibinfo {author} {\bibfnamefont {M.~T.}\ \bibnamefont
  {Randeria}}, \bibinfo {author} {\bibfnamefont {J.}~\bibnamefont {Li}},
  \bibinfo {author} {\bibfnamefont {S.}~\bibnamefont {Jeon}}, \bibinfo {author}
  {\bibfnamefont {Y.}~\bibnamefont {Xie}}, \bibinfo {author} {\bibfnamefont
  {Z.}~\bibnamefont {Wang}}, \bibinfo {author} {\bibfnamefont {I.~K.}\
  \bibnamefont {Drozdov}}, \bibinfo {author} {\bibfnamefont {B.}~\bibnamefont
  {Andrei~Bernevig}},\ and\ \bibinfo {author} {\bibfnamefont {A.}~\bibnamefont
  {Yazdani}},\ }\bibfield  {title} {\bibinfo {title} {High-resolution studies
  of the {Majorana} atomic chain platform},\ }\href
  {https://doi.org/10.1038/nphys3947} {\bibfield  {journal} {\bibinfo
  {journal} {Nature Physics}\ }\textbf {\bibinfo {volume} {13}},\ \bibinfo
  {pages} {286} (\bibinfo {year} {2017})}\BibitemShut {NoStop}%
\bibitem [{\citenamefont {Ruby}\ \emph {et~al.}(2017)\citenamefont {Ruby},
  \citenamefont {Heinrich}, \citenamefont {Peng}, \citenamefont {von Oppen},\
  and\ \citenamefont {Franke}}]{ruby_exploring_2017}%
  \BibitemOpen
  \bibfield  {author} {\bibinfo {author} {\bibfnamefont {M.}~\bibnamefont
  {Ruby}}, \bibinfo {author} {\bibfnamefont {B.~W.}\ \bibnamefont {Heinrich}},
  \bibinfo {author} {\bibfnamefont {Y.}~\bibnamefont {Peng}}, \bibinfo {author}
  {\bibfnamefont {F.}~\bibnamefont {von Oppen}},\ and\ \bibinfo {author}
  {\bibfnamefont {K.~J.}\ \bibnamefont {Franke}},\ }\bibfield  {title}
  {\bibinfo {title} {Exploring a proximity-coupled {Co} chain on {Pb}(110) as a
  possible {M}ajorana platform},\ }\href
  {https://doi.org/10.1021/acs.nanolett.7b01728} {\bibfield  {journal}
  {\bibinfo  {journal} {Nano Letters}\ }\textbf {\bibinfo {volume} {17}},\
  \bibinfo {pages} {4473} (\bibinfo {year} {2017})}\BibitemShut {NoStop}%
\bibitem [{\citenamefont {Wang}\ \emph {et~al.}(2020)\citenamefont {Wang},
  \citenamefont {Rodriguez}, \citenamefont {Jiao}, \citenamefont {Howard},
  \citenamefont {Graham}, \citenamefont {Gu}, \citenamefont {Hughes},
  \citenamefont {Morr},\ and\ \citenamefont {Madhavan}}]{wang_1dmajorana_2020}%
  \BibitemOpen
  \bibfield  {author} {\bibinfo {author} {\bibfnamefont {Z.}~\bibnamefont
  {Wang}}, \bibinfo {author} {\bibfnamefont {J.~O.}\ \bibnamefont {Rodriguez}},
  \bibinfo {author} {\bibfnamefont {L.}~\bibnamefont {Jiao}}, \bibinfo {author}
  {\bibfnamefont {S.}~\bibnamefont {Howard}}, \bibinfo {author} {\bibfnamefont
  {M.}~\bibnamefont {Graham}}, \bibinfo {author} {\bibfnamefont {G.~D.}\
  \bibnamefont {Gu}}, \bibinfo {author} {\bibfnamefont {T.~L.}\ \bibnamefont
  {Hughes}}, \bibinfo {author} {\bibfnamefont {D.~K.}\ \bibnamefont {Morr}},\
  and\ \bibinfo {author} {\bibfnamefont {V.}~\bibnamefont {Madhavan}},\
  }\bibfield  {title} {\bibinfo {title} {Evidence for dispersing 1d majorana
  channels in an iron-based superconductor},\ }\href
  {https://doi.org/10.1126/science.aaw8419} {\bibfield  {journal} {\bibinfo
  {journal} {Science}\ }\textbf {\bibinfo {volume} {367}},\ \bibinfo {pages}
  {104} (\bibinfo {year} {2020})}\BibitemShut {NoStop}%
\bibitem [{\citenamefont {Schneider}\ \emph {et~al.}(2022)\citenamefont
  {Schneider}, \citenamefont {Beck}, \citenamefont {Neuhaus-Steinmetz},
  \citenamefont {Rózsa}, \citenamefont {Posske}, \citenamefont {Wiebe},\ and\
  \citenamefont {Wiesendanger}}]{schneider_precursors_2022}%
  \BibitemOpen
  \bibfield  {author} {\bibinfo {author} {\bibfnamefont {L.}~\bibnamefont
  {Schneider}}, \bibinfo {author} {\bibfnamefont {P.}~\bibnamefont {Beck}},
  \bibinfo {author} {\bibfnamefont {J.}~\bibnamefont {Neuhaus-Steinmetz}},
  \bibinfo {author} {\bibfnamefont {L.}~\bibnamefont {Rózsa}}, \bibinfo
  {author} {\bibfnamefont {T.}~\bibnamefont {Posske}}, \bibinfo {author}
  {\bibfnamefont {J.}~\bibnamefont {Wiebe}},\ and\ \bibinfo {author}
  {\bibfnamefont {R.}~\bibnamefont {Wiesendanger}},\ }\bibfield  {title}
  {\bibinfo {title} {Precursors of majorana modes and their length-dependent
  energy oscillations probed at both ends of atomic shiba chains},\ }\href
  {https://doi.org/10.1038/s41565-022-01078-4} {\bibfield  {journal} {\bibinfo
  {journal} {Nature Nanotechnology}\ }\textbf {\bibinfo {volume} {17}},\
  \bibinfo {pages} {384} (\bibinfo {year} {2022})}\BibitemShut {NoStop}%
\bibitem [{\citenamefont {Liebhaber}\ \emph {et~al.}(2022)\citenamefont
  {Liebhaber}, \citenamefont {Rütten}, \citenamefont {Reecht}, \citenamefont
  {Steiner}, \citenamefont {Rohlf}, \citenamefont {Rossnagel}, \citenamefont
  {von Oppen},\ and\ \citenamefont {Franke}}]{liebhaber_quantum_2022}%
  \BibitemOpen
  \bibfield  {author} {\bibinfo {author} {\bibfnamefont {E.}~\bibnamefont
  {Liebhaber}}, \bibinfo {author} {\bibfnamefont {L.~M.}\ \bibnamefont
  {Rütten}}, \bibinfo {author} {\bibfnamefont {G.}~\bibnamefont {Reecht}},
  \bibinfo {author} {\bibfnamefont {J.~F.}\ \bibnamefont {Steiner}}, \bibinfo
  {author} {\bibfnamefont {S.}~\bibnamefont {Rohlf}}, \bibinfo {author}
  {\bibfnamefont {K.}~\bibnamefont {Rossnagel}}, \bibinfo {author}
  {\bibfnamefont {F.}~\bibnamefont {von Oppen}},\ and\ \bibinfo {author}
  {\bibfnamefont {K.~J.}\ \bibnamefont {Franke}},\ }\bibfield  {title}
  {\bibinfo {title} {Quantum spins and hybridization in
  artificially-constructed chains of magnetic adatoms on a superconductor},\
  }\href {https://doi.org/10.1038/s41467-022-29879-0} {\bibfield  {journal}
  {\bibinfo  {journal} {Nature Communications}\ }\textbf {\bibinfo {volume}
  {13}},\ \bibinfo {pages} {2160} (\bibinfo {year} {2022})}\BibitemShut
  {NoStop}%
\bibitem [{\citenamefont {Mesaros}\ \emph {et~al.}(2024)\citenamefont
  {Mesaros}, \citenamefont {Gu},\ and\ \citenamefont
  {Massee}}]{mesaros2024topologically}%
  \BibitemOpen
  \bibfield  {author} {\bibinfo {author} {\bibfnamefont {A.}~\bibnamefont
  {Mesaros}}, \bibinfo {author} {\bibfnamefont {G.}~\bibnamefont {Gu}},\ and\
  \bibinfo {author} {\bibfnamefont {F.}~\bibnamefont {Massee}},\ }\bibfield
  {title} {\bibinfo {title} {Topologically trivial gap-filling in
  superconducting fe (se, te) by one-dimensional defects},\ }\href
  {https://www.nature.com/articles/s41467-024-48047-0} {\bibfield  {journal}
  {\bibinfo  {journal} {Nature Communications}\ }\textbf {\bibinfo {volume}
  {15}},\ \bibinfo {pages} {3774} (\bibinfo {year} {2024})}\BibitemShut
  {NoStop}%
\bibitem [{\citenamefont {Sau}\ and\ \citenamefont
  {Demler}(2013)}]{Sau_nanow_2013}%
  \BibitemOpen
  \bibfield  {author} {\bibinfo {author} {\bibfnamefont {J.~D.}\ \bibnamefont
  {Sau}}\ and\ \bibinfo {author} {\bibfnamefont {E.}~\bibnamefont {Demler}},\
  }\bibfield  {title} {\bibinfo {title} {Bound states at impurities as a probe
  of topological superconductivity in nanowires},\ }\href
  {https://doi.org/10.1103/PhysRevB.88.205402} {\bibfield  {journal} {\bibinfo
  {journal} {Phys. Rev. B}\ }\textbf {\bibinfo {volume} {88}},\ \bibinfo
  {pages} {205402} (\bibinfo {year} {2013})}\BibitemShut {NoStop}%
\bibitem [{\citenamefont {Ma\ifmmode~\acute{s}\else \'{s}\fi{}ka}\ \emph
  {et~al.}(2017)\citenamefont {Ma\ifmmode~\acute{s}\else \'{s}\fi{}ka},
  \citenamefont {Gorczyca-Goraj}, \citenamefont {Tworzyd\l{}o},\ and\
  \citenamefont {Doma\ifmmode~\acute{n}\else
  \'{n}\fi{}ski}}]{Maska_rashba_2017}%
  \BibitemOpen
  \bibfield  {author} {\bibinfo {author} {\bibfnamefont {M.~M.}\ \bibnamefont
  {Ma\ifmmode~\acute{s}\else \'{s}\fi{}ka}}, \bibinfo {author} {\bibfnamefont
  {A.}~\bibnamefont {Gorczyca-Goraj}}, \bibinfo {author} {\bibfnamefont
  {J.}~\bibnamefont {Tworzyd\l{}o}},\ and\ \bibinfo {author} {\bibfnamefont
  {T.}~\bibnamefont {Doma\ifmmode~\acute{n}\else \'{n}\fi{}ski}},\ }\bibfield
  {title} {\bibinfo {title} {Majorana quasiparticles of an inhomogeneous rashba
  chain},\ }\href {https://doi.org/10.1103/PhysRevB.95.045429} {\bibfield
  {journal} {\bibinfo  {journal} {Phys. Rev. B}\ }\textbf {\bibinfo {volume}
  {95}},\ \bibinfo {pages} {045429} (\bibinfo {year} {2017})}\BibitemShut
  {NoStop}%
\bibitem [{\citenamefont {Legg}\ \emph {et~al.}(2021)\citenamefont {Legg},
  \citenamefont {Loss},\ and\ \citenamefont {Klinovaja}}]{Legg_mzm_2021}%
  \BibitemOpen
  \bibfield  {author} {\bibinfo {author} {\bibfnamefont {H.~F.}\ \bibnamefont
  {Legg}}, \bibinfo {author} {\bibfnamefont {D.}~\bibnamefont {Loss}},\ and\
  \bibinfo {author} {\bibfnamefont {J.}~\bibnamefont {Klinovaja}},\ }\bibfield
  {title} {\bibinfo {title} {Majorana bound states in topological insulators
  without a vortex},\ }\href {https://doi.org/10.1103/PhysRevB.104.165405}
  {\bibfield  {journal} {\bibinfo  {journal} {Phys. Rev. B}\ }\textbf {\bibinfo
  {volume} {104}},\ \bibinfo {pages} {165405} (\bibinfo {year}
  {2021})}\BibitemShut {NoStop}%
\bibitem [{\citenamefont {Sedlmayr}\ and\ \citenamefont
  {Bena}(2021)}]{Sedlmayr_JPCM_2022}%
  \BibitemOpen
  \bibfield  {author} {\bibinfo {author} {\bibfnamefont {N.}~\bibnamefont
  {Sedlmayr}}\ and\ \bibinfo {author} {\bibfnamefont {C.}~\bibnamefont
  {Bena}},\ }\bibfield  {title} {\bibinfo {title} {Instability of majorana
  states in shiba chains due to leakage into a topological substrate},\ }\href
  {https://doi.org/10.1088/1361-648X/ac413f} {\bibfield  {journal} {\bibinfo
  {journal} {Journal of Physics: Condensed Matter}\ }\textbf {\bibinfo {volume}
  {34}},\ \bibinfo {pages} {104004} (\bibinfo {year} {2021})}\BibitemShut
  {NoStop}%
\bibitem [{\citenamefont {Su}\ \emph {et~al.}(1980)\citenamefont {Su},
  \citenamefont {Schrieffer},\ and\ \citenamefont {Heeger}}]{SSH_PRB_1980}%
  \BibitemOpen
  \bibfield  {author} {\bibinfo {author} {\bibfnamefont {W.~P.}\ \bibnamefont
  {Su}}, \bibinfo {author} {\bibfnamefont {J.~R.}\ \bibnamefont {Schrieffer}},\
  and\ \bibinfo {author} {\bibfnamefont {A.~J.}\ \bibnamefont {Heeger}},\
  }\bibfield  {title} {\bibinfo {title} {Soliton excitations in
  polyacetylene},\ }\href {https://doi.org/10.1103/PhysRevB.22.2099} {\bibfield
   {journal} {\bibinfo  {journal} {Phys. Rev. B}\ }\textbf {\bibinfo {volume}
  {22}},\ \bibinfo {pages} {2099} (\bibinfo {year} {1980})}\BibitemShut
  {NoStop}%
\bibitem [{\citenamefont {Bernevig}\ \emph {et~al.}(2006)\citenamefont
  {Bernevig}, \citenamefont {Hughes},\ and\ \citenamefont
  {Zhang}}]{BHZ_Sci_2006}%
  \BibitemOpen
  \bibfield  {author} {\bibinfo {author} {\bibfnamefont {B.~A.}\ \bibnamefont
  {Bernevig}}, \bibinfo {author} {\bibfnamefont {T.~L.}\ \bibnamefont
  {Hughes}},\ and\ \bibinfo {author} {\bibfnamefont {S.-C.}\ \bibnamefont
  {Zhang}},\ }\bibfield  {title} {\bibinfo {title} {Quantum spin hall effect
  and topological phase transition in $\text{HgTe}$ quantum wells},\ }\href
  {https://www.science.org/doi/full/10.1126/science.1133734} {\bibfield
  {journal} {\bibinfo  {journal} {science}\ }\textbf {\bibinfo {volume}
  {314}},\ \bibinfo {pages} {1757} (\bibinfo {year} {2006})}\BibitemShut
  {NoStop}%
\bibitem [{\citenamefont {Qi}\ \emph {et~al.}(2006)\citenamefont {Qi},
  \citenamefont {Wu},\ and\ \citenamefont {Zhang}}]{QWZ_spinlessBHZ_2006}%
  \BibitemOpen
  \bibfield  {author} {\bibinfo {author} {\bibfnamefont {X.-L.}\ \bibnamefont
  {Qi}}, \bibinfo {author} {\bibfnamefont {Y.-S.}\ \bibnamefont {Wu}},\ and\
  \bibinfo {author} {\bibfnamefont {S.-C.}\ \bibnamefont {Zhang}},\ }\bibfield
  {title} {\bibinfo {title} {Topological quantization of the spin hall effect
  in two-dimensional paramagnetic semiconductors},\ }\href
  {https://doi.org/10.1103/PhysRevB.74.085308} {\bibfield  {journal} {\bibinfo
  {journal} {Phys. Rev. B}\ }\textbf {\bibinfo {volume} {74}},\ \bibinfo
  {pages} {085308} (\bibinfo {year} {2006})}\BibitemShut {NoStop}%
\bibitem [{\citenamefont {He}\ \emph {et~al.}(2013)\citenamefont {He},
  \citenamefont {Wang},\ and\ \citenamefont {Xue}}]{He_QAH_2013}%
  \BibitemOpen
  \bibfield  {author} {\bibinfo {author} {\bibfnamefont {K.}~\bibnamefont
  {He}}, \bibinfo {author} {\bibfnamefont {Y.}~\bibnamefont {Wang}},\ and\
  \bibinfo {author} {\bibfnamefont {Q.-K.}\ \bibnamefont {Xue}},\ }\bibfield
  {title} {\bibinfo {title} {Quantum anomalous hall effect},\ }\href
  {https://doi.org/10.1093/nsr/nwt029} {\bibfield  {journal} {\bibinfo
  {journal} {National Science Review}\ }\textbf {\bibinfo {volume} {1}},\
  \bibinfo {pages} {38} (\bibinfo {year} {2013})}\BibitemShut {NoStop}%
\bibitem [{\citenamefont {Liu}\ \emph {et~al.}(2016)\citenamefont {Liu},
  \citenamefont {Zhang},\ and\ \citenamefont {Qi}}]{Liu_QAH_2016}%
  \BibitemOpen
  \bibfield  {author} {\bibinfo {author} {\bibfnamefont {C.-X.}\ \bibnamefont
  {Liu}}, \bibinfo {author} {\bibfnamefont {S.-C.}\ \bibnamefont {Zhang}},\
  and\ \bibinfo {author} {\bibfnamefont {X.-L.}\ \bibnamefont {Qi}},\
  }\bibfield  {title} {\bibinfo {title} {The quantum anomalous hall effect:
  Theory and experiment},\ }\href
  {https://doi.org/https://doi.org/10.1146/annurev-conmatphys-031115-011417}
  {\bibfield  {journal} {\bibinfo  {journal} {Annual Review of Condensed Matter
  Physics}\ }\textbf {\bibinfo {volume} {7}},\ \bibinfo {pages} {301} (\bibinfo
  {year} {2016})}\BibitemShut {NoStop}%
\bibitem [{\citenamefont {Yin}\ \emph {et~al.}(2021)\citenamefont {Yin},
  \citenamefont {Pan},\ and\ \citenamefont {Zahid~Hasan}}]{yin_stminti_2021}%
  \BibitemOpen
  \bibfield  {author} {\bibinfo {author} {\bibfnamefont {J.-X.}\ \bibnamefont
  {Yin}}, \bibinfo {author} {\bibfnamefont {S.~H.}\ \bibnamefont {Pan}},\ and\
  \bibinfo {author} {\bibfnamefont {M.}~\bibnamefont {Zahid~Hasan}},\
  }\bibfield  {title} {\bibinfo {title} {Probing topological quantum matter
  with scanning tunnelling microscopy},\ }\href
  {https://www.nature.com/articles/s42254-021-00293-7} {\bibfield  {journal}
  {\bibinfo  {journal} {Nature Reviews Physics}\ }\textbf {\bibinfo {volume}
  {3}},\ \bibinfo {pages} {249} (\bibinfo {year} {2021})}\BibitemShut {NoStop}%
\bibitem [{\citenamefont {Hsieh}\ \emph {et~al.}(2016)\citenamefont {Hsieh},
  \citenamefont {Ishizuka}, \citenamefont {Balents},\ and\ \citenamefont
  {Hughes}}]{Hsieh_bpte_2016}%
  \BibitemOpen
  \bibfield  {author} {\bibinfo {author} {\bibfnamefont {T.~H.}\ \bibnamefont
  {Hsieh}}, \bibinfo {author} {\bibfnamefont {H.}~\bibnamefont {Ishizuka}},
  \bibinfo {author} {\bibfnamefont {L.}~\bibnamefont {Balents}},\ and\ \bibinfo
  {author} {\bibfnamefont {T.~L.}\ \bibnamefont {Hughes}},\ }\bibfield  {title}
  {\bibinfo {title} {Bulk topological proximity effect},\ }\href
  {https://doi.org/10.1103/PhysRevLett.116.086802} {\bibfield  {journal}
  {\bibinfo  {journal} {Phys. Rev. Lett.}\ }\textbf {\bibinfo {volume} {116}},\
  \bibinfo {pages} {086802} (\bibinfo {year} {2016})}\BibitemShut {NoStop}%
\bibitem [{\citenamefont {Cheng}\ \emph {et~al.}(2019)\citenamefont {Cheng},
  \citenamefont {Klein}, \citenamefont {Plekhanov}, \citenamefont {Sengstock},
  \citenamefont {Aidelsburger}, \citenamefont {Weitenberg},\ and\ \citenamefont
  {Le~Hur}}]{Cheng_proximity_2019}%
  \BibitemOpen
  \bibfield  {author} {\bibinfo {author} {\bibfnamefont {P.}~\bibnamefont
  {Cheng}}, \bibinfo {author} {\bibfnamefont {P.~W.}\ \bibnamefont {Klein}},
  \bibinfo {author} {\bibfnamefont {K.}~\bibnamefont {Plekhanov}}, \bibinfo
  {author} {\bibfnamefont {K.}~\bibnamefont {Sengstock}}, \bibinfo {author}
  {\bibfnamefont {M.}~\bibnamefont {Aidelsburger}}, \bibinfo {author}
  {\bibfnamefont {C.}~\bibnamefont {Weitenberg}},\ and\ \bibinfo {author}
  {\bibfnamefont {K.}~\bibnamefont {Le~Hur}},\ }\bibfield  {title} {\bibinfo
  {title} {Topological proximity effects in a haldane graphene bilayer
  system},\ }\href {https://doi.org/10.1103/PhysRevB.100.081107} {\bibfield
  {journal} {\bibinfo  {journal} {Phys. Rev. B}\ }\textbf {\bibinfo {volume}
  {100}},\ \bibinfo {pages} {081107} (\bibinfo {year} {2019})}\BibitemShut
  {NoStop}%
\bibitem [{\citenamefont {Khanna}\ \emph {et~al.}(2014)\citenamefont {Khanna},
  \citenamefont {Kundu}, \citenamefont {Pradhan},\ and\ \citenamefont
  {Rao}}]{khanna_pisc_2014}%
  \BibitemOpen
  \bibfield  {author} {\bibinfo {author} {\bibfnamefont {U.}~\bibnamefont
  {Khanna}}, \bibinfo {author} {\bibfnamefont {A.}~\bibnamefont {Kundu}},
  \bibinfo {author} {\bibfnamefont {S.}~\bibnamefont {Pradhan}},\ and\ \bibinfo
  {author} {\bibfnamefont {S.}~\bibnamefont {Rao}},\ }\bibfield  {title}
  {\bibinfo {title} {Proximity-induced superconductivity in weyl semimetals},\
  }\href {https://doi.org/10.1103/PhysRevB.90.195430} {\bibfield  {journal}
  {\bibinfo  {journal} {Phys. Rev. B}\ }\textbf {\bibinfo {volume} {90}},\
  \bibinfo {pages} {195430} (\bibinfo {year} {2014})}\BibitemShut {NoStop}%
\bibitem [{\citenamefont {Panas}\ \emph {et~al.}(2020)\citenamefont {Panas},
  \citenamefont {Irsigler}, \citenamefont {Zheng},\ and\ \citenamefont
  {Hofstetter}}]{panas_proximity_2020}%
  \BibitemOpen
  \bibfield  {author} {\bibinfo {author} {\bibfnamefont {J.}~\bibnamefont
  {Panas}}, \bibinfo {author} {\bibfnamefont {B.}~\bibnamefont {Irsigler}},
  \bibinfo {author} {\bibfnamefont {J.-H.}\ \bibnamefont {Zheng}},\ and\
  \bibinfo {author} {\bibfnamefont {W.}~\bibnamefont {Hofstetter}},\ }\bibfield
   {title} {\bibinfo {title} {Bulk topological proximity effect in multilayer
  systems},\ }\href {https://doi.org/10.1103/PhysRevB.102.075403} {\bibfield
  {journal} {\bibinfo  {journal} {Phys. Rev. B}\ }\textbf {\bibinfo {volume}
  {102}},\ \bibinfo {pages} {075403} (\bibinfo {year} {2020})}\BibitemShut
  {NoStop}%
\bibitem [{\citenamefont {Atanov}\ \emph {et~al.}(2024)\citenamefont {Atanov},
  \citenamefont {Tai}, \citenamefont {Xie}, \citenamefont {Ng}, \citenamefont
  {Hammond}, \citenamefont {{Manfred Ho}}, \citenamefont {Koo}, \citenamefont
  {Li}, \citenamefont {Ho}, \citenamefont {Lyu}, \citenamefont {Chong},
  \citenamefont {Zhang}, \citenamefont {Tai}, \citenamefont {Wang},
  \citenamefont {Law}, \citenamefont {Wang},\ and\ \citenamefont
  {Lortz}}]{CIM_proximity}%
  \BibitemOpen
  \bibfield  {author} {\bibinfo {author} {\bibfnamefont {O.}~\bibnamefont
  {Atanov}}, \bibinfo {author} {\bibfnamefont {W.~T.}\ \bibnamefont {Tai}},
  \bibinfo {author} {\bibfnamefont {Y.}~\bibnamefont {Xie}}, \bibinfo {author}
  {\bibfnamefont {Y.~H.}\ \bibnamefont {Ng}}, \bibinfo {author} {\bibfnamefont
  {M.~A.}\ \bibnamefont {Hammond}}, \bibinfo {author} {\bibfnamefont {T.~S.}\
  \bibnamefont {{Manfred Ho}}}, \bibinfo {author} {\bibfnamefont {T.~H.}\
  \bibnamefont {Koo}}, \bibinfo {author} {\bibfnamefont {H.}~\bibnamefont
  {Li}}, \bibinfo {author} {\bibfnamefont {S.~L.}\ \bibnamefont {Ho}}, \bibinfo
  {author} {\bibfnamefont {J.}~\bibnamefont {Lyu}}, \bibinfo {author}
  {\bibfnamefont {S.}~\bibnamefont {Chong}}, \bibinfo {author} {\bibfnamefont
  {P.}~\bibnamefont {Zhang}}, \bibinfo {author} {\bibfnamefont
  {L.}~\bibnamefont {Tai}}, \bibinfo {author} {\bibfnamefont {J.}~\bibnamefont
  {Wang}}, \bibinfo {author} {\bibfnamefont {K.~T.}\ \bibnamefont {Law}},
  \bibinfo {author} {\bibfnamefont {K.~L.}\ \bibnamefont {Wang}},\ and\
  \bibinfo {author} {\bibfnamefont {R.}~\bibnamefont {Lortz}},\ }\bibfield
  {title} {\bibinfo {title} {Proximity-induced quasi-one-dimensional
  superconducting quantum anomalous hall state},\ }\href
  {https://doi.org/https://doi.org/10.1016/j.xcrp.2023.101762} {\bibfield
  {journal} {\bibinfo  {journal} {Cell Reports Physical Science}\ }\textbf
  {\bibinfo {volume} {5}},\ \bibinfo {pages} {101762} (\bibinfo {year}
  {2024})}\BibitemShut {NoStop}%
\bibitem [{\citenamefont {Liang}\ \emph {et~al.}(2023)\citenamefont {Liang},
  \citenamefont {Wei}, \citenamefont {Zhang}, \citenamefont {Wang},
  \citenamefont {Zhang}, \citenamefont {Wang}, \citenamefont {Qi},
  \citenamefont {Liu},\ and\ \citenamefont
  {Zhang}}]{CIM_ultracoldfermion_2023}%
  \BibitemOpen
  \bibfield  {author} {\bibinfo {author} {\bibfnamefont {M.-C.}\ \bibnamefont
  {Liang}}, \bibinfo {author} {\bibfnamefont {Y.-D.}\ \bibnamefont {Wei}},
  \bibinfo {author} {\bibfnamefont {L.}~\bibnamefont {Zhang}}, \bibinfo
  {author} {\bibfnamefont {X.-J.}\ \bibnamefont {Wang}}, \bibinfo {author}
  {\bibfnamefont {H.}~\bibnamefont {Zhang}}, \bibinfo {author} {\bibfnamefont
  {W.-W.}\ \bibnamefont {Wang}}, \bibinfo {author} {\bibfnamefont
  {W.}~\bibnamefont {Qi}}, \bibinfo {author} {\bibfnamefont {X.-J.}\
  \bibnamefont {Liu}},\ and\ \bibinfo {author} {\bibfnamefont {X.}~\bibnamefont
  {Zhang}},\ }\bibfield  {title} {\bibinfo {title} {Realization of qi-wu-zhang
  model in spin-orbit-coupled ultracold fermions},\ }\href
  {https://doi.org/10.1103/PhysRevResearch.5.L012006} {\bibfield  {journal}
  {\bibinfo  {journal} {Phys. Rev. Res.}\ }\textbf {\bibinfo {volume} {5}},\
  \bibinfo {pages} {L012006} (\bibinfo {year} {2023})}\BibitemShut {NoStop}%
\bibitem [{\citenamefont {Wang}\ \emph {et~al.}(2018)\citenamefont {Wang},
  \citenamefont {Lu}, \citenamefont {Sun}, \citenamefont {Chen}, \citenamefont
  {Deng},\ and\ \citenamefont {Liu}}]{Wang_ultracold_2018}%
  \BibitemOpen
  \bibfield  {author} {\bibinfo {author} {\bibfnamefont {B.-Z.}\ \bibnamefont
  {Wang}}, \bibinfo {author} {\bibfnamefont {Y.-H.}\ \bibnamefont {Lu}},
  \bibinfo {author} {\bibfnamefont {W.}~\bibnamefont {Sun}}, \bibinfo {author}
  {\bibfnamefont {S.}~\bibnamefont {Chen}}, \bibinfo {author} {\bibfnamefont
  {Y.}~\bibnamefont {Deng}},\ and\ \bibinfo {author} {\bibfnamefont {X.-J.}\
  \bibnamefont {Liu}},\ }\bibfield  {title} {\bibinfo {title} {Dirac-, rashba-,
  and weyl-type spin-orbit couplings: Toward experimental realization in
  ultracold atoms},\ }\href {https://doi.org/10.1103/PhysRevA.97.011605}
  {\bibfield  {journal} {\bibinfo  {journal} {Phys. Rev. A}\ }\textbf {\bibinfo
  {volume} {97}},\ \bibinfo {pages} {011605} (\bibinfo {year}
  {2018})}\BibitemShut {NoStop}%
\bibitem [{\citenamefont {Sun}\ \emph {et~al.}(2018)\citenamefont {Sun},
  \citenamefont {Wang}, \citenamefont {Xu}, \citenamefont {Yi}, \citenamefont
  {Zhang}, \citenamefont {Wu}, \citenamefont {Deng}, \citenamefont {Liu},
  \citenamefont {Chen},\ and\ \citenamefont {Pan}}]{Sun_ultracold_2018}%
  \BibitemOpen
  \bibfield  {author} {\bibinfo {author} {\bibfnamefont {W.}~\bibnamefont
  {Sun}}, \bibinfo {author} {\bibfnamefont {B.-Z.}\ \bibnamefont {Wang}},
  \bibinfo {author} {\bibfnamefont {X.-T.}\ \bibnamefont {Xu}}, \bibinfo
  {author} {\bibfnamefont {C.-R.}\ \bibnamefont {Yi}}, \bibinfo {author}
  {\bibfnamefont {L.}~\bibnamefont {Zhang}}, \bibinfo {author} {\bibfnamefont
  {Z.}~\bibnamefont {Wu}}, \bibinfo {author} {\bibfnamefont {Y.}~\bibnamefont
  {Deng}}, \bibinfo {author} {\bibfnamefont {X.-J.}\ \bibnamefont {Liu}},
  \bibinfo {author} {\bibfnamefont {S.}~\bibnamefont {Chen}},\ and\ \bibinfo
  {author} {\bibfnamefont {J.-W.}\ \bibnamefont {Pan}},\ }\bibfield  {title}
  {\bibinfo {title} {Highly controllable and robust 2d spin-orbit coupling for
  quantum gases},\ }\href {https://doi.org/10.1103/PhysRevLett.121.150401}
  {\bibfield  {journal} {\bibinfo  {journal} {Phys. Rev. Lett.}\ }\textbf
  {\bibinfo {volume} {121}},\ \bibinfo {pages} {150401} (\bibinfo {year}
  {2018})}\BibitemShut {NoStop}%
\bibitem [{\citenamefont {Liu}\ \emph {et~al.}(2014)\citenamefont {Liu},
  \citenamefont {Law},\ and\ \citenamefont {Ng}}]{CIM_coldatoms_2014}%
  \BibitemOpen
  \bibfield  {author} {\bibinfo {author} {\bibfnamefont {X.-J.}\ \bibnamefont
  {Liu}}, \bibinfo {author} {\bibfnamefont {K.~T.}\ \bibnamefont {Law}},\ and\
  \bibinfo {author} {\bibfnamefont {T.~K.}\ \bibnamefont {Ng}},\ }\bibfield
  {title} {\bibinfo {title} {Realization of 2d spin-orbit interaction and
  exotic topological orders in cold atoms},\ }\href
  {https://doi.org/10.1103/PhysRevLett.112.086401} {\bibfield  {journal}
  {\bibinfo  {journal} {Phys. Rev. Lett.}\ }\textbf {\bibinfo {volume} {112}},\
  \bibinfo {pages} {086401} (\bibinfo {year} {2014})}\BibitemShut {NoStop}%
\bibitem [{\citenamefont {Schindler}\ \emph {et~al.}(2022)\citenamefont
  {Schindler}, \citenamefont {Tsirkin}, \citenamefont {Neupert}, \citenamefont
  {Andrei~Bernevig},\ and\ \citenamefont
  {Wieder}}]{schindler_zddefectin3d_2022}%
  \BibitemOpen
  \bibfield  {author} {\bibinfo {author} {\bibfnamefont {F.}~\bibnamefont
  {Schindler}}, \bibinfo {author} {\bibfnamefont {S.~S.}\ \bibnamefont
  {Tsirkin}}, \bibinfo {author} {\bibfnamefont {T.}~\bibnamefont {Neupert}},
  \bibinfo {author} {\bibfnamefont {B.}~\bibnamefont {Andrei~Bernevig}},\ and\
  \bibinfo {author} {\bibfnamefont {B.~J.}\ \bibnamefont {Wieder}},\ }\bibfield
   {title} {\bibinfo {title} {Topological zero-dimensional defect and flux
  states in three-dimensional insulators},\ }\href
  {https://www.nature.com/articles/s41467-022-33471-x} {\bibfield  {journal}
  {\bibinfo  {journal} {Nature communications}\ }\textbf {\bibinfo {volume}
  {13}},\ \bibinfo {pages} {5791} (\bibinfo {year} {2022})}\BibitemShut
  {NoStop}%
\bibitem [{\citenamefont {Hu}\ and\ \citenamefont
  {Zhang}(2024)}]{hu_dislocation_2024}%
  \BibitemOpen
  \bibfield  {author} {\bibinfo {author} {\bibfnamefont {L.-H.}\ \bibnamefont
  {Hu}}\ and\ \bibinfo {author} {\bibfnamefont {R.-X.}\ \bibnamefont {Zhang}},\
  }\bibfield  {title} {\bibinfo {title} {Dislocation majorana bound states in
  iron-based superconductors},\ }\href
  {https://www.nature.com/articles/s41467-024-46618-9} {\bibfield  {journal}
  {\bibinfo  {journal} {Nature Communications}\ }\textbf {\bibinfo {volume}
  {15}},\ \bibinfo {pages} {2337} (\bibinfo {year} {2024})}\BibitemShut
  {NoStop}%
\bibitem [{\citenamefont {Chang}\ \emph {et~al.}(2013)\citenamefont {Chang},
  \citenamefont {Zhang}, \citenamefont {Feng}, \citenamefont {Shen},
  \citenamefont {Zhang}, \citenamefont {Guo}, \citenamefont {Li}, \citenamefont
  {Ou}, \citenamefont {Wei}, \citenamefont {Wang}, \citenamefont {Ji},
  \citenamefont {Feng}, \citenamefont {Ji}, \citenamefont {Chen}, \citenamefont
  {Jia}, \citenamefont {Dai}, \citenamefont {Fang}, \citenamefont {Zhang},
  \citenamefont {He}, \citenamefont {Wang}, \citenamefont {Lu}, \citenamefont
  {Ma},\ and\ \citenamefont {Xue}}]{CIM_qah_2013}%
  \BibitemOpen
  \bibfield  {author} {\bibinfo {author} {\bibfnamefont {C.}~\bibnamefont
  {Chang}}, \bibinfo {author} {\bibfnamefont {J.}~\bibnamefont {Zhang}},
  \bibinfo {author} {\bibfnamefont {X.}~\bibnamefont {Feng}}, \bibinfo {author}
  {\bibfnamefont {J.}~\bibnamefont {Shen}}, \bibinfo {author} {\bibfnamefont
  {Z.}~\bibnamefont {Zhang}}, \bibinfo {author} {\bibfnamefont
  {M.}~\bibnamefont {Guo}}, \bibinfo {author} {\bibfnamefont {K.}~\bibnamefont
  {Li}}, \bibinfo {author} {\bibfnamefont {Y.}~\bibnamefont {Ou}}, \bibinfo
  {author} {\bibfnamefont {P.}~\bibnamefont {Wei}}, \bibinfo {author}
  {\bibfnamefont {L.}~\bibnamefont {Wang}}, \bibinfo {author} {\bibfnamefont
  {Z.-Q.}\ \bibnamefont {Ji}}, \bibinfo {author} {\bibfnamefont
  {Y.}~\bibnamefont {Feng}}, \bibinfo {author} {\bibfnamefont {S.}~\bibnamefont
  {Ji}}, \bibinfo {author} {\bibfnamefont {X.}~\bibnamefont {Chen}}, \bibinfo
  {author} {\bibfnamefont {J.}~\bibnamefont {Jia}}, \bibinfo {author}
  {\bibfnamefont {X.}~\bibnamefont {Dai}}, \bibinfo {author} {\bibfnamefont
  {Z.}~\bibnamefont {Fang}}, \bibinfo {author} {\bibfnamefont {S.}~\bibnamefont
  {Zhang}}, \bibinfo {author} {\bibfnamefont {K.}~\bibnamefont {He}}, \bibinfo
  {author} {\bibfnamefont {Y.}~\bibnamefont {Wang}}, \bibinfo {author}
  {\bibfnamefont {L.}~\bibnamefont {Lu}}, \bibinfo {author} {\bibfnamefont
  {X.}~\bibnamefont {Ma}},\ and\ \bibinfo {author} {\bibfnamefont
  {Q.}~\bibnamefont {Xue}},\ }\bibfield  {title} {\bibinfo {title}
  {Experimental observation of the quantum anomalous hall effect in a magnetic
  topological insulator},\ }\href {https://doi.org/10.1126/science.1234414}
  {\bibfield  {journal} {\bibinfo  {journal} {Science}\ }\textbf {\bibinfo
  {volume} {340}},\ \bibinfo {pages} {167} (\bibinfo {year}
  {2013})}\BibitemShut {NoStop}%
\bibitem [{\citenamefont {Mogi}\ \emph {et~al.}(2015)\citenamefont {Mogi},
  \citenamefont {Yoshimi}, \citenamefont {Tsukazaki}, \citenamefont {Yasuda},
  \citenamefont {Kozuka}, \citenamefont {Takahashi}, \citenamefont {Kawasaki},\
  and\ \citenamefont {Tokura}}]{CIM_CrBST_2015}%
  \BibitemOpen
  \bibfield  {author} {\bibinfo {author} {\bibfnamefont {M.}~\bibnamefont
  {Mogi}}, \bibinfo {author} {\bibfnamefont {R.}~\bibnamefont {Yoshimi}},
  \bibinfo {author} {\bibfnamefont {A.}~\bibnamefont {Tsukazaki}}, \bibinfo
  {author} {\bibfnamefont {K.}~\bibnamefont {Yasuda}}, \bibinfo {author}
  {\bibfnamefont {Y.}~\bibnamefont {Kozuka}}, \bibinfo {author} {\bibfnamefont
  {K.~S.}\ \bibnamefont {Takahashi}}, \bibinfo {author} {\bibfnamefont
  {M.}~\bibnamefont {Kawasaki}},\ and\ \bibinfo {author} {\bibfnamefont
  {Y.}~\bibnamefont {Tokura}},\ }\bibfield  {title} {\bibinfo {title} {Magnetic
  modulation doping in topological insulators toward higher-temperature quantum
  anomalous hall effect},\ }\href {https://doi.org/10.1063/1.4935075}
  {\bibfield  {journal} {\bibinfo  {journal} {Applied Physics Letters}\
  }\textbf {\bibinfo {volume} {107}},\ \bibinfo {pages} {182401} (\bibinfo
  {year} {2015})}\BibitemShut {NoStop}%
\bibitem [{\citenamefont {Kou}\ \emph {et~al.}(2015)\citenamefont {Kou},
  \citenamefont {Pan}, \citenamefont {Wang}, \citenamefont {Fan}, \citenamefont
  {Choi}, \citenamefont {Lee}, \citenamefont {Nie}, \citenamefont {Murata},
  \citenamefont {Shao}, \citenamefont {Zhang} \emph
  {et~al.}}]{CIM_crbst2_2015}%
  \BibitemOpen
  \bibfield  {author} {\bibinfo {author} {\bibfnamefont {X.}~\bibnamefont
  {Kou}}, \bibinfo {author} {\bibfnamefont {L.}~\bibnamefont {Pan}}, \bibinfo
  {author} {\bibfnamefont {J.}~\bibnamefont {Wang}}, \bibinfo {author}
  {\bibfnamefont {Y.}~\bibnamefont {Fan}}, \bibinfo {author} {\bibfnamefont
  {E.~S.}\ \bibnamefont {Choi}}, \bibinfo {author} {\bibfnamefont
  {W.}~\bibnamefont {Lee}}, \bibinfo {author} {\bibfnamefont {T.}~\bibnamefont
  {Nie}}, \bibinfo {author} {\bibfnamefont {K.}~\bibnamefont {Murata}},
  \bibinfo {author} {\bibfnamefont {Q.}~\bibnamefont {Shao}}, \bibinfo {author}
  {\bibfnamefont {S.}~\bibnamefont {Zhang}}, \emph {et~al.},\ }\bibfield
  {title} {\bibinfo {title} {Metal-to-insulator switching in quantum anomalous
  hall states},\ }\href {https://www.nature.com/articles/ncomms9474} {\bibfield
   {journal} {\bibinfo  {journal} {Nature communications}\ }\textbf {\bibinfo
  {volume} {6}},\ \bibinfo {pages} {8474} (\bibinfo {year} {2015})}\BibitemShut
  {NoStop}%
\bibitem [{\citenamefont {Deng}\ \emph {et~al.}(2020)\citenamefont {Deng},
  \citenamefont {Yu}, \citenamefont {Shi}, \citenamefont {Guo}, \citenamefont
  {Xu}, \citenamefont {Wang}, \citenamefont {Chen},\ and\ \citenamefont
  {Zhang}}]{CIM_MBT_2020}%
  \BibitemOpen
  \bibfield  {author} {\bibinfo {author} {\bibfnamefont {Y.}~\bibnamefont
  {Deng}}, \bibinfo {author} {\bibfnamefont {Y.}~\bibnamefont {Yu}}, \bibinfo
  {author} {\bibfnamefont {M.~Z.}\ \bibnamefont {Shi}}, \bibinfo {author}
  {\bibfnamefont {Z.}~\bibnamefont {Guo}}, \bibinfo {author} {\bibfnamefont
  {Z.}~\bibnamefont {Xu}}, \bibinfo {author} {\bibfnamefont {J.}~\bibnamefont
  {Wang}}, \bibinfo {author} {\bibfnamefont {X.~H.}\ \bibnamefont {Chen}},\
  and\ \bibinfo {author} {\bibfnamefont {Y.}~\bibnamefont {Zhang}},\ }\bibfield
   {title} {\bibinfo {title} {Quantum anomalous hall effect in intrinsic
  magnetic topological insulator $\text{MnBi}_2\text{Te}_4$},\ }\href
  {https://doi.org/10.1126/science.aax8156} {\bibfield  {journal} {\bibinfo
  {journal} {Science}\ }\textbf {\bibinfo {volume} {367}},\ \bibinfo {pages}
  {895} (\bibinfo {year} {2020})}\BibitemShut {NoStop}%
\bibitem [{\citenamefont {Serlin}\ \emph {et~al.}(2020)\citenamefont {Serlin},
  \citenamefont {Tschirhart}, \citenamefont {Polshyn}, \citenamefont {Zhang},
  \citenamefont {Zhu}, \citenamefont {Watanabe}, \citenamefont {Taniguchi},
  \citenamefont {Balents},\ and\ \citenamefont {Young}}]{CIM_moire_2020}%
  \BibitemOpen
  \bibfield  {author} {\bibinfo {author} {\bibfnamefont {M.}~\bibnamefont
  {Serlin}}, \bibinfo {author} {\bibfnamefont {C.~L.}\ \bibnamefont
  {Tschirhart}}, \bibinfo {author} {\bibfnamefont {H.}~\bibnamefont {Polshyn}},
  \bibinfo {author} {\bibfnamefont {Y.}~\bibnamefont {Zhang}}, \bibinfo
  {author} {\bibfnamefont {J.}~\bibnamefont {Zhu}}, \bibinfo {author}
  {\bibfnamefont {K.}~\bibnamefont {Watanabe}}, \bibinfo {author}
  {\bibfnamefont {T.}~\bibnamefont {Taniguchi}}, \bibinfo {author}
  {\bibfnamefont {L.}~\bibnamefont {Balents}},\ and\ \bibinfo {author}
  {\bibfnamefont {A.~F.}\ \bibnamefont {Young}},\ }\bibfield  {title} {\bibinfo
  {title} {Intrinsic quantized anomalous hall effect in a moiré
  heterostructure},\ }\href {https://doi.org/10.1126/science.aay5533}
  {\bibfield  {journal} {\bibinfo  {journal} {Science}\ }\textbf {\bibinfo
  {volume} {367}},\ \bibinfo {pages} {900} (\bibinfo {year}
  {2020})}\BibitemShut {NoStop}%
\bibitem [{\citenamefont {Pan}\ \emph {et~al.}(2022)\citenamefont {Pan},
  \citenamefont {Xie}, \citenamefont {Wu},\ and\ \citenamefont
  {Das~Sarma}}]{CIM_abMote2}%
  \BibitemOpen
  \bibfield  {author} {\bibinfo {author} {\bibfnamefont {H.}~\bibnamefont
  {Pan}}, \bibinfo {author} {\bibfnamefont {M.}~\bibnamefont {Xie}}, \bibinfo
  {author} {\bibfnamefont {F.}~\bibnamefont {Wu}},\ and\ \bibinfo {author}
  {\bibfnamefont {S.}~\bibnamefont {Das~Sarma}},\ }\bibfield  {title} {\bibinfo
  {title} {Topological phases in $\text{AB}$-stacked
  $\text{MoTe}_{2}/\text{WSe}_{2}$: $\mathbb{Z}_{2}$ topological insulators,
  chern insulators, and topological charge density waves},\ }\href
  {https://doi.org/10.1103/PhysRevLett.129.056804} {\bibfield  {journal}
  {\bibinfo  {journal} {Phys. Rev. Lett.}\ }\textbf {\bibinfo {volume} {129}},\
  \bibinfo {pages} {056804} (\bibinfo {year} {2022})}\BibitemShut {NoStop}%
\bibitem [{\citenamefont {Liu}\ \emph {et~al.}(2024)\citenamefont {Liu},
  \citenamefont {He}, \citenamefont {Wang}, \citenamefont {Zhang},
  \citenamefont {Cao},\ and\ \citenamefont {Xiao}}]{CIM_TMD_2024}%
  \BibitemOpen
  \bibfield  {author} {\bibinfo {author} {\bibfnamefont {X.}~\bibnamefont
  {Liu}}, \bibinfo {author} {\bibfnamefont {Y.}~\bibnamefont {He}}, \bibinfo
  {author} {\bibfnamefont {C.}~\bibnamefont {Wang}}, \bibinfo {author}
  {\bibfnamefont {X.}~\bibnamefont {Zhang}}, \bibinfo {author} {\bibfnamefont
  {T.}~\bibnamefont {Cao}},\ and\ \bibinfo {author} {\bibfnamefont
  {D.}~\bibnamefont {Xiao}},\ }\bibfield  {title} {\bibinfo {title}
  {Gate-tunable antiferromagnetic chern insulator in twisted bilayer transition
  metal dichalcogenides},\ }\href
  {https://doi.org/10.1103/PhysRevLett.132.146401} {\bibfield  {journal}
  {\bibinfo  {journal} {Phys. Rev. Lett.}\ }\textbf {\bibinfo {volume} {132}},\
  \bibinfo {pages} {146401} (\bibinfo {year} {2024})}\BibitemShut {NoStop}%
\bibitem [{\citenamefont {Uría-Álvarez}\ and\ \citenamefont
  {Palacios}(2024)}]{CIM_bisb_2024}%
  \BibitemOpen
  \bibfield  {author} {\bibinfo {author} {\bibfnamefont {A.~J.}\ \bibnamefont
  {Uría-Álvarez}}\ and\ \bibinfo {author} {\bibfnamefont {J.~J.}\
  \bibnamefont {Palacios}},\ }\href@noop {} {\bibinfo {title}
  {Amorphization-induced topological and insulator-metal transitions in
  bidimensional $\text{Bi}_x\text{Sb}_{1-x}$ alloys}} (\bibinfo {year}
  {2024}),\ \Eprint {https://arxiv.org/abs/2410.16034} {arXiv:2410.16034
  [cond-mat.dis-nn]} \BibitemShut {NoStop}%
\bibitem [{\citenamefont {Zhang}\ \emph {et~al.}(2018)\citenamefont {Zhang},
  \citenamefont {Zhang}, \citenamefont {Wang},\ and\ \citenamefont
  {Li}}]{CIM_stanene_2018}%
  \BibitemOpen
  \bibfield  {author} {\bibinfo {author} {\bibfnamefont {M.}~\bibnamefont
  {Zhang}}, \bibinfo {author} {\bibfnamefont {C.}~\bibnamefont {Zhang}},
  \bibinfo {author} {\bibfnamefont {P.}~\bibnamefont {Wang}},\ and\ \bibinfo
  {author} {\bibfnamefont {S.}~\bibnamefont {Li}},\ }\bibfield  {title}
  {\bibinfo {title} {Prediction of high-temperature chern insulator with
  half-metallic edge states in asymmetry-functionalized stanene},\ }\href
  {https://doi.org/10.1039/C8NR07503D} {\bibfield  {journal} {\bibinfo
  {journal} {Nanoscale}\ }\textbf {\bibinfo {volume} {10}},\ \bibinfo {pages}
  {20226} (\bibinfo {year} {2018})}\BibitemShut {NoStop}%
\bibitem [{\citenamefont {Guo}\ \emph {et~al.}(2022)\citenamefont {Guo},
  \citenamefont {Liu},\ and\ \citenamefont {Lu}}]{CIM_CrO_2022}%
  \BibitemOpen
  \bibfield  {author} {\bibinfo {author} {\bibfnamefont {P.}~\bibnamefont
  {Guo}}, \bibinfo {author} {\bibfnamefont {Z.}~\bibnamefont {Liu}},\ and\
  \bibinfo {author} {\bibfnamefont {Z.}~\bibnamefont {Lu}},\ }\bibfield
  {title} {\bibinfo {title} {Quantum anomalous hall effect in collinear
  antiferromagnetism},\ }\href
  {https://api.semanticscholar.org/CorpusID:258451432} {\bibfield  {journal}
  {\bibinfo  {journal} {npj Computational Materials}\ }\textbf {\bibinfo
  {volume} {9}},\ \bibinfo {pages} {1} (\bibinfo {year} {2022})}\BibitemShut
  {NoStop}%
\bibitem [{\citenamefont {Wu}\ \emph {et~al.}(2023)\citenamefont {Wu},
  \citenamefont {Song}, \citenamefont {Ji}, \citenamefont {Wang}, \citenamefont
  {Zhang},\ and\ \citenamefont {Zhang}}]{CIM_MoO_2023}%
  \BibitemOpen
  \bibfield  {author} {\bibinfo {author} {\bibfnamefont {B.}~\bibnamefont
  {Wu}}, \bibinfo {author} {\bibfnamefont {Y.}~\bibnamefont {Song}}, \bibinfo
  {author} {\bibfnamefont {W.}~\bibnamefont {Ji}}, \bibinfo {author}
  {\bibfnamefont {P.}~\bibnamefont {Wang}}, \bibinfo {author} {\bibfnamefont
  {S.}~\bibnamefont {Zhang}},\ and\ \bibinfo {author} {\bibfnamefont
  {C.}~\bibnamefont {Zhang}},\ }\bibfield  {title} {\bibinfo {title} {Quantum
  anomalous hall effect in an antiferromagnetic monolayer of $\text{MoO}$},\
  }\href {https://doi.org/10.1103/PhysRevB.107.214419} {\bibfield  {journal}
  {\bibinfo  {journal} {Phys. Rev. B}\ }\textbf {\bibinfo {volume} {107}},\
  \bibinfo {pages} {214419} (\bibinfo {year} {2023})}\BibitemShut {NoStop}%
\bibitem [{\citenamefont {Hafez-Torbati}\ and\ \citenamefont
  {Uhrig}(2024)}]{CIM_afci_2024}%
  \BibitemOpen
  \bibfield  {author} {\bibinfo {author} {\bibfnamefont {M.}~\bibnamefont
  {Hafez-Torbati}}\ and\ \bibinfo {author} {\bibfnamefont {G.~S.}\ \bibnamefont
  {Uhrig}},\ }\bibfield  {title} {\bibinfo {title} {Antiferromagnetic chern
  insulator with large charge gap in heavy transition-metal compounds},\ }\href
  {https://api.semanticscholar.org/CorpusID:267759951} {\bibfield  {journal}
  {\bibinfo  {journal} {Scientific Reports}\ }\textbf {\bibinfo {volume} {14}}
  (\bibinfo {year} {2024})}\BibitemShut {NoStop}%
\bibitem [{\citenamefont {Benalcazar}\ \emph {et~al.}(2017)\citenamefont
  {Benalcazar}, \citenamefont {Bernevig},\ and\ \citenamefont
  {Hughes}}]{benalcazar_HOTI_2017}%
  \BibitemOpen
  \bibfield  {author} {\bibinfo {author} {\bibfnamefont {W.~A.}\ \bibnamefont
  {Benalcazar}}, \bibinfo {author} {\bibfnamefont {B.~A.}\ \bibnamefont
  {Bernevig}},\ and\ \bibinfo {author} {\bibfnamefont {T.~L.}\ \bibnamefont
  {Hughes}},\ }\bibfield  {title} {\bibinfo {title} {Quantized electric
  multipole insulators},\ }\href
  {https://www.science.org/doi/10.1126/science.aah6442} {\bibfield  {journal}
  {\bibinfo  {journal} {Science}\ }\textbf {\bibinfo {volume} {357}},\ \bibinfo
  {pages} {61} (\bibinfo {year} {2017})}\BibitemShut {NoStop}%
\bibitem [{\citenamefont {Schindler}\ \emph {et~al.}(2018)\citenamefont
  {Schindler}, \citenamefont {Cook}, \citenamefont {Vergniory}, \citenamefont
  {Wang}, \citenamefont {Parkin}, \citenamefont {Bernevig},\ and\ \citenamefont
  {Neupert}}]{frank_hoti_2018}%
  \BibitemOpen
  \bibfield  {author} {\bibinfo {author} {\bibfnamefont {F.}~\bibnamefont
  {Schindler}}, \bibinfo {author} {\bibfnamefont {A.~M.}\ \bibnamefont {Cook}},
  \bibinfo {author} {\bibfnamefont {M.~G.}\ \bibnamefont {Vergniory}}, \bibinfo
  {author} {\bibfnamefont {Z.}~\bibnamefont {Wang}}, \bibinfo {author}
  {\bibfnamefont {S.~S.~P.}\ \bibnamefont {Parkin}}, \bibinfo {author}
  {\bibfnamefont {B.~A.}\ \bibnamefont {Bernevig}},\ and\ \bibinfo {author}
  {\bibfnamefont {T.}~\bibnamefont {Neupert}},\ }\bibfield  {title} {\bibinfo
  {title} {Higher-order topological insulators},\ }\href
  {https://doi.org/10.1126/sciadv.aat0346} {\bibfield  {journal} {\bibinfo
  {journal} {Science Advances}\ }\textbf {\bibinfo {volume} {4}},\ \bibinfo
  {pages} {eaat0346} (\bibinfo {year} {2018})}\BibitemShut {NoStop}%
\bibitem [{\citenamefont {Matsugatani}\ and\ \citenamefont
  {Watanabe}(2018)}]{matsugai_hoti_2018}%
  \BibitemOpen
  \bibfield  {author} {\bibinfo {author} {\bibfnamefont {A.}~\bibnamefont
  {Matsugatani}}\ and\ \bibinfo {author} {\bibfnamefont {H.}~\bibnamefont
  {Watanabe}},\ }\bibfield  {title} {\bibinfo {title} {Connecting higher-order
  topological insulators to lower-dimensional topological insulators},\ }\href
  {https://doi.org/10.1103/PhysRevB.98.205129} {\bibfield  {journal} {\bibinfo
  {journal} {Phys. Rev. B}\ }\textbf {\bibinfo {volume} {98}},\ \bibinfo
  {pages} {205129} (\bibinfo {year} {2018})}\BibitemShut {NoStop}%
\bibitem [{\citenamefont {Queiroz}\ \emph {et~al.}(2019)\citenamefont
  {Queiroz}, \citenamefont {Fulga}, \citenamefont {Avraham}, \citenamefont
  {Beidenkopf},\ and\ \citenamefont {Cano}}]{queiroz_hoti_2019}%
  \BibitemOpen
  \bibfield  {author} {\bibinfo {author} {\bibfnamefont {R.}~\bibnamefont
  {Queiroz}}, \bibinfo {author} {\bibfnamefont {I.~C.}\ \bibnamefont {Fulga}},
  \bibinfo {author} {\bibfnamefont {N.}~\bibnamefont {Avraham}}, \bibinfo
  {author} {\bibfnamefont {H.}~\bibnamefont {Beidenkopf}},\ and\ \bibinfo
  {author} {\bibfnamefont {J.}~\bibnamefont {Cano}},\ }\bibfield  {title}
  {\bibinfo {title} {Partial lattice defects in higher-order topological
  insulators},\ }\href {https://doi.org/10.1103/PhysRevLett.123.266802}
  {\bibfield  {journal} {\bibinfo  {journal} {Phys. Rev. Lett.}\ }\textbf
  {\bibinfo {volume} {123}},\ \bibinfo {pages} {266802} (\bibinfo {year}
  {2019})}\BibitemShut {NoStop}%
\bibitem [{\citenamefont {Trifunovic}\ and\ \citenamefont
  {Brouwer}(2019)}]{trifunovic_hoti_2019}%
  \BibitemOpen
  \bibfield  {author} {\bibinfo {author} {\bibfnamefont {L.}~\bibnamefont
  {Trifunovic}}\ and\ \bibinfo {author} {\bibfnamefont {P.~W.}\ \bibnamefont
  {Brouwer}},\ }\bibfield  {title} {\bibinfo {title} {Higher-order
  bulk-boundary correspondence for topological crystalline phases},\ }\href
  {https://doi.org/10.1103/PhysRevX.9.011012} {\bibfield  {journal} {\bibinfo
  {journal} {Phys. Rev. X}\ }\textbf {\bibinfo {volume} {9}},\ \bibinfo {pages}
  {011012} (\bibinfo {year} {2019})}\BibitemShut {NoStop}%
\bibitem [{\citenamefont {Roy}\ and\ \citenamefont {Juri\ifmmode \check{c}\else
  \v{c}\fi{}i\ifmmode~\acute{c}\else \'{c}\fi{}}(2021)}]{roy_hoti_2021}%
  \BibitemOpen
  \bibfield  {author} {\bibinfo {author} {\bibfnamefont {B.}~\bibnamefont
  {Roy}}\ and\ \bibinfo {author} {\bibfnamefont {V.}~\bibnamefont {Juri\ifmmode
  \check{c}\else \v{c}\fi{}i\ifmmode~\acute{c}\else \'{c}\fi{}}},\ }\bibfield
  {title} {\bibinfo {title} {Dislocation as a bulk probe of higher-order
  topological insulators},\ }\href
  {https://doi.org/10.1103/PhysRevResearch.3.033107} {\bibfield  {journal}
  {\bibinfo  {journal} {Phys. Rev. Res.}\ }\textbf {\bibinfo {volume} {3}},\
  \bibinfo {pages} {033107} (\bibinfo {year} {2021})}\BibitemShut {NoStop}%
\bibitem [{\citenamefont {Bombin}(2010)}]{Bombin_twist_2010}%
  \BibitemOpen
  \bibfield  {author} {\bibinfo {author} {\bibfnamefont {H.}~\bibnamefont
  {Bombin}},\ }\bibfield  {title} {\bibinfo {title} {Topological order with a
  twist: Ising anyons from an abelian model},\ }\href
  {https://doi.org/10.1103/PhysRevLett.105.030403} {\bibfield  {journal}
  {\bibinfo  {journal} {Phys. Rev. Lett.}\ }\textbf {\bibinfo {volume} {105}},\
  \bibinfo {pages} {030403} (\bibinfo {year} {2010})}\BibitemShut {NoStop}%
\bibitem [{\citenamefont {You}\ \emph {et~al.}(2013)\citenamefont {You},
  \citenamefont {Jian},\ and\ \citenamefont
  {Wen}}]{Wen_anyoncondensation_2013}%
  \BibitemOpen
  \bibfield  {author} {\bibinfo {author} {\bibfnamefont {Y.-Z.}\ \bibnamefont
  {You}}, \bibinfo {author} {\bibfnamefont {C.-M.}\ \bibnamefont {Jian}},\ and\
  \bibinfo {author} {\bibfnamefont {X.-G.}\ \bibnamefont {Wen}},\ }\bibfield
  {title} {\bibinfo {title} {Synthetic non-abelian statistics by abelian anyon
  condensation},\ }\href {https://doi.org/10.1103/PhysRevB.87.045106}
  {\bibfield  {journal} {\bibinfo  {journal} {Phys. Rev. B}\ }\textbf {\bibinfo
  {volume} {87}},\ \bibinfo {pages} {045106} (\bibinfo {year}
  {2013})}\BibitemShut {NoStop}%
\bibitem [{\citenamefont {Barkeshli}\ \emph
  {et~al.}(2013{\natexlab{a}})\citenamefont {Barkeshli}, \citenamefont {Jian},\
  and\ \citenamefont {Qi}}]{Barkeshli_defcts_2013}%
  \BibitemOpen
  \bibfield  {author} {\bibinfo {author} {\bibfnamefont {M.}~\bibnamefont
  {Barkeshli}}, \bibinfo {author} {\bibfnamefont {C.-M.}\ \bibnamefont
  {Jian}},\ and\ \bibinfo {author} {\bibfnamefont {X.-L.}\ \bibnamefont {Qi}},\
  }\bibfield  {title} {\bibinfo {title} {Theory of defects in abelian
  topological states},\ }\href {https://doi.org/10.1103/PhysRevB.88.235103}
  {\bibfield  {journal} {\bibinfo  {journal} {Phys. Rev. B}\ }\textbf {\bibinfo
  {volume} {88}},\ \bibinfo {pages} {235103} (\bibinfo {year}
  {2013}{\natexlab{a}})}\BibitemShut {NoStop}%
\bibitem [{\citenamefont {Barkeshli}\ \emph
  {et~al.}(2013{\natexlab{b}})\citenamefont {Barkeshli}, \citenamefont {Jian},\
  and\ \citenamefont {Qi}}]{Barkeshli_defectsclass_2013}%
  \BibitemOpen
  \bibfield  {author} {\bibinfo {author} {\bibfnamefont {M.}~\bibnamefont
  {Barkeshli}}, \bibinfo {author} {\bibfnamefont {C.-M.}\ \bibnamefont
  {Jian}},\ and\ \bibinfo {author} {\bibfnamefont {X.-L.}\ \bibnamefont {Qi}},\
  }\bibfield  {title} {\bibinfo {title} {Classification of topological defects
  in abelian topological states},\ }\href
  {https://doi.org/10.1103/PhysRevB.88.241103} {\bibfield  {journal} {\bibinfo
  {journal} {Phys. Rev. B}\ }\textbf {\bibinfo {volume} {88}},\ \bibinfo
  {pages} {241103} (\bibinfo {year} {2013}{\natexlab{b}})}\BibitemShut
  {NoStop}%
\bibitem [{\citenamefont {Mesaros}\ \emph {et~al.}(2013)\citenamefont
  {Mesaros}, \citenamefont {Kim},\ and\ \citenamefont
  {Ran}}]{Andrej_topodefects_2013}%
  \BibitemOpen
  \bibfield  {author} {\bibinfo {author} {\bibfnamefont {A.}~\bibnamefont
  {Mesaros}}, \bibinfo {author} {\bibfnamefont {Y.~B.}\ \bibnamefont {Kim}},\
  and\ \bibinfo {author} {\bibfnamefont {Y.}~\bibnamefont {Ran}},\ }\bibfield
  {title} {\bibinfo {title} {Changing topology by topological defects in
  three-dimensional topologically ordered phases},\ }\href
  {https://doi.org/10.1103/PhysRevB.88.035141} {\bibfield  {journal} {\bibinfo
  {journal} {Phys. Rev. B}\ }\textbf {\bibinfo {volume} {88}},\ \bibinfo
  {pages} {035141} (\bibinfo {year} {2013})}\BibitemShut {NoStop}%
\bibitem [{\citenamefont {Barkeshli}\ \emph
  {et~al.}(2013{\natexlab{c}})\citenamefont {Barkeshli}, \citenamefont {Jian},\
  and\ \citenamefont {Qi}}]{Barkeshli_twist_2013}%
  \BibitemOpen
  \bibfield  {author} {\bibinfo {author} {\bibfnamefont {M.}~\bibnamefont
  {Barkeshli}}, \bibinfo {author} {\bibfnamefont {C.-M.}\ \bibnamefont
  {Jian}},\ and\ \bibinfo {author} {\bibfnamefont {X.-L.}\ \bibnamefont {Qi}},\
  }\bibfield  {title} {\bibinfo {title} {Twist defects and projective
  non-abelian braiding statistics},\ }\href
  {https://doi.org/10.1103/PhysRevB.87.045130} {\bibfield  {journal} {\bibinfo
  {journal} {Phys. Rev. B}\ }\textbf {\bibinfo {volume} {87}},\ \bibinfo
  {pages} {045130} (\bibinfo {year} {2013}{\natexlab{c}})}\BibitemShut
  {NoStop}%
\bibitem [{\citenamefont {Ben-Zion}\ \emph {et~al.}(2016)\citenamefont
  {Ben-Zion}, \citenamefont {Das},\ and\ \citenamefont
  {McGreevy}}]{Diptarka_topoparamagnet_2016}%
  \BibitemOpen
  \bibfield  {author} {\bibinfo {author} {\bibfnamefont {D.}~\bibnamefont
  {Ben-Zion}}, \bibinfo {author} {\bibfnamefont {D.}~\bibnamefont {Das}},\ and\
  \bibinfo {author} {\bibfnamefont {J.}~\bibnamefont {McGreevy}},\ }\bibfield
  {title} {\bibinfo {title} {Exactly solvable models of spin liquids with
  spinons, and of three-dimensional topological paramagnets},\ }\href
  {https://doi.org/10.1103/PhysRevB.93.155147} {\bibfield  {journal} {\bibinfo
  {journal} {Phys. Rev. B}\ }\textbf {\bibinfo {volume} {93}},\ \bibinfo
  {pages} {155147} (\bibinfo {year} {2016})}\BibitemShut {NoStop}%
\bibitem [{\citenamefont {Barkeshli}\ \emph {et~al.}(2019)\citenamefont
  {Barkeshli}, \citenamefont {Bonderson}, \citenamefont {Cheng},\ and\
  \citenamefont {Wang}}]{Barkeshli_defects2_2019}%
  \BibitemOpen
  \bibfield  {author} {\bibinfo {author} {\bibfnamefont {M.}~\bibnamefont
  {Barkeshli}}, \bibinfo {author} {\bibfnamefont {P.}~\bibnamefont
  {Bonderson}}, \bibinfo {author} {\bibfnamefont {M.}~\bibnamefont {Cheng}},\
  and\ \bibinfo {author} {\bibfnamefont {Z.}~\bibnamefont {Wang}},\ }\bibfield
  {title} {\bibinfo {title} {Symmetry fractionalization, defects, and gauging
  of topological phases},\ }\href {https://doi.org/10.1103/PhysRevB.100.115147}
  {\bibfield  {journal} {\bibinfo  {journal} {Phys. Rev. B}\ }\textbf {\bibinfo
  {volume} {100}},\ \bibinfo {pages} {115147} (\bibinfo {year}
  {2019})}\BibitemShut {NoStop}%
\bibitem [{\citenamefont {Wang}\ \emph {et~al.}(2024)\citenamefont {Wang},
  \citenamefont {Liu},\ and\ \citenamefont {Lu}}]{Wang_topodefects_2024}%
  \BibitemOpen
  \bibfield  {author} {\bibinfo {author} {\bibfnamefont {Y.-Q.}\ \bibnamefont
  {Wang}}, \bibinfo {author} {\bibfnamefont {C.}~\bibnamefont {Liu}},\ and\
  \bibinfo {author} {\bibfnamefont {Y.-M.}\ \bibnamefont {Lu}},\ }\bibfield
  {title} {\bibinfo {title} {Theory of topological defects and textures in
  two-dimensional quantum orders with spontaneous symmetry breaking},\ }\href
  {https://doi.org/10.1103/PhysRevB.109.195165} {\bibfield  {journal} {\bibinfo
   {journal} {Phys. Rev. B}\ }\textbf {\bibinfo {volume} {109}},\ \bibinfo
  {pages} {195165} (\bibinfo {year} {2024})}\BibitemShut {NoStop}%
\bibitem [{\citenamefont {Oka}\ and\ \citenamefont
  {Kitamura}(2019)}]{oka2019floquet}%
  \BibitemOpen
  \bibfield  {author} {\bibinfo {author} {\bibfnamefont {T.}~\bibnamefont
  {Oka}}\ and\ \bibinfo {author} {\bibfnamefont {S.}~\bibnamefont {Kitamura}},\
  }\bibfield  {title} {\bibinfo {title} {Floquet engineering of quantum
  materials},\ }\href
  {https://doi.org/https://doi.org/10.1146/annurev-conmatphys-031218-013423}
  {\bibfield  {journal} {\bibinfo  {journal} {Annual Review of Condensed Matter
  Physics}\ }\textbf {\bibinfo {volume} {10}},\ \bibinfo {pages} {387}
  (\bibinfo {year} {2019})}\BibitemShut {NoStop}%
\bibitem [{\citenamefont {Harper}\ \emph {et~al.}(2020)\citenamefont {Harper},
  \citenamefont {Roy}, \citenamefont {Rudner},\ and\ \citenamefont
  {Sondhi}}]{harper2020topology}%
  \BibitemOpen
  \bibfield  {author} {\bibinfo {author} {\bibfnamefont {F.}~\bibnamefont
  {Harper}}, \bibinfo {author} {\bibfnamefont {R.}~\bibnamefont {Roy}},
  \bibinfo {author} {\bibfnamefont {M.~S.}\ \bibnamefont {Rudner}},\ and\
  \bibinfo {author} {\bibfnamefont {S.}~\bibnamefont {Sondhi}},\ }\bibfield
  {title} {\bibinfo {title} {Topology and broken symmetry in floquet systems},\
  }\href
  {https://doi.org/https://doi.org/10.1146/annurev-conmatphys-031218-013721}
  {\bibfield  {journal} {\bibinfo  {journal} {Annual Review of Condensed Matter
  Physics}\ }\textbf {\bibinfo {volume} {11}},\ \bibinfo {pages} {345}
  (\bibinfo {year} {2020})}\BibitemShut {NoStop}%
\bibitem [{\citenamefont {Mondal}\ \emph
  {et~al.}(2023{\natexlab{a}})\citenamefont {Mondal}, \citenamefont {Ghosh},
  \citenamefont {Nag},\ and\ \citenamefont {Saha}}]{mondal_fmbs_2023}%
  \BibitemOpen
  \bibfield  {author} {\bibinfo {author} {\bibfnamefont {D.}~\bibnamefont
  {Mondal}}, \bibinfo {author} {\bibfnamefont {A.~K.}\ \bibnamefont {Ghosh}},
  \bibinfo {author} {\bibfnamefont {T.}~\bibnamefont {Nag}},\ and\ \bibinfo
  {author} {\bibfnamefont {A.}~\bibnamefont {Saha}},\ }\bibfield  {title}
  {\bibinfo {title} {Engineering anomalous floquet majorana modes and their
  time evolution in a helical shiba chain},\ }\href
  {https://doi.org/10.1103/PhysRevB.108.L081403} {\bibfield  {journal}
  {\bibinfo  {journal} {Phys. Rev. B}\ }\textbf {\bibinfo {volume} {108}},\
  \bibinfo {pages} {L081403} (\bibinfo {year}
  {2023}{\natexlab{a}})}\BibitemShut {NoStop}%
\bibitem [{\citenamefont {Mondal}\ \emph
  {et~al.}(2023{\natexlab{b}})\citenamefont {Mondal}, \citenamefont {Ghosh},
  \citenamefont {Nag},\ and\ \citenamefont {Saha}}]{mondal_fmbs2_2023}%
  \BibitemOpen
  \bibfield  {author} {\bibinfo {author} {\bibfnamefont {D.}~\bibnamefont
  {Mondal}}, \bibinfo {author} {\bibfnamefont {A.~K.}\ \bibnamefont {Ghosh}},
  \bibinfo {author} {\bibfnamefont {T.}~\bibnamefont {Nag}},\ and\ \bibinfo
  {author} {\bibfnamefont {A.}~\bibnamefont {Saha}},\ }\bibfield  {title}
  {\bibinfo {title} {Topological characterization and stability of floquet
  majorana modes in rashba nanowires},\ }\href
  {https://doi.org/10.1103/PhysRevB.107.035427} {\bibfield  {journal} {\bibinfo
   {journal} {Phys. Rev. B}\ }\textbf {\bibinfo {volume} {107}},\ \bibinfo
  {pages} {035427} (\bibinfo {year} {2023}{\natexlab{b}})}\BibitemShut
  {NoStop}%
\bibitem [{\citenamefont {Rachel}(2018)}]{Rachel_2018}%
  \BibitemOpen
  \bibfield  {author} {\bibinfo {author} {\bibfnamefont {S.}~\bibnamefont
  {Rachel}},\ }\bibfield  {title} {\bibinfo {title} {Interacting topological
  insulators: a review},\ }\href {https://doi.org/10.1088/1361-6633/aad6a6}
  {\bibfield  {journal} {\bibinfo  {journal} {Reports on Progress in Physics}\
  }\textbf {\bibinfo {volume} {81}},\ \bibinfo {pages} {116501} (\bibinfo
  {year} {2018})}\BibitemShut {NoStop}%
\bibitem [{\citenamefont {Tokura}\ \emph {et~al.}(2019)\citenamefont {Tokura},
  \citenamefont {Yasuda},\ and\ \citenamefont
  {Tsukazaki}}]{tokura2019magnetic}%
  \BibitemOpen
  \bibfield  {author} {\bibinfo {author} {\bibfnamefont {Y.}~\bibnamefont
  {Tokura}}, \bibinfo {author} {\bibfnamefont {K.}~\bibnamefont {Yasuda}},\
  and\ \bibinfo {author} {\bibfnamefont {A.}~\bibnamefont {Tsukazaki}},\
  }\bibfield  {title} {\bibinfo {title} {Magnetic topological insulators},\
  }\href {https://www.nature.com/articles/s42254-018-0011-5} {\bibfield
  {journal} {\bibinfo  {journal} {Nature Reviews Physics}\ }\textbf {\bibinfo
  {volume} {1}},\ \bibinfo {pages} {126} (\bibinfo {year} {2019})}\BibitemShut
  {NoStop}%
\bibitem [{\citenamefont {Agarwala}\ and\ \citenamefont
  {Shenoy}(2017)}]{Agarwala_PRL_2017}%
  \BibitemOpen
  \bibfield  {author} {\bibinfo {author} {\bibfnamefont {A.}~\bibnamefont
  {Agarwala}}\ and\ \bibinfo {author} {\bibfnamefont {V.~B.}\ \bibnamefont
  {Shenoy}},\ }\bibfield  {title} {\bibinfo {title} {Topological insulators in
  amorphous systems},\ }\href {https://doi.org/10.1103/PhysRevLett.118.236402}
  {\bibfield  {journal} {\bibinfo  {journal} {Phys. Rev. Lett.}\ }\textbf
  {\bibinfo {volume} {118}},\ \bibinfo {pages} {236402} (\bibinfo {year}
  {2017})}\BibitemShut {NoStop}%
\bibitem [{\citenamefont {Sahlberg}\ \emph {et~al.}(2020)\citenamefont
  {Sahlberg}, \citenamefont {Weststr\"om}, \citenamefont {P\"oyh\"onen},\ and\
  \citenamefont {Ojanen}}]{Sahlberg_PRR_2020}%
  \BibitemOpen
  \bibfield  {author} {\bibinfo {author} {\bibfnamefont {I.}~\bibnamefont
  {Sahlberg}}, \bibinfo {author} {\bibfnamefont {A.}~\bibnamefont
  {Weststr\"om}}, \bibinfo {author} {\bibfnamefont {K.}~\bibnamefont
  {P\"oyh\"onen}},\ and\ \bibinfo {author} {\bibfnamefont {T.}~\bibnamefont
  {Ojanen}},\ }\bibfield  {title} {\bibinfo {title} {Topological phase
  transitions in glassy quantum matter},\ }\href
  {https://doi.org/10.1103/PhysRevResearch.2.013053} {\bibfield  {journal}
  {\bibinfo  {journal} {Phys. Rev. Res.}\ }\textbf {\bibinfo {volume} {2}},\
  \bibinfo {pages} {013053} (\bibinfo {year} {2020})}\BibitemShut {NoStop}%
\bibitem [{\citenamefont {{Loring, T. A.}}\ and\ \citenamefont {{Hastings, M.
  B.}}(2010)}]{Loring_EPL_2010}%
  \BibitemOpen
  \bibfield  {author} {\bibinfo {author} {\bibnamefont {{Loring, T. A.}}}\ and\
  \bibinfo {author} {\bibnamefont {{Hastings, M. B.}}},\ }\bibfield  {title}
  {\bibinfo {title} {Disordered topological insulators via c*-algebras},\
  }\href {https://doi.org/10.1209/0295-5075/92/67004} {\bibfield  {journal}
  {\bibinfo  {journal} {Eur. Phys. Lett.}\ }\textbf {\bibinfo {volume} {92}},\
  \bibinfo {pages} {67004} (\bibinfo {year} {2010})}\BibitemShut {NoStop}%
\bibitem [{\citenamefont {Hastings}\ and\ \citenamefont
  {Loring}(2011)}]{Hastings_AOP_2011}%
  \BibitemOpen
  \bibfield  {author} {\bibinfo {author} {\bibfnamefont {M.~B.}\ \bibnamefont
  {Hastings}}\ and\ \bibinfo {author} {\bibfnamefont {T.~A.}\ \bibnamefont
  {Loring}},\ }\bibfield  {title} {\bibinfo {title} {Topological insulators and
  c*-algebras: Theory and numerical practice},\ }\href
  {https://www.sciencedirect.com/science/article/pii/S0003491610002277}
  {\bibfield  {journal} {\bibinfo  {journal} {Annals of Physics}\ }\textbf
  {\bibinfo {volume} {326}},\ \bibinfo {pages} {1699} (\bibinfo {year}
  {2011})}\BibitemShut {NoStop}%
\bibitem [{\citenamefont {Houghton}\ and\ \citenamefont
  {Marston}(1993)}]{Houghton_bosonization_1993}%
  \BibitemOpen
  \bibfield  {author} {\bibinfo {author} {\bibfnamefont {A.}~\bibnamefont
  {Houghton}}\ and\ \bibinfo {author} {\bibfnamefont {J.~B.}\ \bibnamefont
  {Marston}},\ }\bibfield  {title} {\bibinfo {title} {Bosonization and fermion
  liquids in dimensions greater than one},\ }\href
  {https://doi.org/10.1103/PhysRevB.48.7790} {\bibfield  {journal} {\bibinfo
  {journal} {Phys. Rev. B}\ }\textbf {\bibinfo {volume} {48}},\ \bibinfo
  {pages} {7790} (\bibinfo {year} {1993})}\BibitemShut {NoStop}%
\bibitem [{\citenamefont {Shankar}(1994)}]{shankar_RG_1994}%
  \BibitemOpen
  \bibfield  {author} {\bibinfo {author} {\bibfnamefont {R.}~\bibnamefont
  {Shankar}},\ }\bibfield  {title} {\bibinfo {title} {Renormalization-group
  approach to interacting fermions},\ }\href
  {https://doi.org/10.1103/RevModPhys.66.129} {\bibfield  {journal} {\bibinfo
  {journal} {Rev. Mod. Phys.}\ }\textbf {\bibinfo {volume} {66}},\ \bibinfo
  {pages} {129} (\bibinfo {year} {1994})}\BibitemShut {NoStop}%
\bibitem [{\citenamefont {Resta}(1998)}]{resta_1998}%
  \BibitemOpen
  \bibfield  {author} {\bibinfo {author} {\bibfnamefont {R.}~\bibnamefont
  {Resta}},\ }\bibfield  {title} {\bibinfo {title} {Quantum-mechanical position
  operator in extended systems},\ }\href
  {https://doi.org/10.1103/PhysRevLett.80.1800} {\bibfield  {journal} {\bibinfo
   {journal} {Phys. Rev. Lett.}\ }\textbf {\bibinfo {volume} {80}},\ \bibinfo
  {pages} {1800} (\bibinfo {year} {1998})}\BibitemShut {NoStop}%
\bibitem [{\citenamefont {Bianco}\ and\ \citenamefont
  {Resta}(2011)}]{Bianco_PRB_2011}%
  \BibitemOpen
  \bibfield  {author} {\bibinfo {author} {\bibfnamefont {R.}~\bibnamefont
  {Bianco}}\ and\ \bibinfo {author} {\bibfnamefont {R.}~\bibnamefont {Resta}},\
  }\bibfield  {title} {\bibinfo {title} {Mapping topological order in
  coordinate space},\ }\href {https://doi.org/10.1103/PhysRevB.84.241106}
  {\bibfield  {journal} {\bibinfo  {journal} {Phys. Rev. B}\ }\textbf {\bibinfo
  {volume} {84}},\ \bibinfo {pages} {241106} (\bibinfo {year}
  {2011})}\BibitemShut {NoStop}%
\bibitem [{\citenamefont {Hegde}\ \emph {et~al.}(2021)\citenamefont {Hegde},
  \citenamefont {Kumar}, \citenamefont {Agarwala},\ and\ \citenamefont
  {Muralidharan}}]{agarwala_ssh_2021}%
  \BibitemOpen
  \bibfield  {author} {\bibinfo {author} {\bibfnamefont {A.}~\bibnamefont
  {Hegde}}, \bibinfo {author} {\bibfnamefont {A.}~\bibnamefont {Kumar}},
  \bibinfo {author} {\bibfnamefont {A.}~\bibnamefont {Agarwala}},\ and\
  \bibinfo {author} {\bibfnamefont {B.}~\bibnamefont {Muralidharan}},\
  }\href@noop {} {\bibinfo {title} {Exploring ideas in topological quantum
  phenomena: A journey through the ssh model}} (\bibinfo {year} {2021}),\
  \Eprint {https://arxiv.org/abs/2108.01460} {arXiv:2108.01460
  [cond-mat.mes-hall]} \BibitemShut {NoStop}%
\bibitem [{\citenamefont {Moustaj}\ \emph {et~al.}(2025)\citenamefont
  {Moustaj}, \citenamefont {Krebbekx},\ and\ \citenamefont
  {Morais~Smith}}]{Cristiane_polE_2025}%
  \BibitemOpen
  \bibfield  {author} {\bibinfo {author} {\bibfnamefont {A.}~\bibnamefont
  {Moustaj}}, \bibinfo {author} {\bibfnamefont {J.}~\bibnamefont {Krebbekx}},\
  and\ \bibinfo {author} {\bibfnamefont {C.}~\bibnamefont {Morais~Smith}},\
  }\bibfield  {title} {\bibinfo {title} {Anomalous polarization in
  one-dimensional aperiodic insulators},\ }\bibfield  {journal} {\bibinfo
  {journal} {Condensed Matter}\ }\textbf {\bibinfo {volume} {10}},\ \href
  {https://doi.org/10.3390/condmat10010003} {10.3390/condmat10010003} (\bibinfo
  {year} {2025})\BibitemShut {NoStop}%
\bibitem [{\citenamefont {Mondal}\ \emph
  {et~al.}(2023{\natexlab{c}})\citenamefont {Mondal}, \citenamefont {Pachhal},\
  and\ \citenamefont {Agarwala}}]{Mondal_PRB_2023}%
  \BibitemOpen
  \bibfield  {author} {\bibinfo {author} {\bibfnamefont {S.}~\bibnamefont
  {Mondal}}, \bibinfo {author} {\bibfnamefont {S.}~\bibnamefont {Pachhal}},\
  and\ \bibinfo {author} {\bibfnamefont {A.}~\bibnamefont {Agarwala}},\
  }\bibfield  {title} {\bibinfo {title} {Percolation transition in a
  topological phase},\ }\href {https://doi.org/10.1103/PhysRevB.108.L220201}
  {\bibfield  {journal} {\bibinfo  {journal} {Phys. Rev. B}\ }\textbf {\bibinfo
  {volume} {108}},\ \bibinfo {pages} {L220201} (\bibinfo {year}
  {2023}{\natexlab{c}})}\BibitemShut {NoStop}%
\bibitem [{\citenamefont {Fidkowski}(2010)}]{Fidkowski_PRL_2010}%
  \BibitemOpen
  \bibfield  {author} {\bibinfo {author} {\bibfnamefont {L.}~\bibnamefont
  {Fidkowski}},\ }\bibfield  {title} {\bibinfo {title} {Entanglement spectrum
  of topological insulators and superconductors},\ }\href
  {https://doi.org/10.1103/PhysRevLett.104.130502} {\bibfield  {journal}
  {\bibinfo  {journal} {Phys. Rev. Lett.}\ }\textbf {\bibinfo {volume} {104}},\
  \bibinfo {pages} {130502} (\bibinfo {year} {2010})}\BibitemShut {NoStop}%
\bibitem [{\citenamefont {Turner}\ \emph {et~al.}(2011)\citenamefont {Turner},
  \citenamefont {Pollmann},\ and\ \citenamefont {Berg}}]{Turner_PRB_2011}%
  \BibitemOpen
  \bibfield  {author} {\bibinfo {author} {\bibfnamefont {A.~M.}\ \bibnamefont
  {Turner}}, \bibinfo {author} {\bibfnamefont {F.}~\bibnamefont {Pollmann}},\
  and\ \bibinfo {author} {\bibfnamefont {E.}~\bibnamefont {Berg}},\ }\bibfield
  {title} {\bibinfo {title} {Topological phases of one-dimensional fermions: An
  entanglement point of view},\ }\href
  {https://doi.org/10.1103/PhysRevB.83.075102} {\bibfield  {journal} {\bibinfo
  {journal} {Phys. Rev. B}\ }\textbf {\bibinfo {volume} {83}},\ \bibinfo
  {pages} {075102} (\bibinfo {year} {2011})}\BibitemShut {NoStop}%
\bibitem [{\citenamefont {Adami}\ and\ \citenamefont
  {Cerf}(1997)}]{Adami_PRA_1997}%
  \BibitemOpen
  \bibfield  {author} {\bibinfo {author} {\bibfnamefont {C.}~\bibnamefont
  {Adami}}\ and\ \bibinfo {author} {\bibfnamefont {N.~J.}\ \bibnamefont
  {Cerf}},\ }\bibfield  {title} {\bibinfo {title} {von neumann capacity of
  noisy quantum channels},\ }\href {https://doi.org/10.1103/PhysRevA.56.3470}
  {\bibfield  {journal} {\bibinfo  {journal} {Phys. Rev. A}\ }\textbf {\bibinfo
  {volume} {56}},\ \bibinfo {pages} {3470} (\bibinfo {year}
  {1997})}\BibitemShut {NoStop}%
\bibitem [{\citenamefont {Wolf}\ \emph {et~al.}(2008)\citenamefont {Wolf},
  \citenamefont {Verstraete}, \citenamefont {Hastings},\ and\ \citenamefont
  {Cirac}}]{Wolf_PRL_2008}%
  \BibitemOpen
  \bibfield  {author} {\bibinfo {author} {\bibfnamefont {M.~M.}\ \bibnamefont
  {Wolf}}, \bibinfo {author} {\bibfnamefont {F.}~\bibnamefont {Verstraete}},
  \bibinfo {author} {\bibfnamefont {M.~B.}\ \bibnamefont {Hastings}},\ and\
  \bibinfo {author} {\bibfnamefont {J.~I.}\ \bibnamefont {Cirac}},\ }\bibfield
  {title} {\bibinfo {title} {Area laws in quantum systems: Mutual information
  and correlations},\ }\href {https://doi.org/10.1103/PhysRevLett.100.070502}
  {\bibfield  {journal} {\bibinfo  {journal} {Phys. Rev. Lett.}\ }\textbf
  {\bibinfo {volume} {100}},\ \bibinfo {pages} {070502} (\bibinfo {year}
  {2008})}\BibitemShut {NoStop}%
\bibitem [{\citenamefont {Peschel}\ and\ \citenamefont
  {Eisler}(2009)}]{Peschel_JPA_2009}%
  \BibitemOpen
  \bibfield  {author} {\bibinfo {author} {\bibfnamefont {I.}~\bibnamefont
  {Peschel}}\ and\ \bibinfo {author} {\bibfnamefont {V.}~\bibnamefont
  {Eisler}},\ }\bibfield  {title} {\bibinfo {title} {Reduced density matrices
  and entanglement entropy in free lattice models},\ }\href
  {https://doi.org/10.1088/1751-8113/42/50/504003} {\bibfield  {journal}
  {\bibinfo  {journal} {Journal of Physics A: Mathematical and Theoretical}\
  }\textbf {\bibinfo {volume} {42}},\ \bibinfo {pages} {504003} (\bibinfo
  {year} {2009})}\BibitemShut {NoStop}%
\end{thebibliography}%
	
\end{document}